%% file: ArxivVol1.tex
\pdfoutput=1
\documentclass[pra,aps,12pt,a4wide,floatfix,notitlepage,nofootinbib,superscriptaddress,longbibliography]{revtex4-2}

\input{macroHK.tex}

\input{Preambule.tex}

\newcommand{\hk}[1]{{#1}}

\usepackage{xcite}
\usepackage{xr-hyper}

\usepackage{geometry}
\usepackage[breaklinks=true]{hyperref}
\hypersetup{
    colorlinks,
    linkcolor={red!50!black},
    citecolor={blue!90},
    urlcolor={blue!80!black}
}

\renewcommand{\theequation}{I.\arabic{equation}}

\begin{document}

\title{A low-energy effective Hamiltonian for Landau quasiparticles:\\ I A unified theory of transport and superfluidity in Fermi liquids}

\author{Pierre-Louis Taillat }
\email{pierre-louis.taillat@sorbonne-universite.fr}
\affiliation{Laboratoire de Physique Théorique de la Matière Condensée, Sorbonne Université, CNRS, 75005, Paris, France}
\author{Hadrien Kurkjian}
\email{hadrien.kurkjian@cnrs.fr}
\affiliation{Laboratoire de Physique Théorique de la Matière Condensée, Sorbonne Université, CNRS, 75005, Paris, France}

\begin{abstract}
We introduce a new renormalisation scheme to construct the Landau quasiparticles of Fermi fluids.
The scheme introduces an energy cutoff $\Lambda$ to remove the resonant couplings, enabling the dressing of the particles into quasiparticles via a unitary transformation.
The dynamics of the quasiparticles is then restricted to low-energy transitions and is
fully determined by an effective Hamiltonian which unifies  the Landau function $f$, the pair interaction $g$ responsible for superfluidity, and the collision amplitude $\mathcal{A}$ responsible for transport and equilibration. Studying the flow equation that results from infinitesimal variations of the cutoff, 
we recover the Bethe-Salpeter relation between $f$ and the forward limit of $\mathcal{A}$, and we demonstrate
an analogue relation between $g$ and the frontal limit of $\mathcal{A}$. 
We show that our effective theory captures all the low-energy phenomena of Fermi liquids, from the equation of state to the
transport properties, both in the normal and in the superfluid phase. We apply it
to the calculation of non-Fermi liquid corrections to the quasiparticle lifetime.
This publication is continued by Ref.~\cite{VolumeII} where we apply the effective
theory to a Fermi fluid of ultracold atoms.
\end{abstract}

\date{\today}
\maketitle

\vspace{\baselineskip}

\tableofcontents

\section*{Introduction}
\label{sec:intro}
\input{heff_IntroEN}

\input{heff_HRG.tex}
\input{heff_flot}
\input{heff_eqtransport}

\input{heff_icoll}
\input{heff_BCS}

\input{heff_thermalcorr}

\section*{Conclusion}
\input{heff_Conclusion}

\section*{Acknowledgements}
Fruitful discussions with Nicolas Dupuis are gratefully acknowledged.
H.K. acknowledges support from the French Agence Nationale de la Recherche (ANR), under grant ANR-23-ERCS-0005 (project DYFERCO).

\renewcommand{\theequation}{I.\Alph{section}\arabic{equation}}
\begin{appendix}
    \input{heff_formulaire.tex}
    \input{heff_appflotEN.tex}
\end{appendix}

\bibliography{biblio.bib}

\end{document}

%% file: macroHK.tex
\newcommand{\be}{\begin{equation}}
\newcommand{\ee}{\end{equation}}
\newcommand{\bea}{\begin{eqnarray}}
\newcommand{\eea}{\end{eqnarray}}

\newcommand{\meanv}[1]{\langle #1 \rangle}

\newcommand{\meanvlr}[1]{\left\langle #1 \right\rangle}

\newcommand{\bb}[1]{\left( #1 \right)}

\newcommand{\bbcro}[1]{\left[ #1 \right]}

\newcommand{\ii}{\textrm{i}}
\newcommand{\dd}{\textrm{d}}
\newcommand{\eee}{\textrm{e}}
\newcommand{\rr}{\textbf{r}}

\newcommand{\qq}{\textbf{q}}

\newcommand{\vv}{\textbf{v}}

\newcommand{\pp}{\textbf{p}}

\newcommand{\zero}{\textbf{0}}
\newcommand{\PP}{\textbf{P}}

\newcommand{\overlinef}[1]{\overline{\vphantom{\pp'}#1}}

\newcommand{\ket}[1]{|#1\rangle}
\newcommand{\bra}[1]{\langle #1|}

\newcommand{\kF}{k_{\rm F}}

\newcommand{\TF}{T_{\rm F}}

\newcommand{\pF}{p_{\rm F}}
\newcommand{\vF}{v_{\rm F}}
\newcommand{\EF}{\epsilon_{\rm F}}
\newcommand{\FS}{\ket{{\rm FS}}}

\newcommand{\upa}{\uparrow}
\newcommand{\dwa}{\downarrow}

\newcommand{\eqqref}[1]{Eq.~\eqref{#1}}
\newcommand{\eqqrefs}[2]{Eqs.~\eqref{#1}--\eqref{#2}}

\usepackage{tikz}
\usetikzlibrary{shapes.geometric, positioning, shadings, arrows.meta}
\usetikzlibrary{calc,patterns,decorations.pathmorphing,decorations.markings}

\tikzset{->-/.style={decoration={
  markings,
  mark=at position .5 with {\arrow{>}}},postaction={decorate}}}
\tikzset{-<-/.style={decoration={
  markings,
  mark=at position .5 with {\arrow{<}}},postaction={decorate}}}
\tikzset{phantom->-/.style={decoration={
  markings,
  mark=at position .5 with {\arrow[scale=2]{>}}},postaction={decorate}}}

\tikzset
  {
    ,my bubble/.style = 
      {
        ,draw=#1!70
        ,fill=#1!10
        ,ellipse
        ,inner sep=1pt
        ,minimum width=2em
        ,minimum height=2em
        ,align=center
      }
    ,my end/.style =
      {
        ,draw=#1!70
        ,top color=#1!10
        ,bottom color=#1!50
        ,minimum height=6em
        ,text width=6em
        ,inner sep=0pt
        ,align=center
      }
    ,my arrow/.style =
      {
        ,>=Stealth
        ,->
        ,draw=black
      }
  }
  
\tikzset{serpent/.style={decoration={snake},postaction={decorate}}}

%% file: Preambule.tex
\usepackage{amsmath}
\usepackage{feynmp-auto}
\usepackage[compat=1.1.0]{tikz-feynman}
\usepackage[mathscr]{euscript}
 \let\mathscr\relax 
\usepackage[scr]{rsfso}
\usepackage{amsthm}
\usepackage{cancel}
\usepackage{amssymb}
\usepackage[normalem]{ulem}
\usepackage{soul}
\soulregister\cite7
\soulregister\ref7

\usetikzlibrary{shapes.geometric, positioning, shadings, arrows.meta}
\usetikzlibrary{calc,patterns,decorations.pathmorphing,decorations.markings}

\tikzset{->-/.style={decoration={
  markings,
  mark=at position .5 with {\arrow{>}}},postaction={decorate}}}
\tikzset{-<-/.style={decoration={
  markings,
  mark=at position .5 with {\arrow{<}}},postaction={decorate}}}
\tikzset{phantom->-/.style={decoration={
  markings,
  mark=at position .5 with {\arrow[scale=2]{>}}},postaction={decorate}}}

\tikzset
  {
    ,my bubble/.style = 
      {
        ,draw=#1!70
        ,fill=#1!10
        ,ellipse
        ,inner sep=1pt
        ,minimum width=2em
        ,minimum height=2em
        ,align=center
      }
    ,my end/.style =
      {
        ,draw=#1!70
        ,top color=#1!10
        ,bottom color=#1!50
        ,minimum height=6em
        ,text width=6em
        ,inner sep=0pt
        ,align=center
      }
    ,my arrow/.style =
      {
        ,>=Stealth
        ,->
        ,draw=black
      }
  }
  
\tikzset{serpent/.style={decoration={snake},postaction={decorate}}}

\counterwithin*{paragraph}{subsection}

%% file: heff_IntroEN.tex
Originally formulated as a phenomenological theory, Fermi liquid theory is based 
on a quadratic action \cite{Landau1956}, in which fermionic quasiparticles are 
described by a semiclassical density field $\delta n$ fluctuating about the 
Fermi sea and interacting through a function $f$.
The physical origin of quasiparticles is not elucidated; their existence is merely justified
by the heuristic assumption that the noninteracting states
can be adiabatically followed when interactions are switched on \cite{NozieresPines1966}.
The semi-classical action expresses the energy of the fluid in terms of $\delta n$
but says nothing about the equilibration of the distribution
and therefore fails to capture the ergodic, dissipative
dynamics. To overcome this serious limitation,
Landau derived \cite{Landau1959} a Bethe-Salpeter equation that expresses 
the probability amplitude $\mathcal{A}$ that quasiparticles collide in the forward direction
(i.e. with a vanishing transferred momentum)
in terms of the interaction function $f$. The Bethe-Salpeter equation
is derived from a microscopic theory \cite{Landau1959,Luttinger1962Formal,Luttinger1962Equilibrium} 
of the Fermi liquid, where quasiparticle properties are related to the particle correlation
functions via a residue.

%

In a more modern perspective, Landau’s theory has been reinterpreted as a 
low-energy effective theory emerging from a renormalization process \cite{Popov1987,Shankar1994,Polchinski1992,Senechal1995,Senechal1998,Dupuis1998}. 
This moves beyond the phenomenological nature of the theory and provides 
it with a fundamental justification. In the renormalization picture, the quasiparticle 
energies and interactions arise from the progressive integration of the high-energy degrees of freedom. 
Although the renormalization group generates in principle a complete effective action for the quasiparticle field \cite{Dupuis1998,Son2022,Ma2024}, in practice 
one rarely includes the collision amplitude in the effective picture, which restricts the description of the dynamics to the collisionless regime.
This limitation is particularly detrimental in three-dimensional (3D) Fermi liquids, where resonant collisions between quasiparticles of the Fermi surface 
depend on two independent angles and thus cannot be integrally recovered through a Bethe-Salpeter equation from the interaction function
$f$  (which depends only on the angle between the quasiparticle momenta $\pp$ and $\pp'$).
In this respect, the 3D case is fundamentally different from its 2D counterpart, where resonant collisions depend on a single angle and thus fall either into the forward channel, or into the pairing channel (where the  angle of incidence approaches $\pi$ and the center-of-mass momentum vanishes) \cite{MacDonald1996,Novikov2006,Hofmann2023,Hofmann2025}.

In fact, a convincing low-energy effective theory should be able to describe, within a unified formalism, all low-energy phenomena, from the low-temperature thermodynamics to the hydrodynamic equations, in both the normal and superfluid phases (provided that superfluidity itself remains a low-energy phenomenon).
In this work, we construct (in Sec.~\ref{sec:HRG}) an effective Hamiltonian that captures the full dynamics of Landau quasiparticles, and thereby the whole low-energy physics of fermionic fluids in which these quasiparticles are well defined. Our formalism relies on a unitary transformation that connects the quasiparticle states to the noninteracting Fock states by excluding from the dressing any quasidegenerate state in an energy band of width $\Lambda$.
This Schrieffer-Wolff transformation,  common in atomic physics \cite{Cohen,VanVleck1929,Wolff1966} when one applies a perturbation to multiplets of quasidegenerate energy levels, band-diagonalizes the Hamiltonian, thereby decoupling levels whose energy separation exceeds $\Lambda$. As the renormalization flow generates
terms of high-order in the quasiparticle field $\hat{\gamma}$, we expand the Hamiltonian in powers of the fluctuations $\delta(\hat{\gamma}^\dagger\hat{\gamma})$ of the density field about its expectation value in the quasiparticle Fermi sea.
The effective Hamiltonian obtained in this way is not limited to specific interaction channels, nor to a gradient 
expansion: its diagonal part in the Fock basis coincides with Landau’s semi-classical Hamiltonian, but its off-diagonal part contains the generic collision amplitude $\mathcal{A}$, as well as a pair interaction $g$.

For infinitesimal variations of the cutoff, we obtain (in Sec.~\ref{sec:flot}) a Continuous Unitary Transformation 
(CUT) \cite{Wolff1966,Wegner2002,Kehrein2006}; as in Ref.~\cite{Wilson1993}, the generator 
of our Schrieffer-Wolff transformation has matrix elements only between states whose energy separation is between $\Lambda$ 
and $\Lambda-\dd\Lambda$. While the flow of the general collision amplitude converges when $\Lambda\ll \EF$,
one must push to $\Lambda\ll\vF q$ in the forward and pairing channels, where
$\vF$ is the Fermi velocity and $q$ the vanishing momentum corresponding to the channel.
The integration of the flow in the forward channel recovers the Bethe-Salpeter
equation of Landau; the flow in the pairing channel yields a second Bethe-Salpeter equation,  
relating the amplitude of frontal collisions to the pairing amplitude $g$, through
an integral equation with a logarithmic angular kernel. This is relation is an original result to
the best of our knowledge.

To conserve the quasiparticle picture of the transformed states, we keep the
cutoff of our Schrieffer-Wolff decomposition well above the quasiparticle damping rate $\Gamma$.
This postpones the study of the quasiparticle dynamics to a second stage.
In a Fermi liquid, the density of excited quasiparticles
vanishes as the temperature $T$ leading to a slow and weakly-correlated quasiparticle dynamics, which
can be described using standard quantum kinetic theory \cite{Bonitz1998,Kira2015}.
This allows us (in Sections.~\ref{sec:demotransport} and \ref{sec:icoll}) to derive the Boltzmann equation—including the collision integral—
directly from the Heisenberg equation of the quasiparticle density field $\hat \gamma^\dagger \hat \gamma$.
We use the Born–Markov approximation to truncate the Bogoliubov–Born–\allowbreak Green–\allowbreak Kirkwood–\allowbreak Yvon 
(BBGKY) hierarchy in the quasiparticle picture, even though the problem is strongly correlated in the particle picture. 
Looking towards the superfluid phase in Sec.~\ref{sec:BCS}, we explain also how to describe the superfluid transition
using the effective Hamiltonian. One needs to properly renormalize the pair interaction $g$ 
to remove a $\text{ln}\,\Lambda$ cancellation \cite{Khalatnikov1957,Popov1987,Zhang1992,Senechal1995}, such that the superfluid observables 
are set by a renormalized pairing strength $G$, independent of $\Lambda$.
Finally, in Sec.~\ref{sec:thermalcorr}, we explain how thermal corrections
to Fermi liquid theory are set by the behavior of the collision amplitude at the
overlap of the forward and pairing channel, that is for processes of the form
$\pp,-\pp\to-\pp,\pp$ or $\pp,-\pp\to\pp,-\pp$. The Bethe-Salpeter equations are thus essential
to the prediction of these thermal corrections.

The present article is the first part of a two-part publication. It is devoted the construction of a general effective theory for Fermi liquids. In the second
part \cite{VolumeII}, we shall apply the effective theory to an ultracold Fermi gas with contact interactions. To guide the reader through the notations of these two parts a table of symbols is available in Appendix \ref{formulaire}.

%% file: heff_HRG.tex
\section{The low-energy effective Hamiltonian of Fermi liquids}
\label{sec:HRG}

\subsection{Generic Hamiltonian of interacting fermions}
We consider a 3D fluid made of $N$ fermionic particles in two distinguishable
states labelled $\upa$ and $\dwa$ by analogy with a spin-$1/2$. The fluid evolves in a volume $\mathcal{V}$ under the Hamiltonian
\be
\hat H =\hat H_0+\hat V
\ee
This section will discuss the construction of the Landau quasiparticles on general grounds, so we make minimal
assumptions on the form of $\hat H$. We write the generic noninteracting Hamiltonian and the generic two-body interaction as
\be
\hat H_0=\sum_{\alpha\in\mathcal{D},\sigma}\omega_{\alpha\sigma} \hat a_{\alpha\sigma}^\dagger \hat a_{\alpha\sigma}
\ee
\be
\hat V=\frac{1}{2}\sum_{\substack{\alpha\beta\gamma\delta\in\mathcal{D}\\\sigma\sigma'=\upa\dwa}} V_{\sigma\sigma'}(\alpha,\beta|\gamma,\delta) \hat a_{\alpha\sigma}^\dagger \hat a_{\beta\sigma'}^\dagger \hat a_{\gamma\sigma'} \hat a_{\delta\sigma} \label{Vnu}
\ee
where $\hat a_{\alpha\sigma}$ annihilates a fermion of spin $\sigma$ in mode $\alpha$, and $\mathcal{D}$ is the
set of modes of $\hat H_0$. The mean density of the gas $\rho=N/\mathcal{V}$ (where $N$ is the total
number of fermions) is fixed by a chemical potential $\mu$, which deviates from the chemical potential 
of the zero-temperature ideal gas, \textit{i.e.}, the Fermi energy $\EF$.

We use $\hbar=k_{\rm B}=1$ throughout this article. This implies that momenta $p$ 
and wavenumber $k$ are not differentiated, in particular $\pF=\kF$ for the Fermi momentum/wavenumber.

Throughout this article, we consider the spin-symmetric case where 
\be \label{spinsym}
V_{\upa\upa}=V_{\dwa\dwa},\quad V_{\dwa\upa}=V_{\upa\dwa}
\ee
The opposite-spin interaction may then be chosen
symmetric under the simultaneous exchange $\alpha\leftrightarrow\beta$ and $\gamma\leftrightarrow\delta$
\be
V_{\upa\dwa}(\alpha,\beta|\gamma,\delta)=V_{\upa\dwa}(\beta,\alpha|\delta,\gamma)
\ee
In the case of a true spin-$1/2$ fluid (such as a ${}^3$He liquid or an electron gas),
the bare potential \eqqref{Vnu} preserves in addition the SU(2) symmetry. 
In our notations, this translates into
\be \label{SU2}
V_{\upa\dwa}(\alpha,\beta|\delta,\gamma)-V_{\upa\dwa}(\alpha,\beta|\gamma,\delta)-V_{\upa\upa}(\alpha,\beta|\gamma,\delta)\underset{\text{SU(2)}}{=}0
\ee
In the case of an effective spin-$1/2$, \textit{e.g.} when $\upa$ and $\dwa$ represent two internal
states of a fermionic atom, the SU(2) symmetry can be violated. It is however preserved
when $V_{\upa\dwa}=$ constant and $V_{\upa\upa}=0$, and hence for the $s$-wave contact potential.

\subsection{Introduction of a cutoff on energy transitions}

The quasiparticles states are often viewed \cite{NozieresPines1966,BaymPethick}
as the states in which the eigenstates of $\hat H_0$ evolve after an adiabatic ramp
of the interactions of the form $\hat V(t)=\lambda(t)\hat V$, with $\lambda(0)=0$ and $\lambda(t_{\rm f})=1$.
It is then argued that the ramping time $t_{\rm f}$ \cite{NozieresPines1966}
should be long enough to ensure an adiabatic evolution, but short enough
to prevent the quasiparticle from decaying.
This picture is problematic since the existence of a finite time $t_{\rm f}$ fulfilling
the adiabatic theorem \cite{Griffiths2004} is questionable in a gapless, strongly-interacting fluid.
Instead, we develop here a rigorous method to construct the quasiparticles states
from the eigenstates of $\hat H_0$, and to continuously
follow them from the non-interacting to the strongly-interacting regime. 

\begin{figure}[htb]
\includegraphics[width=\textwidth]{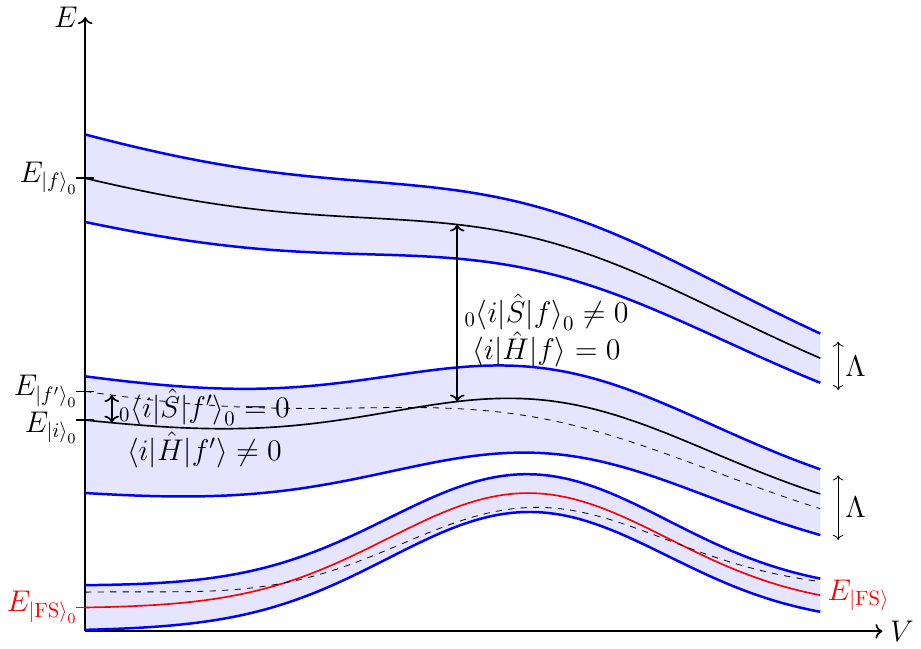}
\caption{\label{schema} Construction of the quasiparticle states through the Schrieffer-Wolff transformation.
An unpertubed Fock state $\ket{i}_0$ is dressed via the operator $\hat S$ by its interactions with the off-resonant states $\ket{f}_0$, 
whose unperturbed energy verifies $|E_{\ket{f}_0}-E_{\ket{i}_0}|>\Lambda$ (here $E_{\ket{\psi}_0}={}_0\bra{\psi}\hat H_0\ket{\psi}_0$).
The dressed state $\ket{i}$ can then be followed adiabatically as the interaction strength $V$ increases. However, due to its incomplete dressing,
it is not an eigenstate of $\hat H$, and it remains coupled to the nearly degenerate states $\ket{f'}$ of energies $|E_{\ket{f'}}-E_{\ket{i}}|<\Lambda$,
(here $E_{\ket{\psi}}=\bra{\psi}\hat H\ket{\psi}$). This construction applies in particular to the particle Fermi sea $\FS_0$, which evolves into
a quasiparticle Fermi sea $\FS$ (red curve).}
\end{figure}
Given an eigenstate $\ket{n}_0$ of $\hat H_0$,
we decompose the rest of the eigenstates of $\hat H_0$ into \textit{quasidegenerate} and \textit{energetically
well-separated} states. An eigenstate $\ket{m}_0$ is quasidegenerate with $\ket{n}_0$ if its energy $E_m^0$ 
is within a narrow energy band $\Lambda$, $|E_n^0-E_m^0|<\Lambda$, and it is well-separated if $|E_n^0-E_m^0|>\Lambda$.
One can then construct the quasiparticle states by a unitary transformation $\ket{n}=\eee^{\hat S}\ket{n}_0$,
where we impose that the hermitian operator $\hat S$ has no matrix elements between quasidegenerate states.
This construction is similar to the van Vleck or Schrieffer-Wolff transformation in quasi-degenerate perturbation theory \cite{Cohen,VanVleck1929,Wolff1966,Redmon1980}.

Rather than an adiabaticity condition, the possibility of such a construction is tied to a \textit{non-crossing}
condition: the only level crossings\footnote{Our acceptation of level crossings is restricted to states $\ket{n}_0$ 
and $\ket{m}_0$ that are significantly coupled by $\hat{V}$. In the low-temperature limit, this includes
only states that differ by a few quasiparticle excitations.} that occur as interactions are increased should be between states $\ket{n}$
and $\ket{m}$ that are already quasidegenerate in the non-interacting state, \textit{i.e.}
\be
|E_n-E_m|\ll\Lambda \iff  |E_n^0-E_m^0|\ll\Lambda
\label{noncrossing}
\ee
This is a reformulation of the usual assumption that the quasiparticles have a
gapless spectrum similar to the spectrum of the particles in the ideal Fermi gas.

The quasiparticle states $\ket{n}$ are not the exact eigenstates of $\hat H$ since there remains
quasi on-shell couplings between them, $\bra{n}\hat H\ket{m}\neq 0$ if $|E_n-E_m|<\Lambda$. These couplings
ensure that the quasiparticle states, which are 
described by the same quantum numbers as the noninteracting 
states (\textit{i.e.} the set of fermionic occupation numbers $\{n_{\alpha\sigma}\}_{\alpha\sigma}$),
decay to the true eigenstates of the ergodic system \cite{Olshanii2008}.
{The quasiparticle gas is thus a nearly-integrable system \cite{Deutsch2018},
with the rare quasiparticle collisions acting as the integrability-breaking
perturbation. In this picture the eigenstates should be viewed as random mixtures 
of all quasiparticle states around the energy $E$, in accordance with the Eigenstate Thermalization Hypothesis.}

{In this formulation of FLT as a low-energy effective theory,
the calculation of the physical observables is conceptually divided
in two steps. First the microscopic parameters are renormalized
into effective parameters, typically an effective mass and an effective
interaction. Unlike other renormalization approaches
\cite{Shankar1994,Polchinski1992,Senechal1995,Senechal1998,Dupuis1998}
which arrive directly to the response functions (although often limited to the collisionless
regime), our Hamiltonian renormalization acts on the quasiparticle states and the matrix
elements of $\hat H$, i.e. purely static objects. Most of the effective parameters will reach a stationary value when
\be
\Lambda\ll \EF
\label{EFLambda}
\ee
This is however not always the case, and $\Lambda$ may still
appear as an infrared cutoff in some quantities, such as the pair interaction
$g_{\sigma\sigma'}$ (see Sections.~\ref{sec:flot} and \ref{sec:BCS}).

After the renormalization, there remains to study the dynamics described by the effective Hamiltonian. 
This second step of the calculation is done in Sec.~\ref{sec:demotransport} using quantum kinetic technics \cite{Bonitz1998}. 
During this phase, the couplings between the quasiparticle states
feel $\Lambda$ as an ultraviolet cutoff.
To remain in the standard case of a Markovian kinetic equation,
this cutoff should be well above the 
frequencies at which the system evolves.
In the case of an isolated system prepared in some excited quasiparticle state, 
the evolution frequencies are set by the intrinsic decay rate $\Gamma_{\rm typ}$ of the quasiparticles.
In the case of a driven system, the external force sets a typical frequency $\omega_{\rm ext}$. Thus,
\be
\Gamma_{\rm typ},\ \omega_{\rm ext}\ll \Lambda
\label{GammaLambda}
\ee
Tuning $\Lambda$ below this lower bound would
result in a non Markovian kinetic equation,
as the renormalized states can no longer
be interpreted as quasiparticle Fock states.

The inequalities \eqqref{EFLambda}--\eqref{GammaLambda}
require a separation of energy scales $\Gamma_{\rm typ},\ \omega_{\rm ext}\ll\EF$, and
constrain the energies at which one
can excite the system without breaking the quasiparticle description.}

\subsection{Partition of the Hilbert space and unitary transformation}
\label{unitaire}

To classify the quasidegenerate and well-separated states, we introduce the projectors
\begin{equation}  \label{PiLambda}
    \hat{P}_\Lambda(E) = \sum_{\ket{n}_0} \Pi_{\Lambda}(E-\hat H_0) \ket{n}_0 \bra{n}_0, \qquad \hat Q_\Lambda(E)=1-\hat{P}_\Lambda(E)
\end{equation}
where the summation runs over the eigenstates $\ket{n}_0$ of $\hat H_0$. 
As long as the filtering function
$\Pi_{\Lambda}(E)$ verifies that $\Pi_{\Lambda} (E\ll\Lambda)=1$
and $\Pi_{\Lambda} (E\gg\Lambda)=0$, its precise shape does not matter. We
shall thus use
\begin{equation} \label{PiLambdaHRG}
    \Pi_{\Lambda}(E) = 
    \begin{cases}
        1 \ \text{ if } \ |E|\leq \Lambda \\
        0 \ \text{ else}
    \end{cases}
\end{equation}


We then construct an antihermitian operator $\hat S$
\be
\hat S^\dagger=-\hat S
\ee
which generates the quasiparticle state
by a unitary transform applied to $\ket{n}_0$:
\be
\ket{n}=\eee^{\hat S}\ket{n}_0 \label{nn0}
\ee
The operator $\hat S$ couples only well-separated states, \textit{i.e.}
its diagonal band vanish
\be
\hat{P}_\Lambda(E) \hat S \hat{P}_\Lambda(E)=0, \label{PSP}
\ee
To construct the off-diagonal elements $\hat P_\Lambda \hat S Q_\Lambda$,
we impose that the couplings between 
transformed states vanish: $\bra{m}\hat H\ket{n}=0$,
if $\ket{n}$ and $\ket{m}$ are off-resonant states,
that is, if $|E_n^0-E_m^0|>\Lambda$.
In other words, we impose that the effective
Hamiltonian
\be
\hat{H}_{\rm eff}=\eee^{-\hat{S}} \hat{H} \eee^{\hat{S}} \label{HeffS}
\ee
is band-diagonal:
\be
\hat{P}_\Lambda(E) \hat{H}_{\rm eff}\hat Q_\Lambda(E)=0, \label{PHP}
\ee

At this stage, we view the effective Hamiltonian as the operator
which, acting on the unperturbed basis, provides the matrix elements
of $\hat H$ in the transformed basis:
\be
\bra{n}\hat H\ket{m}={}_0\bra{n}\hat H_{\rm eff}\ket{m}_0
\ee
Using Baker-Campbell-Hausdorff formula, 
$\hat{H}_{\rm eff}$ is expressed in terms of iterated commutators between $\hat H$ and $\hat S$:
\begin{equation}
   \hat{H}_{\rm eff} = \hat{H}  + [\hat{H},\hat{S}]+\frac{1}{2}\bbcro{[\hat{H},\hat{S}],\hat{S}}+\ldots \label{HeffBCH}
\end{equation}

\subsection{Quasiparticle states and quasiparticle Fermi sea}

The quasiparticle states are now defined as the images of the particle Fock states $\ket{\{n_{\alpha\sigma}\}}_0$ (the eigenstates of $\hat H_0$)
through the unitary transform \eqqref{nn0}
\begin{equation}
    \ket{\{n_{\alpha\sigma}\}} =  \text{e}^{\hat{S}}\ket{\{n_{\alpha\sigma}\}}_0
\end{equation}

Let us apply this transformation to
the particle Fermi sea $\ket{{\rm FS}}_0$:
\be \label{FS0}
\ket{{\rm FS}}_0=\prod_{\substack{\alpha\in\mathcal{D}\\\sigma=\upa,\dwa}} n_{\alpha\sigma}^0 \hat a_{\alpha\sigma}^\dagger\ket{0}\\
\ee
where $\ket{0}$ is the particle vacuum and
\be \label{nalpha0}
n_{\alpha\sigma}^0=\Theta(\EF-\omega_{\alpha\sigma})
\ee
are the Fermi sea occupation numbers.

The {quasiparticle Fermi sea}, which will play a central role in the expansion of the effective Hamiltonian,
is then:
\be \label{FS}
\ket{{\rm FS}}=\eee^{\hat S}\ket{{\rm FS}}_0
\ee
In general $\ket{{\rm FS}}$ is no longer the ground state of $\hat H$; its energy
\be
E_{\rm FS}=\bra{{\rm FS}}\hat H{\FS}
\ee
can be larger than the ground state energy $E_0$.

\subsection{Quasiparticle operators}

Switching to Heisenberg picture, $\hat S$ can 
be used to construct the operators
acting on the transformed basis. 
Consider an operator $\hat{O}$ whose action is known in the unperturbed basis
$\ket{\{n_{\alpha\sigma}\}}_0$. The operator $\hat{O}_\gamma$ having the same action
in the quasiparticle basis $\ket{\{n_{\alpha\sigma}\}}$ is then
\be
  \hat{{O}}_\gamma = \text{e}^{\hat{S}} \hat{{O}} \text{e}^{-\hat{S}}
  \label{OOgamma}
\ee

\paragraph{Annihilation operator} The most straightforward example is the quasiparticle annihilation operator
$\hat\gamma$ which we construct from $\hat a$ through
\be
  \hat\gamma_{\alpha\sigma} = \text{e}^{\hat{S}} \hat{{a}}_{\alpha\sigma} \text{e}^{-\hat{S}}  \label{gammaformelle}
\ee
Since $\hat\gamma$ follows from $\hat a$ through a unitary transformation,
it automatically obeys fermionic anticommutation relations
\be
\{\hat\gamma_{\alpha\sigma},\hat\gamma_{\alpha'\sigma'}^\dagger\}=\delta_{\alpha\alpha'}\delta_{\sigma\sigma'}, \qquad \{\hat\gamma_{\alpha\sigma},\hat\gamma_{\alpha'\sigma'}\}=0
\ee
Similarly, the unitary transformation guarantees that quasiparticle occupation numbers in the quasiparticle Fermi sea $\FS$
follow the Fermi-Dirac distribution \eqref{nalpha0}:
\be
\bra{{\rm FS}}\hat\gamma_{\alpha\sigma}^\dagger \hat\gamma_{\alpha\sigma}\FS=n_{\alpha\sigma}^0
\ee

\paragraph{Hamiltonian}  A second example is the operator $\hat H_\gamma$ which acts on the quasiparticle states as $\hat H$ acts on the particle states:
\be
\hat H_\gamma=\eee^{\hat S}  \hat H \eee^{-\hat S} =\sum_{\alpha\sigma}\omega_{\alpha\sigma}\hat \gamma_{\alpha\sigma}^\dagger \hat \gamma_{\alpha\sigma}+\frac{1}{2}\sum_{\substack{\alpha\beta\gamma\delta\in\mathcal{D}\\\sigma\sigma'=\upa\dwa}} V_{\sigma\sigma'}(\alpha,\beta|\gamma,\delta) \hat \gamma_{\alpha\sigma}^\dagger \hat \gamma_{\beta\sigma'}^\dagger \hat \gamma_{\gamma\sigma'} \hat \gamma_{\delta\sigma} \label{Hgamma}
\ee
The quasiparticle states, like any other state, do not evolve under the Hamiltonian $\hat H_\gamma$ but under
the true Hamiltonian $\hat H$. Inverting \eqqref{Hgamma} to express $\hat H$ in terms of $\hat H_\gamma$ allows us
to reinterpret the effective Hamiltonian \eqqref{HeffBCH}: 
\be
\hat H=\eee^{-\hat S} \hat H_\gamma \eee^{\hat S}=\hat H_{\rm eff,\gamma} \label{Heffgamma}
\ee
In other words, $\hat H$ is  written in terms of the $\hat \gamma$ operators exactly like $\hat H_{\rm eff}$ is written in terms of the $\hat a$ operators.

\paragraph{Number operator} A third example is the quasiparticle number operator
\be
\hat N_\gamma=\eee^{\hat S}  \hat N \eee^{-\hat S}=\sum_{\alpha\sigma} \hat \gamma_{\alpha\sigma}^\dagger \hat \gamma_{\alpha\sigma} \quad \text{ with }\quad  \hat N=\sum_{\alpha\sigma} \hat a_{\alpha\sigma}^\dagger \hat a_{\alpha\sigma}
\ee
This case is special since $\hat N$ commutes separately with $\hat H_0$ and $\hat V$. One can then show (order-by-order in $\hat V$)
that it commutes with $\hat S$. We recover in this way the Luttinger theorem
\be
\hat N_\gamma=\hat N \label{NNgamma}
\ee

\paragraph{Projectors} Finally, in the quasiparticle picture, the projector onto an energy-shell becomes
\begin{equation}
    \hat{P}_{\Lambda,\gamma}(E) = \eee^{\hat S}  \hat{P}_\Lambda (E) \eee^{-\hat S}= \sum_{\{n_{\alpha\sigma}\}} \Pi_{\Lambda}(E-\hat H_{0,\gamma}) \ket{\{n_{\alpha\sigma}\}} \bra{\{n_{\alpha\sigma}\}}
\end{equation}
The operators $\hat{P}_\gamma$ thus project the quasiparticle states $\ket{\{n_{\alpha\sigma}\}}$ according to their \textit{unperturbed
energy} $\sum_{\alpha\sigma} \omega_{\alpha\sigma} n_{\alpha\sigma}$, rather than their full energy
$\bra{\{n_{\alpha\sigma}\}}\hat H\ket{\{n_{\alpha\sigma}\}}$. In a generic many-fermion system,
this would render this van Vleck transformation useless. In a Fermi liquid, this
limitation is lifted by the weak-crossing
condition \eqqref{noncrossing}, which we may rewrite as 
\be
{\left\vert\bra{n}\hat H\ket{n}-\bra{m}\hat H\ket{m}\right\vert}\ll\Lambda  \iff  {\left\vert{}_0\bra{n}\hat H_0\ket{n}_0-{}_0\bra{m}\hat H_0\ket{m}_0\right\vert}\ll\Lambda 
\ee

\subsection{Energy, residue and interaction functions of the quasiparticles}

In Ref.~\cite{devvisco}, we related the energy of the quasiparticles to the average
value of $\hat H$ in quasiparticle states with one or two excitations above the Fermi sea. 
Let us here generalize this definition to any quasiparticle
Fock state $\ket{\psi}$. To this aim, we introduce states with either one quasiparticle or one quasihole (depending
on whether $n_{\alpha\sigma}^{\ket{\psi}}=\bra{\psi} \hat\gamma^\dagger_{\alpha\sigma} \hat\gamma_{\alpha\sigma}\ket{\psi}=$  0 or 1)
added to $\ket{\psi}$ in mode $\alpha\sigma$
\begin{align}
\ket{\alpha\sigma,\psi}\equiv
\begin{cases}
\ket{\psi} &\text{ if } n_{\alpha\sigma}^{\ket{\psi}}=1 \\
\hat\gamma^\dagger_{\alpha\sigma}\ket{\psi} &\text{ else}
\end{cases}
&\qquad \qquad
\ket{\overline{\alpha\sigma},\psi}\equiv
\begin{cases}
\ket{\psi} &\text{ if } n_{\alpha\sigma}^{\ket{\psi}}=0 \\
\hat\gamma_{\alpha\sigma}\ket{\psi} &\text{ else}
\end{cases}
\label{alphaalphabar}
\end{align}
The energy $\epsilon_{\alpha\sigma}^{\ket{\psi}}$ of the quasiparticle $\alpha\sigma$
is then a functional of $\ket{\psi}$ (more precisely of its occupation numbers in modes $\alpha'\sigma'\neq\alpha\sigma$):
\be
\epsilon_{\alpha\sigma}^{\ket{\psi}}\equiv\bra{\alpha\sigma,\psi} \hat H \ket{\alpha\sigma,\psi}-
\bra{\overline{\alpha\sigma},\psi} \hat H \ket{\overline{\alpha\sigma},\psi} \label{epsilonpsi}
\ee
 
To define the interaction functions $f$ in an arbitrary state $\ket{\psi}$,
one should iterate the notation \eqqref{alphaalphabar} to allow for the creation or annihilation of
two (or more) quasiparticles\footnote{
Together with this piling rule, the states obey a fermionic permutation rule: $\ket{\overline{\alpha'\sigma'},\alpha\sigma}=-\ket{\alpha\sigma,\overline{\alpha'\sigma'}}$.}
\begin{multline}
\ket{\alpha\sigma,\alpha_1\sigma_1,\ldots,\alpha_n\sigma_n,\overline{\beta_1\sigma_1},\ldots,\overline{\beta_m\sigma_m},\psi}\\
=\begin{cases}
\ket{\alpha_1\sigma_1,\ldots,\alpha_n\sigma_n,\overline{\beta_1\sigma_1},\ldots,\overline{\beta_m\sigma_m},\psi} &\text{if } n_{\alpha\sigma}^{\ket{\psi}}=1 \\
\hat\gamma^\dagger_{\alpha\sigma} \ket{\alpha_1\sigma_1,\ldots,\alpha_n\sigma_n,\overline{\beta_1\sigma_1},\ldots,\overline{\beta_m\sigma_m},\psi} &\text{else}
\end{cases}
\end{multline}
We can now define the interaction functions as functionals of $\ket{\psi}$:
\be
\frac{f_{\sigma\sigma'}^{\ket{\psi}}(\alpha,\beta)}{\mathcal{V}}=E_{\ket{\alpha\sigma,\beta\sigma'\psi}} + E_{\ket{\overlinef{\alpha\sigma},\overlinef{\beta\sigma'}\psi}}-E_{\ket{\overlinef{\alpha\sigma},\beta\sigma'\psi}}-E_{\ket{{\alpha\sigma},\overlinef{\beta\sigma'}\psi}} \label{fpsi}
\ee
where $E_{\ket{\psi}}=\bra{\psi}\hat H\ket{\psi}$. The volume factor $\mathcal{V}$ makes sure that $f_{\sigma\sigma'}$ has a finite thermodynamic limit.
Our definition is a quantum version of the semi-classical definition of $f$ as a second derivative 
$f_{\sigma\sigma'}^{\ket{\psi}}(\alpha,\beta)/\mathcal{V}=\partial^2 E_{\ket{\psi}}/\partial n_{\alpha\sigma} \partial n_{\beta\sigma'}$

In the same spirit, one can define the residue of the quasiparticle as the variation
of the number $\hat{a}_{\alpha\sigma}^\dagger \hat{a}_{\alpha\sigma}$ of \textit{particle} in mode $\alpha\sigma$ when the \textit{quasiparticle}
$\alpha\sigma$ is added to the fluid:
\be
Z_{\alpha\sigma}^{\ket{\psi}}\equiv\bra{\alpha\sigma,\psi}\hat{a}_{\alpha\sigma}^\dagger \hat{a}_{\alpha\sigma} \ket{\alpha\sigma,\psi}-\bra{\overline{\alpha\sigma},\psi}\hat{a}_{\alpha\sigma}^\dagger \hat{a}_{\alpha\sigma}\ket{\overline{\alpha\sigma},\psi}
\label{defZ}
\ee
Although conceptually important to identify the origin of the quasiparticles, the residue 
breaks the low-energy effective description, as it involves measuring a microscopic quantity $\hat{a}_{\alpha\sigma}^\dagger \hat{a}_{\alpha\sigma}$,
unlike \textit{e.g.} $\epsilon_{\alpha\sigma}$ which involves only the energy.
Thus, the low-energy properties will preferably be formulated without resorting to $Z_{\alpha\sigma}$.
In the case of an homogeneous system (where $\alpha$ stands for the wavevector $\pp$), 
we will relate this definition of the residue to the discontinuity of the momentum distribution
at the Fermi level.

Also note that using the unitary transformation of the operators \eqqref{OOgamma}, 
there is a dual interpretation of the residue as the variation of the 
number of {quasiparticle} in $\alpha\sigma$ upon adding the corresponding particle:
\be
Z_{\alpha\sigma}^{\ket{\psi}}={}_0\bra{\alpha\sigma,\psi}\hat{\gamma}_{\alpha\sigma}^\dagger \hat{\gamma}_{\alpha\sigma} \ket{\alpha\sigma,\psi}_0-{}_0\bra{\overline{\alpha\sigma},\psi}\hat{\gamma}_{\alpha\sigma}^\dagger \hat{\gamma}_{\alpha\sigma}\ket{\overline{\alpha\sigma},\psi}_0
\ee



\subsection{Collision amplitudes}
\label{collisionamplitudes}
The effective description of the Fermi fluid
is not exhaustive if we restrict ourselves to the eigenenergy $\epsilon_{\alpha\sigma}$
and interaction functions $f_{\sigma\sigma'}$ defined above.
In fact these two quantities characterize only the diagonal elements
of $\hat{H}$, while
for many equilibrium and dynamical properties, a knowledge of the 
off-diagonal elements is required:
\be \label{Aif}
\mathcal{A}_{i\to f}\equiv
\bra{f}\hat H \ket{i}
\ee

Just like the interaction functions $f_{\sigma\sigma'}$, the $2\leftrightarrow2$ transitions amplitudes\footnote{Even though it is restricted to $2\leftrightarrow2$ particle collisions, $\hat H$ can generate high-order collisions between quasiparticles. From \eqqref{HeffBCH}, one can easily count that there are up to $n+1\leftrightarrow n+1$ quasiparticle collisions if $\hat H_{\rm eff}$ is truncated to order $\hat V^n$. 
However, we shall see that $2\leftrightarrow2$ quasiparticle collisions remain the most likely
if excited quasiparticles are confined to a low-energy shell about the Fermi
level. \label{note:collisionsn}} 
depend on the reference state $\ket{\psi}$
in which we compute them. Let $\ket{i}=\ket{\overline{\alpha\sigma},\overline{\beta\sigma'},\gamma\sigma',\delta\sigma,\psi}$
be a reference state in which we made sure that quasiparticles are absent in $\alpha\sigma, \beta\sigma'$ and present in $\gamma\sigma',\delta\sigma$. Then let
\be
\ket{f}=\hat\gamma_{\alpha\sigma}^\dagger \hat\gamma_{\beta\sigma'}^\dagger \hat\gamma_{\gamma\sigma'} \hat\gamma_{\delta\sigma}\ket{i}
\qquad (\alpha\sigma,\beta\sigma')\neq(\delta\sigma,\gamma\sigma'),\,(\gamma\sigma',\delta\sigma)
\ee
be the final state not proportional to $\ket{i}$. We then define
the collision amplitude  $\mathcal{A}_{\sigma\sigma'}$ between $\sigma$ and $\sigma'$ quasiparticles through
\be
\mathcal{A}_{i\to f}\equiv\frac{\mathcal{A}_{\sigma\sigma'}^{\ket{i}}(\alpha\beta|\gamma\delta)}{\mathcal{V}} \label{Apsi} 
\ee
From the fermionic commutation relations and the hermiticity of the Hamiltonian, these amplitudes verify the relations
\bea
\mathcal{A}_{\sigma\sigma'}^{\ket{i}}(\delta,\gamma|\beta,\alpha)&=&\mathcal{A}_{\sigma\sigma'}^{\ket{i}}(\alpha,\beta|\gamma,\delta) \label{pteA2}\\
\mathcal{A}_{\upa\upa}^{\ket{i}}(\beta,\alpha|\gamma,\delta)&=&\mathcal{A}_{\upa\upa}^{\ket{i}}(\alpha,\beta|\delta,\gamma)=-\mathcal{A}_{\upa\upa}^{\ket{i}}(\alpha,\beta|\gamma,\delta) \label{pteA3}
\eea
Using the spin-exchange symmetry \eqqref{spinsym} for $\mathcal{A}_{\upa\dwa}$, they verify in addition
\be
\mathcal{A}_{\sigma\sigma'}^{\ket{i}}(\beta\alpha|\delta\gamma)=\mathcal{A}_{\sigma\sigma'}^{\ket{i}}(\alpha\beta|\gamma\delta)\label{pteA1}
\ee
Remember that our van Vleck transformation a priori restricts $\mathcal{A}_{\sigma\sigma'}^{\ket{i}}(\alpha\beta|\gamma\delta)$ to transitions
between quasi-degenerate states $|E_f-E_i|<\Lambda$. Comparing $E_i=\bra{i}\hat H\ket{i}$ and $E_f=\bra{f}\hat H\ket{f}$,
this can be turned in the thermodynamic limit\footnote{One can show that
$E_{{f}}-E_{{i}}=\epsilon_{\alpha\sigma}^{\ket{i}}+\epsilon_{\beta\sigma'}^{\ket{i}}-\epsilon_{\gamma\sigma'}^{\ket{i}}-\epsilon_{\delta\sigma'}^{\ket{i}}+O\bb{\frac{1}{\mathcal{V}}}$} into a $\Lambda$-resonance condition on the energies (in $\ket{i}$) of the colliding quasiparticles:
\be
\left\vert\epsilon_{\alpha\sigma}^{\ket{i}}+\epsilon_{\beta\sigma'}^{\ket{i}}-\epsilon_{\gamma\sigma'}^{\ket{i}}-\epsilon_{\delta\sigma}^{\ket{i}}\right\vert<\Lambda
\ee

\subsection{Low-energy effective Hamiltonian in the vicinity of the quasiparticle Fermi sea}

So far, we have described the matrix elements of $\hat H$ between arbitrary
quasiparticle states, noticing that even if we restrict to few-quasiparticle transitions
the matrix elements retain a dependence on the reference state $\ket{\psi}$.
This can be seen as a consequence of \eqqref{Heffgamma}, where the expression of $\hat H$
 (at strong coupling) contains an infinite number of $\hat\gamma$.

One can however derive a tractable truncation of $\hat H$, containing few operators
$\hat\gamma$, and valid for quasiparticle states $\ket{\{n_{\alpha\sigma}\}}$
that deviate from the Fermi sea $\FS$ only at low energy.
In the followings, we assume that $\FS$ is the default reference state and we drop the superscript in the effective functions: 
$\epsilon_{\alpha\sigma}\equiv\epsilon_{\alpha\sigma}^{\FS}$,
$f_{\sigma\sigma'}\equiv f_{\sigma\sigma'}^{\FS}$ and 
$\mathcal{A}_{\sigma\sigma'}\equiv \mathcal{A}_{\sigma\sigma'}^{\FS}$.

As the main result of this section we write the truncation of the Hamiltonian
to the vicinity of the quasiparticle Fermi sea, including the collision
amplitudes between nearly-resonant states.
The truncation is written in terms of the fluctuations of the quasiparticle-hole operator,
\bea
\delta(\hat\gamma_{\alpha\sigma}^\dagger \hat\gamma_{\beta\sigma})&\equiv&\hat\gamma_{\alpha\sigma}^\dagger \hat\gamma_{\beta\sigma}-n_{\alpha\sigma}^{0}\delta_{\alpha\beta} \label{deltagammagamma}
\\
\delta\hat n_{\alpha\sigma}
&\equiv&\delta(\hat\gamma_{\alpha\sigma}^\dagger \hat\gamma_{\alpha\sigma}) \label{deltagammagamma2}
\eea
The fluctuation of the quasiparticle number $\delta \hat n$
can be viewed as the quantum version of the classical field $\delta n$ (the semi-classical ``number of quasiparticles'')
in which Fermi liquid theory is usually formulated.
Restricting to terms quadratic in $\delta(\hat\gamma^\dagger \hat\gamma)$, we can write
\be
\hat H=E_{\rm FS}+\sum_{\alpha\sigma}\epsilon_{\alpha\sigma}\delta \hat n_{\alpha\sigma} 
+\frac{1}{2\mathcal{V}} \sum_{\substack{\alpha\beta\gamma\delta\in\mathcal{D}\\\sigma\sigma'=\upa\dwa}}
 {\mathcal{B}}_{\sigma\sigma'}(\alpha\beta|\gamma{\delta}) \delta(\hat\gamma_{\alpha\sigma}^\dagger \hat\gamma_{{\delta}\sigma})\delta(\hat\gamma_{\beta\sigma'}^\dagger \hat\gamma_{\gamma\sigma'})+O(\delta(\hat\gamma^\dagger \hat \gamma)^3) \label{HFS}
\ee
This second quantization writing of the Hamiltonian must reproduce the diagonal
and off-diagonal matrix elements \eqqrefs{fpsi}{Aif}, which fixes the coefficients $\mathcal{B}_{\sigma\sigma'}$.
The function $\mathcal{B}_{\upa\dwa}$ is straighforwardly related to $f_{\upa\dwa}$ and $\mathcal{A}_{\upa\dwa}$ by
\begin{align}
\mathcal{B}_{\upa\dwa}(\alpha,\beta|\beta,\alpha) &= {f}_{\upa\dwa}(\alpha,\beta)  \label{Bud1}\\
\mathcal{B}_{\upa\dwa}(\alpha,\beta|\gamma,\delta) &= \mathcal{A}_{\upa\dwa}(\alpha,\beta|\gamma,\delta),\ \alpha\neq\delta  \label{Bud2}
\end{align}
For the $\sigma\sigma$ collision, one would naively choose
a coefficient $\mathcal{B}_{\sigma\sigma}$ obeying the fermionic
antisymmetry relations. However, we have broken the $\alpha\leftrightarrow\beta$ antisymmetry
by choosing a specific pairing of the operators $\hat\gamma$, via the expansion
$\delta(\hat\gamma^\dagger\hat\gamma)$.
There remains a symmetry with respect to the full exchange $\alpha\leftrightarrow\beta$, $\gamma\leftrightarrow\delta$ if we impose:
\be
\mathcal{B}_{\sigma\sigma}(\beta,\alpha|\delta,\gamma)= \mathcal{B}_{\sigma\sigma}(\alpha,\beta|\gamma,\delta) \label{contrainteB2}
\ee
However, to ensure that the average value of the interaction term in $\FS$ is zero
while reproducing the correct quasiparticle interaction $f_{\sigma\sigma}$, 
$\mathcal{B}_{\sigma\sigma}$ cannot be antisymmetric with respect to the exchange $\alpha\leftrightarrow\beta$. We must have
\bea
\mathcal{B}_{\sigma\sigma}(\alpha,\beta|\alpha,\beta)&=&0 \label{contrainteB1}\\
\mathcal{B}_{\sigma\sigma}(\alpha,\beta|\beta,\alpha) &=&f_{\sigma\sigma}(\alpha,\beta) \label{Bss1} 
\eea
Note that this ensures that the particle-hole operators $\delta(\hat\gamma_{\alpha\upa}^\dagger \hat\gamma_{{\delta}\upa})$ and $\delta(\hat\gamma_{\beta\upa}^\dagger \hat\gamma_{\gamma\upa})$ commute in \eqqref{HFS}.
Then, the collision amplitude is set by the antisymmetric part of $\mathcal{B}_{\sigma\sigma}$
\be
 \mathcal{B}_{\sigma\sigma}(\alpha,\beta|\gamma,\delta)-\mathcal{B}_{\sigma\sigma}(\beta,\alpha|\gamma,\delta) = \mathcal{A}_{\sigma\sigma}(\alpha,\beta|\gamma,\delta), \ \ \alpha\neq\gamma,\delta  \label{Bss2}
\ee
The symmetric component of $\mathcal{B}_{\sigma\sigma}$ can be chosen arbitrarily as long as \eqqrefs{contrainteB1}{Bss1} are fulfilled.
In Ref.~\cite{VolumeII}, we will illustrate how a function $\mathcal{B}_{\sigma\sigma}$ fulfilling conditions \eqref{contrainteB2}--\eqref{Bss2} arises naturally 
from the perturbative calculation of the truncated Hamiltonian. 

The truncated Hamiltonian \eqqref{HFS} is exact (by definition) for the matrix elements between $\FS$ and states connected to $\FS$ by up to 4
operators $\hat \gamma$. It remains valid up to corrections in $O(\epsilon_0)$
for states $\ket{\psi}$, whose excited quasiparticles are contained in a low-energy shell, that is:
\be
\bra{\psi}\delta\hat n_{\alpha\sigma}\ket{\psi}=0 \text{ if } |\epsilon_{\alpha\sigma}-\mu|>\epsilon_0 \label{fenetre}
\ee
with
\be
\epsilon_0\ll\EF
\ee
The omission of terms cubic or higher in $\delta(\hat\gamma^\dagger \hat \gamma)$ in \eqqref{HFS} 
then leads to errors in the energy and transition amplitudes controlled by $\epsilon_0/\mu$.

We recover the usual semi-classical Hamiltonian of Fermi liquid theory
\be
E=E_{\rm FS}+\sum_{\alpha\sigma}\epsilon_{\alpha\sigma}\delta n_{\alpha\sigma} 
+\frac{1}{2\mathcal{V}} \sum_{\substack{\alpha\beta\in\mathcal{D}\\\sigma\sigma'=\upa\dwa}} f_{\sigma\sigma'}(\alpha,\beta) \delta n_{\alpha\sigma}  \delta n_{\beta\sigma'}+O(\delta n)^3 
\ee
as the restriction of \eqqref{HFS} to terms $\alpha=\delta$, \textit{i.e.} to terms diagonal in the basis of quasiparticle Fock states. The off-diagonal elements
$\alpha\neq\delta$ in \eqqref{HFS} are however crucial to accurately describe quasiparticle collisions.
Thus, unless we are interested only in the collisionless dynamics of the Fermi liquid,
our effective theory should specify not only $f_{\sigma\sigma'}(\alpha,\beta)$,
but also $\mathcal{A}_{\sigma\sigma'}(\alpha,\beta|\gamma,\delta)$ for $\alpha\neq\delta$.

In fluids where the index $\alpha$ describes a continuous sets of modes,
one may think, looking at Eqs.~\eqref{Bud1} and \eqref{Bss1}, that the Landau functions $f_{\sigma\sigma'}$ are continuously
connected to the amplitude $\mathcal{A}_{\sigma\sigma'}(\alpha,\beta|\beta-\dd\alpha,\alpha+\dd\alpha)$ as $\dd\alpha\to0$.
However, we will see in the next section that the energy cutoff $\Lambda$ separates two limits:
\bea
\underset{\substack{\dd\alpha\to 0\\|\epsilon_{\alpha}-\epsilon_{\alpha+\dd\alpha}| \gg\Lambda}}{{\text{lim}}}\mathcal{A}_{\sigma\sigma'}(\alpha,\beta|\beta-\dd\alpha,\alpha+\dd\alpha)&\equiv& \mathcal{A}_{\sigma\sigma'}^{\rm fwd}(\alpha,\beta) \label{limdalphaBforward}\\
\underset{\substack{\dd\alpha\to 0\\|\epsilon_{\alpha}-\epsilon_{\alpha+\dd\alpha}|\ll\Lambda}}{{\text{lim}}}\mathcal{A}_{\sigma\sigma'}(\alpha,\beta|\beta-\dd\alpha,\alpha+\dd\alpha)&=&f_{\sigma\sigma'}(\alpha,\beta) \label{limdalphaf}
\eea
The amplitude $\mathcal{A}_{\sigma\sigma'}^{\rm forward}$ obtained for energy transfer $\epsilon_{\alpha}-\epsilon_{\alpha+\dd\alpha}$
large compared to $\Lambda$ but small compared to $\EF$ is called the forward-scattering amplitude. 
The forward-scattering approximation, common in the literature on ${}^3$He, consists in replacing the full amplitude $\mathcal{A}(\alpha,\beta|\gamma,\delta)$
by its forward value $\mathcal{A}_{\sigma\sigma'}^{\rm fwd}(\alpha,\beta)$. 
This uncontrolled approximation compensates the lack of knowledge on the collision amplitude.
In ${}^3$He where the Landau function $f$ has a large isotropic component ($F_{l=0}^+\geq10$), the Bethe-Salpeter equation
is particularly useful to obtain a correct order of magnitude of the scattering amplitudes.

\subsection{Effective Hamiltonian in homogeneous space}

We assume now that the fluid evolves in a homogeneous cubic volume $\mathcal{V}=L^3$, with $L\to+\infty$ in the thermodynamic limit. 
The one-body Hamiltonian $\hat H_0$ then restricts to the kinetic part:
\be
\hat{H}_0=\sum_{\pp\in\mathcal{D},\sigma} \omega_\pp \hat a_{\pp\sigma}^\dagger \hat a_{\pp\sigma} \label{H0homogene}
\ee
where $\mathcal{D}=(2\pi \mathbb{Z}/L)^3$ is the set of 3D momenta $\pp$ and $\omega_\pp=p^2/2m$ is the kinetic energy.

The particle Fermi sea is now delimited by the sphere of radius $\pF$
\be
\ket{{\rm FS}}_0=\prod_{\substack{\pp\in\mathcal{D}\\\sigma=\upa,\dwa}} \Theta(\pF-p) \hat a_{\pp\sigma}^\dagger\ket{0}_0
\ee
where 
\be
\pF=(3\pi^2\rho)^{1/3}
\ee is the Fermi momentum. The occupation numbers of $\FS$ and $\FS_0$ are 
given by the Heaviside function:
\be
n_\pp^0=\Theta(\pF-p), \qquad \bar n_\pp^0=1-n_\pp^0=\Theta(p-\pF)
\ee

The Luttinger theorem \eqref{NNgamma}, together with the thermodynamic relation
$\dd E_{\FS}/\dd\rho=\mu$, ensures that the quasiparticle
energy at the Fermi level matches $\mu$:
\be
\epsilon_{\pF}=\mu
\ee
Since the quasiparticle dynamics is restricted to the vicinity of the Fermi level,
this eigenenergy can be expanded in powers of $p-\pF$.
\be
\epsilon_{\pp}-\mu=\frac{\pF}{m^*}(p-\pF)+O(p-\pF)^2
\ee
This reduces the effective function $\epsilon_{\pp}$ to a single effective parameter: the effective mass $m^*$.
The effective Hamiltonian is now given by \eqqref{HFS} through the replacement of the indices $\alpha,\beta,\gamma,\delta$ 
by 4 momenta constrained by momentum conservation: $\pp_\alpha+\pp_\beta=\pp_\gamma+\pp_\delta$. 
In the next sections, we will need its decomposition into terms quadratic and quartic in $\hat\gamma$:
\be
\hat H=E_{\rm FS}+\hat H_2+\hat H_4^{\rm d}+\hat H_4^{\rm x}+O\bb{\delta(\hat\gamma^\dagger \hat\gamma)^3} \label{expHp}
\ee
where 
\be
\hat H_2=\sum_{\pp\in\mathcal{D},\sigma}\epsilon_\pp \delta\hat n_{\pp\sigma} \label{H2}
\ee
and we have split the terms quartic in $\hat\gamma$
into diagonal and off-diagonal parts
\begin{align}
\hat H_4^{\rm d}&=\frac{1}{2L^3}\sum_{\substack{\pp\pp'\in\mathcal{D}\\\\\sigma,\sigma'=\upa,\dwa}}f_{\sigma\sigma'}(\pp,\pp')\delta \hat n_{\pp\sigma} \delta \hat n_{\pp'\sigma'} \label{H4d}\\
\hat H_4^{\rm x}&=\frac{1}{2L^3} \sum_{\substack{(\pp_\alpha,\pp_\beta)\neq(\pp_\delta,\pp_\gamma)\\\sigma,\sigma'=\upa,\dwa}}
\delta_{\pp_\alpha+\pp_\beta}^{\pp_\gamma+\pp_{\delta}} {\mathcal{B}}_{\sigma\sigma'}(\pp_\alpha\pp_\beta|\pp_\gamma\pp_{\delta}) \hat \gamma_{\pp_\alpha\sigma}^\dagger \hat \gamma_{\pp_\beta\sigma'}^\dagger \hat\gamma_{\pp_\gamma\sigma'} \hat\gamma_{\pp_{\delta}\sigma} \label{H4x}
\end{align}

To compare our approach with effective theories based on a gradient expansion, let us rewrite \eqqref{expHp} in real space by performing a Wigner transform of the quasiparticle distribution
\be
\delta\hat n_{\pp\sigma}(\rr)\equiv\sum_{\qq}\eee^{\ii\qq\cdot\rr} \delta(\hat\gamma_{\pp-\frac{\qq}{2}\sigma}^\dagger \hat\gamma_{\pp+\frac{\qq}{2}\sigma})
\ee
In terms of $\delta\hat n(\rr)$, we have
\begin{multline}
\hat H=E_{\rm FS}+\sum_{\pp\sigma}\int\frac{\dd^3 r }{\mathcal{V}}\epsilon_{\pp\sigma}\delta \hat n_{\pp\sigma} (\rr)
\\+ \frac{1}{2} \sum_{\substack{\pp\pp'\in\mathcal{D}\\\sigma\sigma'=\upa\dwa}} \int\frac{\dd^3 r_1 \dd^3 r_2}{\mathcal{V}^2} \mathcal{B}_{\sigma\sigma'}(\pp,\pp'|\rr_1-\rr_2) \delta\hat n_{\pp\sigma}(\rr_1) \delta\hat n_{\pp'\sigma'}(\rr_2) +O(\delta\hat n)^3
\label{HFSreel}
\end{multline}
The Wigner transform of the amplitude $\mathcal{B}$ acts as a finite-range interaction potential between the $\pp\sigma$ and $\pp'\sigma'$ quasiparticles:
\be 
\mathcal{B}_{\sigma\sigma'}(\pp,\pp'|\rr_1-\rr_2)=\frac{1}{\mathcal{V}}\sum_{\qq}\eee^{\ii\qq\cdot(\rr_1-\rr_2)} \mathcal{B}_{\sigma\sigma'}\bb{\pp-\frac{\qq}{2},\pp'+\frac{\qq}{2}\Big\vert\pp'-\frac{\qq}{2},\pp+\frac{\qq}{2}}
\ee 
A gradient expansion \cite{Son2022} would replace this finite-range interaction by a short-range one.

Finally, let us note that our definition \eqqref{defZ} of the residue (with $\ket{\psi}=\FS$ as the reference state)
can be reinterpreted\footnote{The link between the two definitions \eqref{Zpsigma2} and \eqref{defZ} follows from the continuity of the residue at the Fermi level $Z_{\pp_+\sigma}=Z_{\pp_-\sigma}$, and the relations 
\be
\bra{\pp_+\sigma,{\rm FS}} \hat{a}_{\pp_+\sigma}^\dagger \hat{a}_{\pp_+\sigma} \ket{\pp_+\sigma,{\rm FS}}=\bra{{\rm FS}} \hat{a}_{\pp_-\sigma}^\dagger \hat{a}_{\pp_-\sigma} \ket{{\rm FS}}
\ee
\be
\bra{\overline{\pp_-\sigma},{\rm FS}} \hat{a}_{\pp_-\sigma}^\dagger \hat{a}_{\pp_-\sigma} \ket{\overline{\pp_-\sigma},{\rm FS}}=\bra{{\rm FS}} \hat{a}_{\pp_+\sigma}^\dagger \hat{a}_{\pp_+\sigma} \ket{{\rm FS}}
\ee.}, in an homogeneous system, as the discontinuity
of the momentum distribution at $\pF$:
\bea
Z_{\pp\sigma}&=&\bra{\pp\sigma,{\rm FS}}\hat{a}_{\pp\sigma}^\dagger \hat{a}_{\pp\sigma} \ket{\pp\sigma,{\rm FS}}-\bra{\overline{\pp\sigma},{\rm FS}}\hat{a}_{\pp\sigma}^\dagger \hat{a}_{\pp\sigma}\ket{\overline{\pp\sigma},{\rm FS}} \label{Zpsigma1}\\
&=&\bra{{\rm FS}}\hat{a}_{\pp_+\sigma}^\dagger \hat{a}_{\pp_+\sigma} - \hat{a}_{\pp_-\sigma}^\dagger \hat{a}_{\pp_-\sigma}\ket{{\rm FS}} \label{Zpsigma2}
\eea
where $p_\pm=\pF\pm 0^+$.

%% file: heff_flot.tex
\section{Flow of the effective Hamiltonian}
\label{sec:flot}
\subsection{Continuous unitary transformation}
The Hamiltonian renormalization procedure introduced in Sec.~\ref{sec:HRG} can be carried out continuously through a succession of infinitesimal unitary transformations. This  defines a renormalization group, from which we derive flow equations for the parameters of the effective Hamiltonian, in particular for the scattering amplitudes $\mathcal{A}$.

Let $\hat{H}(\Lambda)$ denote the effective Hamiltonian at the scale $\Lambda$
\be
\hat{H}(\Lambda)=\eee^{-\hat S(\Lambda)}\hat H \eee^{\hat S(\Lambda)}
\label{Hlambda}
\ee
and write its variation from $\Lambda$ to $\Lambda-\dd\Lambda$:
\be
\hat{H}(\Lambda-\dd\Lambda)=\eee^{-\hat S_\Lambda^{\Lambda-\dd\Lambda}} \hat H(\Lambda) \eee^{\hat S_\Lambda^{\Lambda-\dd\Lambda}}
\label{varHlambda}
\ee
There appears here the generator $\eee^{\hat S_\Lambda^{\Lambda-\dd\Lambda}} = \eee^{-\hat S(\Lambda)} \eee^{\hat S(\Lambda-\dd\Lambda)}$
of the renormalization group\footnote{The expression of the generator in terms of $\hat S(\Lambda)$ and $\dd\hat S=\hat S(\Lambda-\dd\Lambda)-\hat S(\Lambda)$ 
involves the Baker-Campbell-Hausdorff formula, which we fortunately will not need.}. As before,
we choose an operator $\hat S$ that does not contain any low-energy transition
\be
\bra{f}  \hat S_\Lambda^{\Lambda-\dd\Lambda}\ket{i}=0 \text{ if } |E_i-E_f|<\Lambda-\dd\Lambda
\ee  
We then decompose the initial Hamiltonian $\hat H(\Lambda)$
into a diagonal block grouping transitions at energy $0\leq E\leq \Lambda-\dd\Lambda$
and off-diagonal blocks grouping transitions at
$\Lambda-\dd\Lambda\leq E\leq \Lambda$, which are to be removed from $\hat H(\Lambda-\dd\Lambda)$:
\be
\hat H(\Lambda)=\hat H_d(\Lambda)+\hat H_x(\Lambda),
\ee
\be
\hat P_{\Lambda-\dd\Lambda}(E)\hat H_d(\Lambda) \hat Q_{\Lambda-\dd\Lambda}(E)=0 \qquad
\hat H_x(\Lambda) =\hat H(\Lambda)-\hat H_d(\Lambda)
\ee

We use the block-diagonal structure of $\hat H(\Lambda-\dd\Lambda)$ to compute
the matrix elements of the generator\footnote{Again, we neglect terms that introduce
at least one intermediate state $\ket{I}$:
\be
\bra{a}\hat S_{\Lambda}^{\Lambda-\dd\Lambda}\ket{b}-\frac{ \bra{a}\hat H_x(\Lambda)\ket{b}}{E_b-E_a}=\frac{1}{E_b-E_a}\sum_{\ket{{\rm I}}}\frac{\bra{a}\hat H_d(\Lambda)\ket{{\rm I}}\bra{{\rm I}}\hat H_x(\Lambda)\ket{i}}{E_b-E_{\rm I}}-\frac{\bra{a}\hat H_x(\Lambda)\ket{{\rm I}}\bra{{\rm I}}\hat H_d(\Lambda)\ket{b}}{E_{\rm I}-E_a}+O(\dd\Lambda)^2
\ee
The intermediate state $\ket{{\rm I}}$ is confined to an energy shell of width $\dd\Lambda$ around $E_a$ or $E_b$. Its contribution to the matrix element from $\ket{b}$ to $\ket{a}$
(these being themselves separated by $\Lambda$ up to $\dd\Lambda$ corrections) is therefore negligible.
}:
\be
\hat P_{\Lambda-\dd\Lambda} \hat H(\Lambda-\dd\Lambda) \hat Q_{\Lambda-\dd\Lambda}=0 \implies \bra{a}\hat S_{\Lambda}^{\Lambda-\dd\Lambda}\ket{b}=\frac{ \bra{a}\hat H_x(\Lambda)\ket{b}}{E_b-E_a}+O(\dd\Lambda)
\ee
To first order in $\dd\Lambda$, the variation \eqref{varHlambda} of the diagonal blocks of the effective Hamiltonian can be rewritten \cite{Redmon1980}
\be
\hat P_{\Lambda-\dd\Lambda} \hat H(\Lambda-\dd\Lambda) \hat P_{\Lambda-\dd\Lambda}=\hat P_{\Lambda-\dd\Lambda}\bb{ \hat H(\Lambda)  +\frac{1}{2}\bbcro{\hat H(\Lambda),\hat S_{\Lambda}^{\Lambda-\dd\Lambda}}}\hat P_{\Lambda-\dd\Lambda}+O(\dd\Lambda)^2
\label{PHlambadP}
\ee
The matrix elements on the mass shell are therefore coupled either directly
through the old Hamiltonian $\hat H(\Lambda)$, or indirectly via an intermediate state
$\ket{\rm I}$ confined to an energy band of width $\dd\Lambda$ (see Figure \ref{fig:bande}).
Higher-order contributions in $\hat S_{\Lambda}^{\Lambda-\dd\Lambda}$ in equation \eqref{PHlambadP}
involve multiple intermediate states confined to such an energy band; they are therefore subleading in $\dd\Lambda$.

\begin{figure}
\begin{center}
\begin{tikzpicture}[x=2cm,y=3cm]

\def\xL{0}
\def\xR{4}

\def\Ei{0.0}
\def\dL{0.35}        
\def\L{1.2}          
\def\EI{\Ei+\L-0.6*\dL/2}

\draw[->] (\xL,-0.2) -- (\xL,2.0) node[above] {$E$};

\fill[gray!30] (\xL,\Ei) rectangle (\xR,\Ei+\L-\dL);


\fill[gray!15]
  (\xL,\Ei+\L) rectangle (\xR,1.9);

\node at (\xR/2,\Ei+\L+0.4)
  {$\langle f|\hat H(\Lambda)|{\rm I}\rangle = 0$};

\node at (\xR/2,\L/2-0.2)
  {$\langle {\rm I}|\hat S_\Lambda^{\Lambda-\dd\Lambda}|i\rangle = 0$};

\draw[thick] (\xL,\Ei) -- (\xR,\Ei);
\node[right] at (\xR,\Ei) {$E_i = E_f$};

\draw[thick,dashed]
  (\xL,\Ei+\L-\dL) -- (\xR,\Ei+\L-\dL);
\node[right] at (\xR,\Ei+\L-\dL)
  {$E_i+\Lambda-\mathrm \dd\Lambda$};
  
\draw[thick]
  (\xL,\EI) -- (\xR,\EI);
\node[right] at (\xR,\EI)
  {$E_{\rm I}$};

\draw[thick,dashed] (\xL,\Ei+\L) -- (\xR,\Ei+\L);
\node[above right] at (\xR,\Ei+\L) {$E_i+\Lambda$};

\draw[->,red,thick]
  (0.3,\Ei) -- node[sloped,above]   {$\langle I|\hat S_\Lambda^{\Lambda-\dd\Lambda}|i\rangle$} (2.0,\EI);

\draw[->,red,thick] (2.0,\EI) -- node[sloped,above]   {$\langle f|\hat H(\Lambda)|I\rangle$} (3.7,\Ei);
  
\end{tikzpicture}
\caption{Contribution of second-order processes to the transition amplitude from $\ket{i}$ to $\ket{f}$,
via an intermediate state $\ket{{\rm I}}$ confined to an energy shell $\dd\Lambda$. \label{fig:bande}}
\end{center}
\end{figure}

\subsection{Flow of the scattering amplitudes}

We can now express the variation of the transition amplitude $\mathcal{A}_{i\to f}^\Lambda$
between two states $\ket{i}$ and $\ket{f}$ on the mass shell ($E_i=E_f$):
\be
\mathcal{A}_{i\to f}^{\Lambda-\dd\Lambda}=\mathcal{A}_{i\to f}^{\Lambda}+\sum_{\ket{{\rm I}}}{\mathcal{A}^\Lambda_{i\to{\rm I}}\mathcal{A}^\Lambda_{I\to{\rm f}}}\frac{\Pi_{\Lambda}^{\Lambda-\dd\Lambda}(E_i-E_{\rm I})}{E_i-E_{\rm I}}
\label{flot}
\ee
where the function
\be
\Pi_{\Lambda}^{\Lambda-\dd\Lambda}(E)=
\begin{cases} 
1 \text{ if } \Lambda-\dd\Lambda<|E|<\Lambda\\
0 \text{ otherwise }
\end{cases}
\ee
implements the energy constraint on the intermediate state.

The variation of $\mathcal{A}$ with $\dd\Lambda$ thus comes solely from second-order diagrams in $\hat H(\Lambda)$,
which may seem extremely simple.
However, $\hat H(\Lambda)$ is a much more complex object than the bare interaction Hamiltonian
$\hat V$, since the interaction terms of $\hat H(\Lambda)$ are not restricted
to two-body processes. Thus, even if we restrict to
$2\leftrightarrow2$ transitions for which
\be
\ket{f}=\hat\gamma_{\alpha\sigma}^\dagger \hat\gamma_{\beta\sigma'}^\dagger \hat\gamma_{\gamma\sigma'} \hat\gamma_{\delta\sigma}\ket{i}
\label{ketf}
\ee
intermediate states $\ket{\rm I}$ containing an arbitrary number of virtual quasiparticles
are possible; we give in Figure \ref{diag3corps} a diagrammatic representation of the transition $\ket{i}\to\ket{f}$ via
the intermediate state $\ket{{\rm I}}= \hat \gamma_{\pp_a}^\dagger \hat \gamma_{\pp_b}^\dagger \hat \gamma_{\pp_c}^\dagger \hat \gamma_{\pp_d} \hat\gamma_{\gamma} \hat\gamma_{\delta} \ket{i}$.
Even though it is of second order in $\hat H(\Lambda)$, this process involves 4 virtual quasiparticles and three-body scattering amplitudes.

\begin{figure}
\begin{center}
\begin{tikzpicture}[xscale=5.3,yscale=4]
\coordinate (A) at (-0.5,0);
\coordinate (B) at (0.5,0);
\def\xA{-0.5}
\def\yA{0}
\def\xB{0.5}
\def\yB{0}
\draw (A) node{\textbullet};
\draw (B) node{\textbullet};
\draw[thin,dashed] (\xA,1.1) -- (\xA,-1);
\draw[thin,dashed] (\xB,1.1) -- (\xB,-1);
\draw[->-] (A) arc (180:0:{0.5});
\draw[->-] (A) arc (135:45:{1/sqrt(2.)});
\draw[-<-] (A) arc (-180:0:{0.5});
\draw[->-] (A) arc (-135:-45:{1/sqrt(2.)});
\draw (0,0.55) node {$\pp_a$};
\draw (0,0.3) node {$\pp_b$};
\draw (0,-0.3) node {$\pp_c$};
\draw (0,-0.55) node {$\pp_d$};
\draw[->-] (\xA-0.5,\yA+0.5) -- node[above=0.1cm]{$\gamma$} (A);
\draw[->-] (\xA-0.5,\yA-0.5) -- node[below=0.1cm]{$\delta$} (A);
\draw[->-] (B) -- node[above=0.1cm]{$\alpha$} (\xB+0.5,\yB+0.5);
\draw[->-] (B) -- node[below=0.1cm]{$\beta$} (\xB+0.5,\yB-0.5);
\draw (0,1) node { {$\ket{{\rm I}}=\hat \gamma_{\pp_a}^\dagger \hat \gamma_{\pp_b}^\dagger \hat \gamma_{\pp_c}^\dagger \hat \gamma_{\pp_d} \hat\gamma_{\gamma} \hat\gamma_{\delta} \ket{i}$}};
\draw (-0.75,1) node { {$\ket{i}$}};
\draw (0.75,1) node { {$\ket{f}$}};
\end{tikzpicture}
\end{center}
\caption{\label{diag3corps} Diagram representing the transition $\gamma,\delta\,\to\,\alpha,\beta$
via an intermediate state $\ket{I}$ with 4 virtual quasiparticles (curved lines representing
the momenta $\pp_a$, $\pp_b$, $\pp_c$, and $\pp_d$). This diagram is of second order in $\hat H(\Lambda)$
(the dots \textbullet\ represent the two three-body scattering amplitudes). It is nevertheless omitted in
Eq.~\eqref{dAud}, where quasiparticles are assumed to be weakly correlated.}
\end{figure}

Thus, in the general case, the flow equation \eqref{flot} couples two-body scattering amplitudes to $p$-body amplitudes ($p\geq 3$), and thereby contains the full complexity of the many-body problem. It becomes tractable only if this hierarchy of couplings can be truncated, either by a weak-coupling argument ($V\ll\EF$), or by invoking the low density of quasiparticles available for scattering. This is the argument used in a Fermi liquid to integrate the final stage of the renormalization flow, from
\be
\Lambda_0\ll\EF
\ee
to $\Lambda_f<\Lambda_0$.
If we assume that the incoming and outgoing quasiparticles are contained in a low-energy shell
($|\epsilon_{i}-\EF|<\epsilon_0<\Lambda_0\ll\EF$ for $i=\alpha\sigma,\beta\sigma',\gamma\sigma'$ and $\delta\sigma$),
then any sum over a virtual quasiparticle (unconstrained by momentum conservation) is reduced by a factor $\Lambda_0/\EF$. Thus, the three-body diagram (Fig.~\ref{diag3corps}),
which contains 3 independent virtual quasiparticles, is smaller by a factor $(\Lambda_0/\EF)^2$
than the two-body diagrams of Fig.~\ref{diagrammesupdw}. 

One can then restrict \eqref{flot} to two-body scattering, and the calculation becomes analogous to second-order perturbation theory \cite{devvisco}.
For a $2\leftrightarrow2$ transition (see \eqqref{ketf}),
there are 6 intermediate states, which can be grouped in pairs (by time reversal) to form the Hartree (V and VI), Fock (III and IV), and Bogoliubov (I and II) contributions.
We give a diagrammatic representation of these intermediate states in Fig.~\ref{diagrammesupdw}.
By choosing the Fermi sea as the reference state (so that $n_{\pp\sigma}=n_p^0$ for $\pp\sigma\neq\alpha\sigma,\beta\sigma',\gamma\sigma',\delta\sigma$),
we obtain the flow of the scattering amplitude $\upa\dwa$:
\setlength{\belowdisplayskip}{0pt}
\setlength{\abovedisplayskip}{0pt}
\begin{multline}
\frac{\dd\mathcal{A}_{\sigma\sigma'}^\Lambda(\alpha,\beta|\gamma,\delta)}{\dd\Lambda}=\frac{1}{\Lambda L^3}\sum_{\pp}\sum_{\sigma_a\sigma_b}\Bigg[\notag\\
\mathcal{A}_{\sigma\sigma_a}^\Lambda(\alpha,\pp-\qq_{\alpha\delta}|\pp,\delta)(\mathcal{S}^{\rm h})^{\sigma\sigma'}_{\sigma_a\sigma_b}  \mathcal{A}_{\sigma_b\sigma'}^\Lambda(\pp,\beta|\gamma,\pp-\qq_{\alpha\delta})
[{n^0_{\pp}-n^0_{\pp-\qq_{\alpha\delta}}}]\delta_\Lambda\bb{\epsilon_{\delta}+\epsilon_{\pp}-\epsilon_{\alpha}-\epsilon_{\pp-\qq_{\alpha\delta}}} \notag\\{\small \text{ (Hartree)}}\notag
\end{multline}
\begin{multline}
-\mathcal{A}_{\sigma\sigma_a}^\Lambda(\alpha,\pp-\qq_{\alpha\gamma}|\gamma,\pp)(\mathcal{S}^{\rm f})^{\sigma\sigma'}_{\sigma_a\sigma_b}  \mathcal{A}_{\sigma_b\sigma'}^\Lambda(\pp,\beta|\pp-\qq_{\alpha\gamma},\delta)
[{n^0_{\pp}-n^0_{\pp-\qq_{\alpha\gamma}}}]\delta_\Lambda\bb{\epsilon_{\gamma}+\epsilon_{\pp}-\epsilon_{\alpha}-\epsilon_{\pp-\qq_{\alpha\gamma}}} \notag\\ {\small \text{ (Fock)}}\notag
\end{multline}
\begin{multline}
+\mathcal{A}_{\sigma\sigma_a}^\Lambda(\alpha,\beta|2\PP_{\alpha\beta}-\pp,\pp)(\mathcal{S}^{\rm b})^{\sigma\sigma'}_{\sigma_a\sigma_b}\mathcal{A}_{\sigma_b\sigma'}^\Lambda(\pp,2\PP_{\alpha\beta}-\pp|\gamma,\delta)\\
\times\bbcro{1- n^0_{\pp}- n^0_{2\PP_{\alpha\beta}-\pp}}\delta_\Lambda\bb{\epsilon_{\alpha}+\epsilon_{\beta}-\epsilon_{\pp}-\epsilon_{2\PP_{\alpha\beta}-\pp}}\Bigg]  {\small \text{ (Bogoliubov)}} \label{dAud}
\end{multline}
\abovedisplayskip=11pt plus 3pt minus 6pt
\belowdisplayskip=11pt plus 3pt minus 6pt
with
\be
\qq_{\alpha\gamma}=\alpha-\gamma,\qquad \qq_{\alpha\delta}=\alpha-\delta,\quad\text{ and }\quad \PP_{\alpha\beta}=\frac{\alpha+\beta}{2}
\ee
The energy-shell constraint has been replaced by a Dirac delta function:
\be
\frac{\Pi_{\Lambda-\dd\Lambda}^\Lambda(\epsilon)}{\epsilon}=\frac{\dd\Lambda}{\Lambda}\delta_{\Lambda}(\epsilon) \text{ with } \delta_{\Lambda}(\epsilon)=\delta(\epsilon-\Lambda)-\delta(\epsilon+\Lambda)
\ee
We treat the spin structure via a matrix $\mathcal{S}$ specific to each diagram:
\bea
(\mathcal{S}^{\rm h})^{\sigma\sigma'}_{\sigma_a\sigma_b}&=&\delta_{\sigma_a\sigma_b} \\
(\mathcal{S}^{\rm f})^{\sigma\sigma'}_{\sigma_a\sigma_b}&=&\delta_{\sigma\sigma_a,\sigma'\sigma_b}(1-\delta_{\sigma}^{-\sigma'}\delta_{\sigma}^{\sigma_a})\\
(\mathcal{S}^{\rm b})^{\sigma\sigma'}_{\sigma_a\sigma_b}&=& \delta_{\sigma'\sigma_a}\delta_{\sigma\sigma_b}
\eea
The Fock diagram involves a sum over the spin of the intermediate quasiparticle only for scattering between indistinguishable quasiparticles: $\mathcal{A}_{\sigma\sigma_a}\mathcal{A}_{\sigma_b\sigma'}\to\mathcal{A}_{\upa\upa}\mathcal{A}_{\upa\upa}+\mathcal{A}_{\upa\dwa}\mathcal{A}_{\upa\dwa}$ for $\sigma=\sigma'=\upa$, whereas $\mathcal{A}_{\sigma\sigma_a}\mathcal{A}_{\sigma_b\sigma'}\to\mathcal{A}_{\upa\dwa}\mathcal{A}_{\upa\dwa}$ for $\sigma\neq\sigma'$.
In this way, the renormalization flow preserves SU(2) symmetry:
\begin{multline}
{\mathcal{A}}_{\upa\dwa}^\Lambda(\alpha,\beta|\gamma,\delta)-{\mathcal{A}}_{\upa\dwa}^\Lambda(\alpha,\beta|\delta,\gamma)-{\mathcal{A}}_{\upa\upa}^\Lambda(\alpha,\beta|\gamma,\delta)=0\\\implies\frac{\dd}{\dd\Lambda}\bbcro{{\mathcal{A}}_{\upa\dwa}^\Lambda(\alpha,\beta|\gamma,\delta)-{\mathcal{A}}_{\upa\dwa}^\Lambda(\alpha,\beta|\delta,\gamma)-{\mathcal{A}}_{\upa\upa}^\Lambda(\alpha,\beta|\gamma,\delta)}=0
\end{multline}

\subsection{Flow equations at non-zero angles} \label{anglesnonnuls}
When studying transport at low temperature or low energy,
the only excited quasiparticles have momenta close to $\pF$.
To first order in $T$ (or more generally in the maximal excitation energy $\epsilon_0$),
one can then approximate the scattering amplitude for $\gamma,\delta\to\alpha,\beta$ by its value on the Fermi surface, i.e. for
\be
\epsilon_\alpha=\epsilon_\beta=\epsilon_\gamma=\epsilon_\delta=\mu
\label{resstricte}
\ee

The flow equation \eqref{dAud} obtained above has the particular property of leaving the neighborhood of the Fermi surface invariant:
if $\alpha,\beta,\gamma,\delta$ lie at the Fermi level, then all intermediate momenta (namely $p$ and $|\pp-\qq_{ij}|$ in the Hartree-Fock terms,
$p$ and $|2\PP_{\alpha\beta}-\pp|$ in the Bogoliubov term)
remain at most at a distance $\Lambda/\vF$ from $\pF$. This follows from the combined effect
of the $\Lambda$-resonance condition (imposed by the $\delta_\Lambda$ functions)
and the constraint of quasiparticle/quasihole availability
(through the factors $n^0_{\pp}-n^0_{\pp-\qq_{ij}}$
in Hartree-Fock and $1- n^0_{\pp}- n^0_{2\PP_{\alpha\beta}-\pp}$
in Bogoliubov).

In the generic case, i.e. when none of the vectors $\alpha$, $\beta$, $\gamma$, $\delta$ are collinear,
the sum over $\pp$ in \eqqref{dAud}, in addition to its restriction to $p-\pF\approx\Lambda$,
is also restricted to a narrow angular shell of width $\Lambda$ (see Figs.~\ref{figHartree} and \ref{figBogo} in Appendix \ref{app:flot}).
We then show that the integration of the flow at energy scales below $\Lambda_0$ produces no significant change in $\mathcal{A}$:
\be \label{varAanglesnonnuls}
{{\mathcal{A}}^{\Lambda_0}_{\sigma\sigma'}(\alpha\beta|\gamma\delta)}-{{\mathcal{A}}^{\Lambda_f}_{\sigma\sigma'}(\alpha\beta|\gamma\delta)}=O(\Lambda_0), \quad\text{if}\quad s_{\alpha\delta},s_{\alpha\gamma},c_{\alpha\beta}>\Lambda_0
\ee
with
\be
s_{\alpha\delta}=\sin\frac{\theta_{\alpha\delta}}{2}, \qquad s_{\alpha\gamma}=\sin\frac{\theta_{\alpha\gamma}}{2}, \qquad c_{\alpha\beta}=\cos\frac{\theta_{\alpha\beta}}{2}
\ee
and $\theta_{i,j}=(\widehat{i,j})$.
For this non-collinear amplitude, one can therefore consider that the renormalization is complete for $\Lambda_0\ll\vF\pF$.

\subsection{Bethe-Salpeter equations at small angles} \label{BSsec}

We now consider the case where the momenta are pairwise quasi-collinear.
Three cases are possible:
\begin{alignat}{4}
&\text{Forward scattering:} \qquad &\alpha\approx\delta,&\ \beta\approx\gamma, &\qquad s_{\alpha\delta}\ll 1,    \\
&\text{Exchange scattering:}  \qquad &\alpha\approx\gamma,&\ \beta\approx\delta, &\qquad s_{\alpha\gamma}\ll 1,  \\
&\text{Frontal scattering:}  \qquad &\alpha\approx-\beta,&\ \gamma\approx-\delta, &\qquad c_{\alpha\beta}\ll 1,  
\end{alignat}
The transferred momentum $q_{ij}$ or the center-of-mass momentum
$P_{\alpha\beta}$ is then small compared to $\pF$. Let us generically denote this small momentum by $\qq$
\be
\qq=\qq_{\alpha\delta},\qq_{\alpha\gamma}, \text{ or } \PP_{\alpha\beta}, 
\ee
respectively for forward, exchange, or frontal scattering.
This momentum $q$ defines
a new energy scale $\vF q$, small compared to $\EF$, to which the renormalization flow is sensitive. 

To capture this rapid flow, we choose the initial and final cutoffs, $\Lambda_0$ and $\Lambda_f$, respectively
above and below the small energy scale:
\be
{\Lambda_f} \ll \vF q \lesssim {\Lambda_{0}}\ll\vF\pF \qquad \qquad
\bar\Lambda_f \ll s_{\alpha\delta},\, s_{\alpha\gamma},\text{ or } c_{\alpha\beta}\lesssim\bar\Lambda_0\ll1 \label{ineqLambda}
\ee
where we have introduced the notation $\bar{.}$ for the rescaling
by Fermi units:
\be
\bar\Lambda=\frac{\Lambda}{\pF\vF},\qquad \bar p=\frac{p}{\pF}, \qquad \bar q=\frac{q}{\pF}, \qquad \bar\epsilon_\pp=\frac{\epsilon_\pp}{\vF\pF}, \qquad \bar{\mathcal{A}}=\frac{m^\ast\pF}{(2\pi)^3}\mathcal{A}
\ee
Note that $\bar q= 2s_{\alpha\delta},\, 2s_{\alpha\gamma},\text{ or } 2c_{\alpha\beta}$.

In the energy conservation relations (Eqs.~\eqref{EHartree}--\eqref{EBogolioubov} in the Appendix),
there is no longer any dominant term in $O(\Lambda^0)$ constraining
the angular variations. This results in large variations of $\mathcal{A}$ within a narrow energy window.
These large variations are due to a specific diagram: the Hartree diagram for forward scattering,
the Fock diagram for exchange scattering, and the Bogoliubov diagram for frontal scattering;
the two remaining diagrams are in each case negligible, contributing only $O(\Lambda_0)$ corrections
over the entire flow. One can thus speak of a Hartree, Fock, and Bogoliubov channel 
for the renormalization of each of the three small-angle limits of $\mathcal{A}$.

\paragraph{Hartree channel} \label{BSHartree}
We consider here the forward scattering limit, i.e.
\be
q\to0\iff \begin{cases}\delta\to\alpha,\\ \gamma\to\beta\end{cases}
\ee
Note that $\qq$ is then orthogonal to the incoming and outgoing vectors $\alpha,\beta,\gamma,\delta$
to ensure energy conservation (i.e. $\epsilon_\alpha=\epsilon_{\alpha-\qq}=\mu$). The vector
$\qq$ defines the $\textbf{e}_z$ axis of a spherical frame adapted to the Hartree channel (see the 3D illustration in Fig.~\ref{figHartree}),
and we introduce the polar angles $\theta_{i}=(\widehat{i,\qq})$. The external vectors $\alpha$, $\beta$, $\alpha-\qq$, and $\beta+\qq$
lie on the equator
\be \label{thetapi2Hartree}
\text{lim}_{q\to0}\theta_\alpha, \theta_\beta=\frac{\pi}{2}
\ee
while the intermediate momentum $\pp$ will move from the poles to the equator during the flow.
The matrix element of a forward transition between vectors $\alpha$ and $\beta$ on the Fermi sphere depends only on
the angle $\varphi=(\widehat{\alpha,\beta})$ between these vectors. We therefore define
\be
\mathcal{A}^{\rm h}(\cos\varphi,\lambda)=\underset{\substack{q\to0\\\epsilon_\alpha=\epsilon_\beta=\mu\\(\widehat{\alpha,\beta})=\varphi}}{\text{lim}}\mathcal{A}^{\Lambda}(\alpha,\beta+\qq|\beta,\alpha-\qq)
\ee
where we introduced the reduced variable
\be
\lambda=\frac{\Lambda}{\vF q}
\ee

The flow of $\mathcal{A}^{\rm h}$, which had reached a plateau for $\vF q<\Lambda\ll\vF\pF$, accelerates
abruptly at lower energy scales. This is due to a sudden increase in the phase-space volume
contributing to the flow: while this volume is $O(\dd\Lambda)$ for $\Lambda>\vF q$ (cf. the discussion after
\eqref{EBogolioubov}), it becomes $O(\dd\Lambda/\bar q)$ for $\Lambda<\vF q$; the flow from $\Lambda_0$
to $\Lambda_f$ then produces an order-one change in $\mathcal{A}^{{\rm h}}$, even though
both scales $\Lambda_0$ and $\Lambda_f$ are small compared to the Fermi energy.
At small $q$, the $\Lambda$-conservation constraint for the intermediate quasiparticle
with momentum $\pp$, namely ${\epsilon_{\pp-\qq}-\epsilon_\pp}=\pm\Lambda$ (cf. the second line of \eqref{dAud}),
can be rewritten as
\be
\bar q { \cos\theta}=\pm{\bar\Lambda}+O(q^2)\label{EHartreecolin}
\ee
The constraint acts only on the polar angle $\theta=(\widehat{\pp,\qq})$ (the integral
over the energy $\epsilon_\pp-\mu$ becomes trivial in this small-angle regime),
and allows it to vary far beyond a band of width $\Lambda_0$,
up to covering the entire interval $[0,\pi]$ as $\Lambda$ decreases from $\vF q$ to 0.
Since the flow of $\mathcal{A}^{\rm h}$ remains slow at energy scales above $\vF q$
(the resonance in \eqref{EHartreecolin} not being possible yet), we choose $\Lambda_0=\vF q$ to sit at the onset of the rapid flow.
We can now distinguish the two limits of $\mathcal{A}^{{\rm h}}$,
depending on whether the low-energy limit ($\Lambda\to0$) is taken before or after the limit $q\to0$:
\bea
f_{\sigma\sigma'}(\cos\varphi)&=&\lim_{\bar\Lambda\to0}\lim_{\bar q\to0}\mathcal{A}^{\Lambda}_{\sigma\sigma'}(\alpha,\beta+\qq|\beta,\alpha-\qq)=\lim_{\substack{\lambda\to\infty}}\mathcal{A}^{{\rm h}}_{\sigma\sigma'}(\cos\varphi,\lambda) \label{limlamlimq}\\
\mathcal{A}^{{\rm fwd}}_{\sigma\sigma'}(\cos\varphi)&=&\lim_{\bar q\to0}\lim_{\bar \Lambda\to0}\mathcal{A}^{\Lambda}_{\sigma\sigma'}(\alpha,\beta+\qq|\beta,\alpha-\qq)=\lim_{\substack{\lambda\to0}}\mathcal{A}^{{\rm h}}_{\sigma\sigma'}(\cos\varphi,\lambda) \label{limqlimlam}
\eea

For $q$ tending to zero faster than $\Lambda$, one obtains the interaction function
$f_{\sigma\sigma'}$, i.e. the coefficient of the diagonal terms of the effective Hamiltonian, which enters
in particular in the equilibrium properties of the fluid. Conversely,
for $\Lambda$ tending to zero faster than $q$, one obtains the forward scattering amplitude $\mathcal{A}^{\rm fwd}$,
which enters (together with the large-angle scattering amplitude) in the hydrodynamic properties of the fluid.

We now relate these two quantities by solving the flow equation.
To do so, we parameterize the intermediate momentum $\pp$ in the Hartree diagram by the angles $\theta$ and $\phi$
choosing $\alpha$ as origin of azimuthal angles\footnote{We approximate
the transition amplitude $(\alpha+\qq,\pp)\to(\alpha,\pp-\qq)$
by the forward amplitude $\mathcal{A}^{\rm h}(\cos(\widehat{\alpha,\pp}),\lambda)$, with $\cos(\widehat{\alpha,\pp})=\sin\theta\cos\phi$.
Similarly for the transition amplitude $(\pp-\qq,\beta)\to(\pp,\beta+\qq)$ approximated by $\mathcal{A}^{\rm h}(\cos(\widehat{\beta,\pp}),\lambda)$ with $\cos(\widehat{\beta,\pp})=\sin\theta\cos(\varphi-\phi)$.
}.
Up to $O(q)$ terms, the flow equation then takes the form
\begin{multline}
\frac{\dd\bar{\mathcal{A}}^{\rm h}_{\sigma\sigma'}}{\dd\lambda}(\cos\varphi,\lambda)
=\sum_{\sigma_a} \int_0^{\pi}\sin\theta\dd\theta\int_0^{2\pi}\dd\phi \bar{\mathcal{A}}^{\rm h}_{\sigma\sigma_a}(\sin\theta\cos\phi,\lambda)\bar{\mathcal{A}}^{\rm h}_{\sigma_a\sigma'}(\sin\theta\cos(\varphi-\phi),\lambda)\\\times \bbcro{\delta\bb{\lambda-\cos\theta}+\delta\bb{\lambda+\cos\theta}}
\end{multline}

This quadratic partial-differential equation on $\mathcal{A}^{\rm h}$ can be solved into a linear integral equation,
where the initial function $\mathcal{A}^{\rm h}(\lambda_0)$ plays the role of the integral kernel, and where the function $\delta(\lambda-\cos\theta)+\delta(\lambda+\cos\theta)$
is replaced by its primitive\footnote{In general, one can write, for angular functions
$u(\theta,\phi,\varphi)$ and $v(\theta,\phi,\varphi)$ and a function $f(\lambda)$ admitting $F(\lambda)$ as a primitive:
\begin{multline}
\frac{\dd\mathcal{A}(\varphi,\lambda)}{\dd\lambda}=\int_{0}^{\pi} \sin\theta\dd\theta \int_{0}^{2\pi}\dd \phi  \mathcal{A}(u(\theta,\phi,\varphi),\lambda) \mathcal{A}(v(\theta,\phi,\varphi),\lambda) f(\lambda) \\
\implies\mathcal{A}(\varphi,\lambda)=\mathcal{A}(\varphi,\lambda_0)+\int_{0}^{\pi} \sin\theta\dd\theta \int_{0}^{2\pi}\dd \phi \mathcal{A}(u(\theta,\phi,\varphi),\lambda_0) \mathcal{A}(v(\theta,\phi,\varphi),\lambda) \bbcro{F(\lambda)-F(\lambda_0)}
\end{multline}
We apply this relation to $f(\lambda)=\delta(\lambda\pm\cos\theta)$ \label{primitiveflot}}
${\Theta(\lambda_f-|\cos\theta|)-\Theta(\lambda_0-|\cos\theta|)}$,
which acts as a restriction of the integration domain over $\theta$.
We thus obtain
\be
\begin{split}
{\bar{\mathcal{A}}^{{\rm h}}_{\sigma\sigma'}}(\cos\varphi,\lambda_f)&={\bar{\mathcal{A}}^{{\rm h}}_{\sigma\sigma'}}(\cos\varphi,\lambda_0) \\ & -\sum_{\sigma_a} \int_0^{2\pi}\dd\phi \int_{\mathcal{D_\theta}}\sin\theta\dd\theta
\bar{\mathcal{A}}^{\rm h}_{\sigma\sigma_a}(\sin\theta\cos\phi,\lambda_0)\bar{\mathcal{A}}^{\rm h}_{\sigma_a\sigma'}(\sin\theta\cos(\varphi-\phi),\lambda_{f}) 
\end{split}
\ee
The integration domain over the polar angle $\mathcal{D}_\theta=[-1,-\lambda_f]\cap[\lambda_f,1]$ excludes a band around the equator of the Fermi sphere,
an equator on which the external momenta $\alpha$ and $\beta$ lie. For $\lambda_f\to0$, the contribution of this excluded band becomes negligible. Identifying $\mathcal{A}^{\rm h}_{\sigma\sigma'}(\lambda_0)$ with ${f}_{\sigma\sigma'}$ and
$\mathcal{A}^{\rm h}_{\sigma\sigma'}(\lambda_f)$ with ${\mathcal{A}}^{{\rm fwd}}_{\sigma\sigma'}$, one finally obtains the Bethe-Salpeter equation
\be
\bar{\mathcal{A}}^{{\rm fwd}}_{\sigma\sigma'}(\cos\varphi)=\bar{f}_{\sigma\sigma'}(\cos\varphi)-\sum_{\sigma_a} \int_0^{2\pi}\dd\phi \int_{0}^{\pi}\sin\theta\dd\theta
\bar{f}_{\sigma\sigma_a}(\sin\theta\cos\phi)\bar{\mathcal{A}}^{\rm fwd}_{\sigma_a\sigma'}(\sin\theta\cos(\varphi-\phi)) \label{BS}
\ee

\paragraph{Fock channel} \label{BSFock}
We now turn to exchange scattering, i.e. the limit
\be
q\to0\iff \begin{cases}\delta\to\beta,\\ \gamma\to\alpha \end{cases}
\ee
For scattering between indistinguishable quasiparticles,
this channel simply enforces the antisymmetry of the forward scattering amplitude $\mathcal{A}_{\upa\upa}(\alpha,\beta|\alpha,\beta)
=-\mathcal{A}_{\upa\upa}(\alpha,\beta|\beta,\alpha)$, as a consequence of the Pauli principle. For $\upa\dwa$ scattering, however,
it determines the amplitude $\mathcal{A}_{\upa\dwa}(\alpha,\beta|\alpha,\beta)$
where opposite-spin fermions exchange their momenta. In the presence of an SU(2) symmetry breaking,
this parameter is independent of the forward amplitudes $\mathcal{A}^{\rm fwd}_{\upa\dwa}$ and $\mathcal{A}^{\rm fwd}_{\upa\upa}$.
As in the Hartree channel, we define the two low-energy limits of the exchange amplitude:
\bea
f_{\upa\dwa}^{\rm ex}(\cos\varphi)&=&\lim_{\bar\Lambda\to0}\lim_{\bar q\to0}\mathcal{A}^{\Lambda}_{\upa\dwa}(\alpha,\beta+\qq|\alpha-\qq,\beta)\\
\mathcal{A}^{\rm ex}_{\upa\dwa}(\cos\varphi)&=&\lim_{\bar q\to0}\lim_{\bar \Lambda\to0}\mathcal{A}^{\Lambda}_{\upa\dwa}(\alpha,\beta+\qq|\alpha-\qq,\beta)
\eea
By analogy with \eqref{limlamlimq}, we denote $f_{\upa\dwa}^{\rm ex}$ the amplitude obtained for $\vF q\ll\Lambda$. This is the coefficient
of the exchange operator $\hat\gamma_{\pp\upa}^\dagger \hat\gamma_{\pp'\dwa}^\dagger \hat\gamma_{\pp\dwa} \hat\gamma_{\pp'\upa}$
in the effective Hamiltonian. Note that, unlike the Landau interaction function $f_{\upa\dwa}$, this is not a diagonal coefficient. 
The exchange amplitude $\mathcal{A}^{\rm ex}_{\upa\dwa}$ instead enters the collision integral; it is related to $f^{\rm ex}_{\upa\dwa}$ by the integral equation:
\be
\bar{\mathcal{A}}^{\rm ex}_{\upa\dwa}(\cos\varphi)=\bar{f}_{\upa\dwa}^{\rm ex}(\cos\varphi) + \int_0^{2\pi}\dd\phi \int_{0}^{\pi}\sin\theta\dd\theta
\bar{f}_{\upa\dwa}^{\rm ex}(\sin\theta\cos\phi)\bar{\mathcal{A}}^{\rm ex}_{\upa\dwa}(\sin\theta\cos(\varphi-\phi)) 
\ee
In an SU(2)-symmetric Fermi liquid, this equation is redundant with the Bethe-Salpeter equations \eqref{BS}
for $\mathcal{A}_{\upa\dwa}^{\rm fwd}$ and $\mathcal{A}_{\upa\upa}^{\rm fwd}$ since
\bea
{f}_{\upa\dwa}^{\rm ex}&\underset{\text{SU(2)}}{=}&f_{\upa\dwa}-f_{\upa\upa}\\
\mathcal{A}_{\upa\dwa}^{\rm ex}&\underset{\text{SU(2)}}{=}&\mathcal{A}_{\upa\dwa}^{\rm fwd}-\mathcal{A}_{\upa\upa}^{\rm fwd}
\eea
The relation between the exchange amplitude and the forward amplitude provided by the SU(2) symmetry
is cleverly used by Dy and Pethick \cite{DyPethick1969} to construct an approximation of the scattering amplitude at arbitrary angles:
\bea
\mathcal{A}_{\upa\dwa}(\alpha,\beta|\gamma,\delta)&\approx&\mathcal{A}_{\upa\dwa}^{\rm fwd}(\cos(\widehat{\alpha,\beta}))\cos^2\frac{\phi_{\alpha\to\delta}}{2}+{\mathcal{A}_{\upa\dwa}^{\rm ex}(\cos(\widehat{\alpha,\beta}))} \sin^2\frac{\phi_{\alpha\to\delta}}{2} \\
\mathcal{A}_{\upa\upa}(\alpha,\beta|\gamma,\delta)&\approx&\mathcal{A}_{\upa\upa}^{\rm fwd}(\cos(\widehat{\alpha,\beta}))\cos{\phi_{\alpha\to\delta}}
\eea
where $\phi_{\alpha\to\delta}$ is the azimuthal angle between $\alpha$ and $\delta$ in the spherical frame\footnote{$\phi_{\alpha\to\delta}=\phi_\delta^{\rm b}-\phi_\alpha^{\rm b}$ in the notation of Fig.~\ref{figBogo}} where $\textbf{e}_z\parallel(\alpha+\beta)$.
Unlike the forward-scattering approximation [$\mathcal{A}_{\sigma\sigma'}(\alpha,\beta|\gamma,\delta)\approx\mathcal{A}_{\sigma\sigma'}^{\rm fwd}(\cos(\widehat{\alpha,\beta}))$],
this approximation respects the Pauli principle. Although uncontrolled, it is often remarkably effective, as we have checked for the contact gas.

\paragraph{Bogoliubov channel} \label{BSBogolioubov}
We finally turn to frontal scattering:
\be
q\to0\iff \begin{cases}\alpha\to-\beta,\\ \gamma\to-\delta \end{cases}
\ee
Once again, the incoming and outgoing vectors become orthogonal to $q$ (cf. Fig.~\ref{figBogo}).
\be \label{thetapi2Bog}
\text{lim}_{q\to0}\theta_\alpha, \theta_\delta=\frac{\pi}{2}
\ee
with $\theta_i=(\widehat{i,\qq})$ (recall that $\qq=(\alpha+\beta)/2$ here).
This time, the amplitude $\mathcal{A}$ does not admit a finite non-zero limit when $q$ and $\Lambda$ simultaneously go to zero.
We therefore introduce
\be
\mathcal{A}^{\rm b}_{\sigma\sigma'}(\cos\varphi,q,\Lambda)=\mathcal{A}^{\Lambda}_{\sigma\sigma'}(\alpha,-\alpha+\qq|-\delta+\qq,\delta)
\ee
At this stage, this is simply a reparametrization in terms of $\varphi=(\widehat{\alpha,\delta})$ and $q$ of the scattering amplitude for four quasiparticles at the Fermi level ($\epsilon_\alpha=\epsilon_{-\alpha+\qq}=\epsilon_{-\delta+\qq}=\epsilon_\delta=\mu$). The limit of this amplitude for $q\to0$ at fixed $\Lambda$ defines the \textit{pairing function}
\be \label{gssP}
g_{\sigma\sigma'}(\cos\varphi,\Lambda)=\underset{q\to 0}{\text{lim}}\mathcal{A}^{\rm b}_{\sigma\sigma'}(\cos\varphi,\Lambda,q)
\ee
which can be seen as the Bogoliubov-channel counterpart of the Hartree interaction function $f_{\sigma\sigma'}$. Unlike $f_{\sigma\sigma'}$,
the function $g_{\sigma\sigma'}$ retains a logarithmic dependence on $\Lambda$ in the limit $\Lambda\to0$. However, once this dependence is regularized, $g_{\sigma\sigma'}$ provides effective parameters describing equilibrium pairing physics, notably the gap $\Delta$ and the superfluid critical temperature $T_c$.

The limit $\Lambda\to0$ at fixed $q$ yields instead the scattering amplitude
\be
\mathcal{A}^{\rm frt}_{\sigma\sigma'}(\cos\varphi,q)=\underset{\Lambda\to0}{\text{lim}}\mathcal{A}^{\rm b}_{\sigma\sigma'}(\cos\varphi,\Lambda,q)
\ee
whose behavior at $q\to0$, i.e. for quasi-frontal scattering, will be studied in more detail below.

We return to the flow equation \eqqref{dAud}. At small $q$, it is dominated by the Bogoliubov channel, and the energy conservation relation ($\epsilon_{\qq-\pp}+\epsilon_\pp-2\mu=\pm\Lambda$) can be rewritten as
\be \label{rezBogo}
\bar q { \cos\theta}=(2e\pm1)\bar\Lambda
\ee
where
\be \label{energiereduite}
e\equiv\frac{\epsilon_\pp-\mu}{\Lambda}=\frac{\bar\epsilon}{\bar\Lambda}
\ee
is the energy of the intermediate quasiparticle measured in units of $\Lambda$.
Thanks to the dependence on $e$ on the right-hand side of \eqqref{rezBogo},
the resonance is possible already for $\Lambda\gg\vF q$, and the number of $\Lambda$-resonant intermediate states
grows like $1/\Lambda$, as long as $\Lambda> \vF q$.
After integration over the reduced energy $e$, the flow equation becomes, up to $O(q)$ terms\footnote{We approximate the transition amplitudes
to the intermediate state by pairing amplitudes $\mathcal{A}^{\Lambda}(\alpha,-\alpha+\qq|-\pp+\qq,\pp)\simeq\mathcal{A}^{\rm b}(\cos(\widehat{\alpha,\pp}),q,\Lambda)$ with
$\cos(\widehat{\alpha,\pp})=\sin\theta\cos\phi$
and $\mathcal{A}^{\Lambda}(\pp,-\pp+\qq|-\delta+\qq,\delta)\simeq\mathcal{A}^{\rm b}(\cos(\widehat{\delta,\pp}),q,\Lambda)$ with
$\cos(\widehat{\delta,\pp})=\sin\theta\cos(\varphi-\phi)$. The angles
are taken in a spherical frame with $\textbf{e}_z\parallel\qq$ and $\textbf{e}_x\parallel\alpha$.}:
\begin{multline}
\frac{\dd\bar{\mathcal{A}}^{\rm b}_{\sigma\sigma'}}{\dd\bar\Lambda}(\cos\varphi,q,\Lambda)=
\int_0^{2\pi}\dd\phi\int_0^\pi\sin\theta\dd\theta \bar{\mathcal{A}}^{\rm b}_{\sigma\sigma'}(\sin\theta\cos\phi,q,\Lambda)\bar{\mathcal{A}}^{\rm b}_{\sigma\sigma'}(\sin\theta\cos(\varphi-\phi),q,\Lambda)\\\times \frac{\Theta\bb{\bar\Lambda/\bar q-|\cos\theta|}}{\bar\Lambda}
\label{dApaire}
\end{multline}
This can be solved (see note \ref{primitiveflot}) using the primitive of $\Theta({\bar\Lambda}/{\bar q}-|u|)/\bar \Lambda$ (with $u=\cos\theta$):
\begin{multline}
\bar{\mathcal{A}}^{\rm b}_{\sigma\sigma'}(\cos\varphi,q,\Lambda_f)=\bar{\mathcal{A}}^{\rm b}_{\sigma\sigma'}(\cos\varphi,q,\Lambda_0)+
\int_0^{2\pi}\dd\phi\int_{-1}^1 \dd u 
\bar{\mathcal{A}}^{\rm b}_{\sigma\sigma'}(\sin\theta\cos\phi,q,\Lambda_0)
\\\times
\bar{\mathcal{A}}^{\rm b}_{\sigma\sigma'}(\sin\theta\cos(\varphi-\phi),q,\Lambda_f)
\bbcro{\text{ln}\frac{\bar\Lambda_f}{\bar q |u|}\Theta\bb{{\bar\Lambda_f}/{\bar q}-|u|}-\text{ln}\frac{\bar\Lambda_0}{\bar q |u|}\Theta\bb{{\bar\Lambda_0}/{\bar q}-|u|}}
\label{flotApaire}
\end{multline}
We can distinguish two regimes in this flow: a first regime with $\bar\Lambda_f,\bar\Lambda_0>\bar q$ describing the logarithmic flow
of $g_{\sigma\sigma'}(\Lambda)$. A second regime with $\bar\Lambda_0\approx\bar q$ and $\Lambda_f\ll\bar q$ converts $g_{\sigma\sigma'}$ into the frontal scattering amplitude $\mathcal{A}^{\rm frt}$.

\subparagraph{Logarithmic flow of $g_{\sigma\sigma'}(\Lambda)$}
For $\bar\Lambda_f,\bar\Lambda_0>\bar q$, the Heaviside functions in \eqqref{flotApaire} can be replaced by 1. Thus,
\be
\begin{split}
\bar g_{\sigma\sigma'}(\cos\varphi,\Lambda_f)&=\bar g_{\sigma\sigma'}(\cos\varphi,\Lambda_0)\\ & +\text{ln}\frac{\Lambda_f}{\Lambda_0}
\int_0^{2\pi}\dd\phi\int_0^\pi\sin\theta\dd\theta \bar g_{\sigma\sigma'}(\sin\theta\cos\phi,\Lambda_0) \bar g_{\sigma\sigma'}(\sin\theta\cos(\varphi-\phi),\Lambda_f)
\end{split}
\ee
We introduce the dimensionless rescaling of $g_{\sigma\sigma'}$ in Fermi units:
\be \label{adimg}
\bar g_{\sigma\sigma'}=\frac{m^\ast \pF}{(2\pi)^3} g_{\sigma\sigma'}
\ee
Assuming the initial condition $g_{\sigma\sigma'}(\cos\varphi,\Lambda_0)$ is known, this integral equation on the function $\varphi\mapsto g_{\sigma\sigma'}(\cos\varphi,\Lambda_f)$ is solved
by expanding in Legendre polynomials,
\be \label{gl}
\bar g_{\sigma\sigma'}(\cos\varphi,\Lambda)=\sum_{l=0}^{+\infty} \bar g_{\sigma\sigma'}^l(\Lambda) P_l(\cos\varphi)
\ee
and using the addition theorem of spherical harmonics\footnote{We move to a spherical frame with axis $\textbf{e}_z\parallel\alpha$ where $\delta/\pF=(\sin\varphi,0,\cos\varphi)$ and $\pp/\pF=(\sin\tilde\theta\cos\tilde\phi,\sin\tilde\theta\sin\tilde\phi,\cos\tilde\theta)$:
\begin{multline}
\bar g_{\sigma\sigma'}(\cos\varphi,\Lambda_f)-\bar g_{\sigma\sigma'}(\cos\varphi,\Lambda_0)\\=\text{ln}\frac{\Lambda_f}{\Lambda_0}\int_0^{2\pi}\dd\tilde\phi\int_0^\pi\sin\tilde\theta\dd\tilde\theta \bar g_{\sigma\sigma'}(\cos\tilde\theta,\Lambda_0)\bar g_{\sigma\sigma'}(\cos\tilde\theta\cos\tilde\theta'+\sin\tilde\theta\sin\tilde\theta'\cos(\varphi-\tilde\phi)),\Lambda_f) \\
=\text{ln}\frac{\Lambda_f}{\Lambda_0}\sum_{l=0}^{+\infty}\frac{4\pi}{2l+1}\bar g_{\sigma\sigma'}^l(\Lambda_0) \bar g_{\sigma\sigma'}^l(\Lambda_f) P_l(\cos\varphi)
\end{multline}}:
\be
\bar g_{\sigma\sigma'}^l(\Lambda_f)=\frac{\bar g_{\sigma\sigma'}^l(\Lambda_0)}{1-\frac{4\pi}{2l+1}\bar g_{\sigma\sigma'}^l(\Lambda_0)\text{ln}\frac{\Lambda_f}{\Lambda_0}}
\label{gfg0}
\ee
This yields a logarithmic suppression of the pairing function $g_{\sigma\sigma'}\underset{\Lambda_f\to0}{\propto}1/\text{ln}\Lambda_f$.
This suppression comes from the arbitrary separation between high 
and low energies at $\Lambda$; in low-energy physical observables such as the superfluid critical temperature $T_c$, this logarithmic (infrared) suppression
is systematically compensated by an ultraviolet logarithmic divergence. The physics is thus only sensitive to the residual part of the pairing function:
\be
\frac{1}{G_{\sigma\sigma'}^l}\equiv\underset{\Lambda\to0}{\text{lim}}\bbcro{{\frac{1}{\bar g_{\sigma\sigma'}^l(\Lambda)}+\frac{4\pi}{2l+1}\text{ln}\frac{2\Lambda}{\vF\pF}}}
\label{tildeg}
\ee
Eq.~\eqref{gfg0} shows that $G_{\sigma\sigma'}$ is a conserved quantity under the final stage of the renormalization flow: $\dd G_{\sigma\sigma'}^l/\dd\Lambda=0$ for $\Lambda\ll\vF\pF$.
This logarithmic behavior of $1/g_{\sigma\sigma'}(\Lambda)$ was first postulated by Popov \cite{Popov1987-III14},
in 3D and for the s- and p-wave channels corresponding to the superfluid phases of Helium-3.
It was later proven (in 2D and for all Fourier components) by Chitov and Sénéchal \cite{Senechal1995}
using the flow equations of the functional renormalization group.

\subparagraph{Bethe-Salpeter equation for the frontal scattering amplitude}
As in the Hartree channel, the nature of the flow changes abruptly when $\Lambda$ reaches $\vF q$. We place ourselves at the onset of this fast flow with $\Lambda_0=\vF q$,
so that $\bar{\mathcal{A}}^{\rm b}_{\sigma\sigma'}(\Lambda_0)$ still coincides with the function $g_{\sigma\sigma'}$, and let $\bar\Lambda_f/\bar q \to 0$ so that $\bar{\mathcal{A}}^{\rm b}_{\sigma\sigma'}(\Lambda_f)$ becomes the head-on scattering amplitude $\mathcal{A}^{\rm frt}_{\sigma\sigma'}$. The flow equation \eqref{flotApaire} immediately gives a relation between the two:
\be
\begin{split}
\bar{\mathcal{A}}^{\rm frt}_{\sigma\sigma'}(\cos\varphi,q)=&\bar g_{\sigma\sigma'}(\cos\varphi,\vF q)\\
& +
\int_0^{2\pi}\dd\phi\int_{-1}^1 \dd u 
\bar g_{\sigma\sigma'}(\sin\theta\cos\phi,\vF q)
\bar{\mathcal{A}}^{\rm frt}_{\sigma\sigma'}(\sin\theta\cos(\varphi-\phi),q)
\text{ln} |u|
\end{split}
\label{BetheSalpeterpaire}
\ee
This Bethe-Salpeter equation is the main result of this section; note that it no longer depends on any cutoff, since the function $\bar g_{\sigma\sigma'}$, evaluated at $\Lambda=\vF q$, is fully determined by the renormalized function $G_{\sigma\sigma'}$:
\be
\bar g_{\sigma\sigma'}^l(\vF q)=\frac{1}{\frac{1}{G_{\sigma\sigma'}^l}-\frac{4\pi}{2l+1}\text{ln}2\bar q}
\ee
Eq.~\eqref{BetheSalpeterpaire} allows one to deduce
the head-on component of the scattering amplitude from measurements of equilibrium quantities.
A measurement of $\Delta$ or $T_c$ can thus provide (partial) information on $\mathcal{A}^{\rm frt}_{\sigma\sigma'}$,
exactly as the Bethe-Salpeter equation in the Hartree channel
allowed, in helium-3, an estimate of the forward scattering amplitude $\mathcal{A}^{\rm fwd}_{\sigma\sigma'}$
from thermodynamic measurements of the Landau parameters $f_{\sigma\sigma'}^l$.

The logarithmic angular kernel (the $\text{ln} |u|$ in \eqref{BetheSalpeterpaire}) however complicates the relation between ${\mathcal{A}}^{\rm frt}_{\sigma\sigma'}$
and $G_{\sigma\sigma'}$. As a concrete example, assume $G_{\sigma\sigma'}$ is angle-independent
(that is $G_{\sigma\sigma'}^l=G_{\sigma\sigma'}\delta_{l0}$). In this case, $\bar{\mathcal{A}}^{\rm frt}_{\sigma\sigma'}$
and $\bar g_{\sigma\sigma'}$ are also angle-independent and given by
\be
\frac{1}{\bar g_{\sigma\sigma'}(\Lambda)}={\frac{1}{G_{\sigma\sigma'}}-{4\pi}\text{ln}2\bar \Lambda}
\ee
\be
\frac{1}{\mathcal{A}^{\rm frt}_{\sigma\sigma'}(q)}={\frac{1}{G_{\sigma\sigma'}}-4\pi\bb{\text{ln}2\bar q-1}}
\ee

\begin{figure}
\begin{center}

\rule{\textwidth}{0.5pt}

\vspace{0.3cm}
{\Large Bogolioubov}
\vspace{0.1cm}
\makebox[\textwidth]{\dotfill}

\begin{tabular}[b]{cc}

\begin{tikzpicture}[xscale=3.7,yscale=2.5]
\coordinate (A) at (-0.5,0);
\coordinate (B) at (0.5,0);
\def\xA{-0.5}
\def\yA{0}
\def\xB{0.5}
\def\yB{0}
\draw (A) node{\textbullet};
\draw (B) node{\textbullet};
\draw[thin,dashed] (\xB,0.9) -- (\xB,-0.55);
\draw[thin,dashed] (\xA,0.9) -- (\xA,-0.55);
\draw[->-] (A) arc (135:45:{1/sqrt(2.)}) ;
\draw (0,0.4) node {$\pp_a\sigma$};
\draw (0,-0.4) node {$\pp_b\sigma'$};
\draw[->-] (A) arc (-135:-45:{1/sqrt(2.)});
\draw[->-] (\xA-0.3,\yA+0.5) -- node[above=0.1cm]{$\gamma\sigma'$} (A);
\draw[->-] (\xA-0.3,\yA-0.5) -- node[below=0.1cm]{$\delta\sigma$} (A);
\draw[->-] (B) -- node[above=0.1cm]{$\alpha\sigma$} (\xB+0.3,\yB+0.5);
\draw[->-] (B) -- node[below=0.1cm]{$\beta\sigma'$} (\xB+0.3,\yB-0.5);
\draw (0,0.8) node { {\scriptsize $\ket{n_{\rm I}}=\hat\gamma_{\pp_a\sigma}^\dagger \hat\gamma_{\pp_b\sigma'}^\dagger \hat\gamma_{\gamma\sigma'} \hat\gamma_{\delta\sigma}\ket{i}$}};
\draw (-0.75,0.8) node { {$\ket{i}$}};
\draw (0.75,0.8) node { {$\ket{f}$}};
\end{tikzpicture}
& \hspace{0.5cm}
\begin{tikzpicture}[xscale=3,yscale=2.5]
\coordinate (A) at (1,0);
\coordinate (B) at (-1,0);
\def\xA{-0.5}
\def\yA{0}
\def\xB{0.5}
\def\yB{0}
\draw (A) node{\textbullet};
\draw (B) node{\textbullet};
\draw[-<-] (B) arc (120:60:{2}) ;
\draw (0,0.4) node {$\pp_a\sigma$};
\draw (0,-0.4) node {$\pp_b\sigma'$};
\draw[-<-] (B) arc (-120:-60:{2});
\draw[->-] (\xA+0.7,\yA+0.5) -- node[above=0.1cm]{$\gamma\sigma'$} (A);
\draw[->-] (\xA+0.7,\yA-0.5) -- node[below=0.1cm]{$\delta\sigma$} (A);
\draw[->-] (B) -- node[above=0.1cm]{$\alpha\sigma$} (\xB-0.7,\yB+0.5);
\draw[->-] (B) -- node[below=0.1cm]{$\beta\sigma'$} (\xB-0.7,\yB-0.5);
\draw (0,0.8) node {\scriptsize $\ket{n_{\rm II}}=\hat\gamma_{\alpha\sigma}^\dagger \hat\gamma_{\beta\sigma'}^\dagger \hat\gamma_{\pp_b\sigma'} \hat\gamma_{\pp_a\sigma}\ket{i}$};
\end{tikzpicture}
\end{tabular}

\rule{\textwidth}{0.5pt}

\vspace{0.3cm}
\noindent
\makebox[\textwidth]{%
\hfill
\makebox[0pt][c]{{\Large Fock}}%
\hfill
\makebox[0pt][r]{\scriptsize $\begin{cases} \sigma_a=\sigma,\ \sigma_b=\sigma'\ \ \, \text{ if }\sigma\neq\sigma'\\\sigma_a=\sigma_b=\upa \text{ and } \dwa \text{ if }\sigma=\sigma'\end{cases}$}%
}
\makebox[\textwidth]{\dotfill}

\hspace{-0.4cm}
\begin{tabular}[b]{cc}
\begin{tikzpicture}[xscale=3.7,yscale=1.7]
\coordinate (A) at (-0.5,0.5);
\coordinate (B) at (0.5,-0.5);
\def\xA{-0.5}
\def\yA{0.5}
\def\xB{0.5}
\def\yB{-0.5}
\draw (A) node{\textbullet};
\draw (B) node{\textbullet};
\draw[thin,dashed] (\xB,1.5) -- (\xB,-1.1);
\draw[thin,dashed] (\xA,1.5) -- (\xA,-1.1);
\draw (0.25,0.4) node {$\pp_a\sigma_a$};
\draw (-0.25,-0.4) node {$\pp_b\sigma_b$};
\draw[-<-] (A) arc (90:0:{1});
\draw[->-] (A) arc (-180:-90:{1});
\draw[->-] (\xA-0.3,\yA+0.5) -- node[below=0.2cm]{$\gamma\sigma'$} (A);
\draw[->-] (\xB-0.3,\yB-0.5) -- node[below=0.1cm]{$\delta\sigma$} (B);
\draw[->-] (A) -- node[above=0.1cm]{$\alpha\sigma$} (\xA+0.3,\yA+0.5);
\draw[->-] (B) -- node[right=0.1cm]{$\beta\sigma'$} (\xB+0.3,\yB-0.5);
\draw (0,1.4)  node {\scriptsize $\ket{n_{\rm III}}=\hat\gamma_{\alpha\sigma}^\dagger \hat\gamma_{\pp_b\sigma_b}^\dagger \hat\gamma_{\gamma\sigma'} \hat\gamma_{\pp_a\sigma_a}\ket{i}$};
\draw (-0.75,1.4) node { {$\ket{i}$}};
\draw (0.75,1.4) node { {$\ket{f}$}};
\end{tikzpicture}
& \hspace{0.5cm}
\begin{tikzpicture}[xscale=3.7,yscale=1.7]
\coordinate (A) at (0.5,0.5);
\coordinate (B) at (-0.5,-0.5);
\def\xA{0.5}
\def\yA{0.5}
\def\xB{-0.5}
\def\yB{-0.5}
\draw (A) node{\textbullet};
\draw (B) node{\textbullet};
\draw (0.5,-0.3) node {$\pp_a\sigma_a$};
\draw (-0.5,0.3) node {$\pp_b\sigma_b$};
\draw[-<-] (A) arc (0:-90:{1});
\draw[->-] (A) arc (90:180:{1});
\draw[->-] (\xA-0.3,\yA+0.5) -- node[left=0.2cm]{$\gamma\sigma'$} (A);
\draw[->-] (\xB-0.3,\yB-0.5) -- node[below=0.1cm]{$\delta\sigma$} (B);
\draw[->-] (A) -- node[above=0.1cm]{$\alpha\sigma$} (\xA+0.3,\yA+0.5);
\draw[->-] (B) -- node[right=0.1cm]{$\beta\sigma'$} (\xB+0.3,\yB-0.5);
\draw (0,1.4)  node {\scriptsize $\ket{n_{\rm IV}}=\hat\gamma_{\pp_a\sigma_a}^\dagger \hat\gamma_{\beta\sigma'}^\dagger \hat\gamma_{\pp_b\sigma_b} \hat\gamma_{\delta\sigma}\ket{i}$};
\end{tikzpicture}
\end{tabular}

\rule{\textwidth}{0.5pt}

\vspace{0.3cm}
\noindent
\makebox[\textwidth]{%
\hfill
\makebox[0pt][c]{{\Large Hartree}}%
\hfill
\makebox[0pt][r]{$\sigma_a=\upa$ \text{ and } $\dwa$}%
}
\makebox[\textwidth]{\dotfill}

\hspace{-0.4cm}
\begin{tabular}[b]{cc}
\begin{tikzpicture}[xscale=3.7,yscale=1.7]
\coordinate (A) at (-0.5,0.5);
\coordinate (B) at (0.5,-0.5);
\def\xA{-0.5}
\def\yA{0.5}
\def\xB{0.5}
\def\yB{-0.5}
\draw (A) node{\textbullet};
\draw (B) node{\textbullet};
\draw[thin,dashed] (\xB,1.5) -- (\xB,-1.1);
\draw[thin,dashed] (\xA,1.5) -- (\xA,-1.1);
\draw (0.25,0.4) node {$\pp_a\sigma_a$};
\draw (-0.25,-0.4) node {$\pp_b\sigma_a$};
\draw[-<-] (A) arc (90:0:{1});
\draw[->-] (A) arc (-180:-90:{1});
\draw[->-] (\xA-0.3,\yA+0.5) -- node[below]{$\delta\sigma$} (A);
\draw[->-] (\xB-0.3,\yB-0.5) -- node[below=0.1cm]{$\gamma\sigma'$} (B);
\draw[->-] (A) -- node[above=0.1cm]{$\alpha\sigma$} (\xA+0.3,\yA+0.5);
\draw[->-] (B) -- node[right=0.1cm]{$\beta\sigma'$} (\xB+0.3,\yB-0.5);
\draw (0,1.4)  node {\scriptsize $\ket{n_{\rm V}}=\hat\gamma_{\alpha\sigma}^\dagger \hat\gamma_{\pp_b\sigma_a}^\dagger  \hat\gamma_{\pp_a\sigma_a} \hat\gamma_{\delta\sigma}\ket{i}$};
\draw (-0.75,1.4) node { {$\ket{i}$}};
\draw (0.75,1.4) node { {$\ket{f}$}};
\end{tikzpicture}
& \hspace{0.5cm}
\begin{tikzpicture}[xscale=3.7,yscale=1.7]
\coordinate (A) at (0.5,0.5);
\coordinate (B) at (-0.5,-0.5);
\def\xA{0.5}
\def\yA{0.5}
\def\xB{-0.5}
\def\yB{-0.5}
\draw (A) node{\textbullet};
\draw (B) node{\textbullet};
\draw (0.5,-0.3) node {$\pp_a\sigma_a$};
\draw (-0.5,0.3) node {$\pp_b\sigma_a$};
\draw[-<-] (A) arc (0:-90:{1});
\draw[->-] (A) arc (90:180:{1});
\draw[->-] (\xA-0.3,\yA+0.5) -- node[left=0.1cm]{$\delta\sigma$} (A);
\draw[->-] (\xB-0.3,\yB-0.5) -- node[below=0.1cm]{$\gamma\sigma'$} (B);
\draw[->-] (A) -- node[above=0.1cm]{$\alpha\sigma$} (\xA+0.3,\yA+0.5);
\draw[->-] (B) -- node[right=0.1cm]{$\beta\sigma'$} (\xB+0.3,\yB-0.5);
\draw (0,1.4)  node {\scriptsize $\ket{n_{\rm VI}}=\hat\gamma_{\pp_a\sigma_a}^\dagger \hat\gamma_{\beta\sigma'}^\dagger \hat\gamma_{\gamma\sigma'} \hat\gamma_{\pp_b\sigma_a} \ket{i}$};
\end{tikzpicture}
\end{tabular}
\end{center}
\caption{\label{diagrammesupdw} Diagrams representing the transition $\delta\sigma,\gamma\sigma'\ \to\ \alpha\sigma,\beta\sigma'$
via intermediate states with two virtual quasiparticles (curved lines representing the momenta $\pp_a$ and $\pp_b$), and to second order in $\hat H(\Lambda)$ (the two effective two-body interaction amplitudes are represented by the dots \textbullet).
These diagrams dominate the renormalization of the scattering amplitude $\mathcal{A}_{\sigma\sigma'}$ in the limit of weakly correlated quasiparticles.
The Hartree diagrams (V and VI), as well as the Fock diagrams (III and IV) for collisions of indistinguishable quasiparticles ($\sigma=\sigma'$), involve a sum over the spin ($\sigma_a$) of the intermediate quasiparticle.}
\end{figure}

%% file: heff_eqtransport.tex
\section{Derivation of the Fermi liquid kinetic equations}
\label{sec:demotransport}

Knowing the energy and transition amplitudes of the quasiparticle fluid,
we can now describe its dynamics by a kinetic equation. 
Our formulation of Fermi liquid theory is conceptually divided in two steps: 
after the renormalization flow used to encapsulate high-energy processes into the effective
Hamiltonian $\hat H(\Lambda)$, there remains to study 
the low-energy physics described by $\hat H(\Lambda)$,
which includes the long-wavelength/low-energy dynamics.
This point of view is natural from the perspective of low-energy effective theories, 
but not for other RG approaches where the flow directly obtains the dynamical response functions
\cite{Shankar1994,Polchinski1992,Senechal1995,Senechal1998,Dupuis1998}.

The main advantage of the effective picture
is to convert the dynamics of the fluid into a weakly-correlated
problem, where the collisional dynamics in particular can be treated under the Born-Markov
approximation.
We recall that Fermi systems at intermediate temperatures $T\approx \TF$
and strong interactions \cite{vtemp}
do not obey a kinetic equation, as there is no separation of timescales
to break the BBGKY hierarchy.
By introducing the long-lived states $\ket{\psi}=\eee^{\hat S} \ket{\psi}_0$,
the quasiparticle description manages to overcome this limitation in the low temperature
limit $T\ll \TF$, regardless of the interaction strength (as long as the quasiparticle picture
holds).

This section uses the effective Hamiltonian \eqqref{HFS} to derive 
the kinetic equation, and discuss its domain of validity.
In Sec.~\ref{sec:lifetime}, we consider the case of a homogeneous system
with an out-of-equilibrium quasiparticle distribution. We show that if the cloud of excited quasiparticles
is contained in a low-energy shell $\epsilon_0\ll\EF$, one can treat the evolution of the quasiparticle
distribution in the Born-Markov approximation. This results in a nonlinear kinetic equation, from
which we extract the thermal lifetime of the quasiparticles. 

In Sec.~\ref{sec:transportT} and \ref{sec:transportT0}, we study
transport phenomena, where the quasiparticle gas is excited by a
perturbation periodic in space and time, at frequency $\omega$ and wavenumber $q$.
We assume that the corresponding energy scales are comparable and small compared to 
the Fermi energy, such that we can insert the cutoff $\Lambda$ between the two scales:
\be
\vF q\ \approx\ \omega\ll\Lambda\ll\EF
\ee

In the presence of the dynamical parameters $q$ and $\omega$, there exist several ways to take the low temperature limit.
In Sec.~\ref{sec:transportT}, we derive the collisional transport equation
in the limit
\be
\frac{T}{\TF}\to 0 \text{ at fixed } \vF q\tau,\ \omega\tau
\ee
where  the mean collision time $\tau$ scales, as
we shall see, as $1/T^2$. In this regime, all the lower bounds on $\Lambda$ in \eqqref{GammaLambda}
are of order $T^2$:
\be
\vF q,\ \omega,\ \Gamma_{\rm typ}\approx T^2 \label{Gammaomega}
\ee
Varying the parameter $\omega\tau$ (after the limit $T\to0$ is taken), 
this regime describes the crossover from hydrodynamic ($\omega\tau\ll1$) to
collisionless transport ($\omega\tau\gg1$).

In Sec.~\ref{sec:transportT0} instead, we take the limit:
\be
\frac{T}{\TF}\to 0 \text{ at fixed } \vF q,\ \omega
\ee
In this regime, the excitation energy $\vF q$
sets the high-energy tail of the quasiparticle distribution 
and the lower bound on $\Lambda$.
The collision integral vanishes as $(\vF q/\EF)^2$, such that
transport is collisionless to leading order in $q$.

\subsection{Kinetic equation in a spatially homogeneous state}
\label{sec:lifetime}

\paragraph{Equation of motion of the quasiparticle distribution} 
We assume that the initial state $\hat\varrho$ of the system
describes an uncorrelated distribution of quasiparticles
\be
\hat \varrho(0)=\prod_{\alpha\sigma}\hat\varrho_{\alpha\sigma}
\ee
The reduced density matrix $\hat\varrho_{\alpha\sigma}$ is a function of $\hat \gamma_{\alpha\sigma}$ and
$\hat \gamma_{\alpha\sigma}^\dagger$ only which defines the occupation of mode $\alpha\sigma$
\be
n_{\alpha\sigma}=\text{Tr}(\hat\varrho_{\alpha\sigma} \hat n_{\alpha\sigma})
\ee 
We further assume that the excited quasiparticles are contained in a low-energy shell of width $p_0$
\be
 n_{\pp\sigma}=n_\pp^0\quad \text{ for }\quad |p-\pF|>p_0 \label{conditionBE}
\ee
Note that this is more restrictive than just assuming a ``low density of excitation'' ${\sum_{\alpha\sigma} \delta n_{\alpha\sigma}}\ll{N}$.
In fact, exciting a small energy density in highly energetic modes would result in a breakdown of the quasiparticle picture.

We describe the evolution of $n_{\pp\sigma}$ in the Heisenberg picture using the effective Hamiltonian \eqqref{HFS}.
The equation of motion is triggered only by the off-diagonal part $\hat H_4^{\rm x}$ (see the decomposition of $\hat H$
in \eqqref{expHp}):
\be
\ii\partial_t   n_{\pp\sigma}=\frac{1}{2L^3} \sum_{\substack{\pp_2,\pp_3,\pp_4\\\sigma'=\upa\dwa}} 
     \Bigg[s_{\sigma\sigma'}\mathcal{A}_{\sigma\sigma'}\bb{\pp,\pp_2|\pp_3,\pp_4}
    \delta_{\pp+\pp_2}^{\pp_3+\pp_4} 
    \meanv{\hat{\gamma}^{\dagger}_{\pp, \sigma}\hat{\gamma}^{\dagger}_{\pp_2 \sigma'} \hat{\gamma}_{\pp_3\sigma'}\hat{\gamma}_{\pp_4\sigma}}-\text{c.c.}\Big]+O(\delta(\hat\gamma^\dagger\hat\gamma))^3 \label{dtnp} 
\ee
where $\meanv{\hat O}\equiv \text{Tr}(\hat\varrho \hat O)$ and $s_{\upa\dwa}=1$ and $s_{\upa\upa}=1/2$ is a counting factor. Notice that the matrix elements $\mathcal{B}_{\sigma\sigma'}$ have been combined to form the amplitudes $\mathcal{A}_{\sigma\sigma'}$. 

\paragraph{Born-Markov approximation} 
\eqqref{dtnp} is not a closed system, due to the presence of terms quartic in $\hat\gamma$. 
To perform a Born-Markov approximation on the dynamics of those quartic terms, similar 
to the classical ``molecular chaos hypothesis'',  
we introduce the quartic cumulants
\bea
\hat Q_{\alpha\beta\gamma\delta}^{\sigma\sigma'} &\equiv&   ({ \hat{\gamma}^{\dagger}_{\alpha \sigma}\hat{\gamma}^{\dagger}_{\beta \sigma'} \hat{\gamma}_{\gamma\sigma'}\hat{\gamma}_{\delta\sigma}})_c
    \\
    \bigl(\hat{a}^\dagger\hat{b}^\dagger\hat{c}\hat{d}\bigr)_{c} &\equiv &
    \hat{a}^\dagger\hat{b}^\dagger\hat{c}\hat{d}
     -\hat{a}^\dagger\hat{d}\langle \hat{b}^\dagger\hat{c} \rangle
    - \hat{b}^\dagger\hat{c}\langle \hat{a}^\dagger\hat{d} \rangle
    + \hat{a}^\dagger\hat{c}\langle \hat{b}^\dagger\hat{d} \rangle
    +\hat{b}^\dagger\hat{d}\langle \hat{a}^\dagger\hat{c} \rangle\notag
     \\&+& \meanv{\hat{a}^\dagger\hat{d}}\meanv{ \hat{b}^\dagger\hat{c} } 
     - \meanv{\hat{a}^\dagger\hat{c}}\meanv{ \hat{b}^\dagger\hat{d} }
    \label{expcumulant}
\eea
In the equation of motion of an hermitian operator such as $\delta \hat n_{\pp\sigma}$, the contracted terms 
\\ ${\hat{\gamma}^{\dagger}\hat{\gamma}^{\dagger} \hat{\gamma}\hat{\gamma}}-({\hat{\gamma}^{\dagger}\hat{\gamma}^{\dagger} \hat{\gamma}\hat{\gamma}})_c$ drop out\footnote{In \eqqref{dtnp} this is seen by cancelling the contraction $\pp=\pp_3$ or $\pp=\pp_4$ with its complex conjugate)}. This shows that $n_{\pp\sigma}$ is a conserved quantity and the equation of motion are integrable in the absence of collisions.

The quartic cumulant evolves according to the Heisenberg equation of motion
\be
\ii\partial_t \hat Q_{\alpha\beta\gamma\delta}^{\sigma\sigma'}=\bb{\hat \epsilon_{\delta\sigma}+\hat \epsilon_{\gamma\sigma'}-\hat\epsilon_{\beta\sigma'}-\hat\epsilon_{\alpha\sigma}}\hat Q_{\alpha\beta\gamma\delta}^{\sigma\sigma'}+ \hat S_{\alpha\beta\gamma\delta}^{\sigma\sigma'}(t)\label{dtQ}
\ee
where
\be
\hat S_{\alpha\beta\gamma\delta}^{\sigma\sigma'}\equiv[\hat Q_{\alpha\beta\gamma\delta}^{\sigma\sigma'},\hat H_4^{\rm x}]=\frac{1}{2L^3}\sum_{abcd}\mathcal{B}_{\sigma_a\sigma_b}(dc|ba) \delta_{a+b}^{c+d}[(\hat{\gamma}^{\dagger}_{\alpha \sigma}\hat{\gamma}^{\dagger}_{\beta \sigma'} \hat{\gamma}_{\gamma\sigma'}\hat{\gamma}_{\delta\sigma})_c,(\hat{\gamma}^{\dagger}_{d \sigma_a}\hat{\gamma}^{\dagger}_{c \sigma_b} \hat{\gamma}_{b\sigma_b}\hat{\gamma}_{a\sigma_a})_c] \label{commS}
\ee 
is the source term, and the instantaneous quasiparticle energy appears as an operator:
\be
\hk{\hat\epsilon_{\pp\sigma}=\epsilon_{\pp}+\frac{1}{L^3}\sum_{\pp'\sigma'}f_{\sigma\sigma'}(\pp,\pp')(\hat n_{\pp'\sigma'}-n_{p'}^0)} \label{energielocale}
\ee
The deviation
from the Fermi sea eigenenergy originates from $\hat H_4^{\rm d}$ (the diagonal part of the interaction Hamiltonian):
$[\hat \gamma_{\pp\sigma},\hat H_4^{\rm d}]=(\hat\epsilon_{\pp\sigma}-\epsilon_\pp)\hat \gamma_{\pp\sigma}$.

In the Born approximation, we assume that the correlations among quasiparticle modes remain small at all times.
We then replace $\meanv{\hat S}$ by its Wick contraction
\be
\meanvlr{\hat S_{\alpha\beta\gamma\delta}^{\sigma\sigma'}}=
      \frac{\mathcal{A}_{\sigma\sigma'}(\alpha\beta|\gamma\delta)}{L^3}\bbcro{n_{\alpha\sigma} n_{\beta\sigma'} \bar n_{\gamma\sigma'} \bar n_{\delta\sigma}-\bar n_{\alpha\sigma} \bar n_{\beta\sigma'}  n_{\gamma\sigma'}  n_{\delta\sigma}}  + O(p_0/\pF)^2 \label{Born}
\ee 
where $n_{\pp\sigma}(t)\equiv \meanv{\hat\gamma_{\pp\sigma}^\dagger \hat\gamma_{\pp\sigma}(t)}$, and we use the short-hand notations $\bar n=1-n$.
The contractions have imposed $(\sigma_a,\sigma_b)=(\sigma,\sigma')$ or $(\sigma_a,\sigma_b)=(\sigma',\sigma)$ and removed all the summations over momenta in \eqqref{commS}. Corrections to \eqqref{Born} beyond the Born approximation would involve at least two summations and are thus smaller a priori by a factor $(p_0/\pF)^2$.

With the Born approximation, the source term $\hat S$ becomes independent of $\hat Q$, which allows to time integrate \eqqref{dtQ}. Taking the average value in $\hat\varrho(t)$,
\hk{and omitting to time-dependence of the perturbed energies $\epsilon_{\pp\sigma}=\meanv{\hat\epsilon_{\pp\sigma}}$, we obtain}
\be
\meanvlr{ \hat Q_{\alpha\beta\gamma\delta}^{\sigma\sigma'}(t)}=-\ii\int_{- t}^0\dd t' \eee^{-\ii({\epsilon_{\delta\sigma}}+\epsilon_{\gamma\sigma'}-\epsilon_{\beta\sigma'}-\epsilon_{\alpha\sigma})t'}  \meanvlr{\hat S_{\alpha\beta\gamma\delta}^{\sigma\sigma'}(t+t') }
\ee
Modelling the slow time-dependence of $\hat S$ as $ \hat S_{\alpha\beta\gamma\delta}^{\sigma\sigma'}(t+t')= \hat S_{\alpha\beta\gamma\delta}^{\sigma\sigma'}(t)\eee^{\eta t'}$, and sending the initial condition $-\eta t\to-\infty$, we obtain the Markovian approximation of $\hat Q$:
\be
\meanvlr{ \hat Q_{\alpha\beta\gamma\delta}^{\sigma\sigma'}(t)}=-\frac{\meanvlr{\hat S_{\alpha\beta\gamma\delta}^{\sigma\sigma'}(t)}}{{\epsilon_{\delta\sigma}+\epsilon_{\gamma\sigma'}-\epsilon_{\beta\sigma'}-\epsilon_{\alpha\sigma}}+\ii\eta } \label{integrationformelle}
 \ee
 Anticipating on \eqqref{Icollmagn} which gives the time scale at which $n_{\alpha\sigma}(t)$ and $\hat S(t)$ vary, one can estimate $\eta=O(p_0/\pF)^2$.
 
\paragraph{Nonlinear kinetic equation}
Replacing the expression of $\meanv{\hat Q}$ in the kinetic equation \eqref{dtnp} and using the Plemelj formula $1/(x+\ii0^+)=\mathcal{P}(1/x)-\ii\pi\delta(x)$, we obtain
a non-linear kinetic equation
\begin{multline}
\partial_t   n_{\pp\sigma}= I_{\pp\sigma}[n]\equiv\frac{2\pi}{L^6}\sum_{\pp_2,\pp_3,\pp_4\in\mathcal{D}}\delta_{\pp+\pp_2}^{\pp_3+\pp_4}  \\
\Bigg(\delta(\epsilon_{\pp\sigma}+\epsilon_{\pp_2,-\sigma}-\epsilon_{\pp_3,-\sigma}-\epsilon_{\pp_4\sigma})
 W_{\upa\dwa}(\pp,\pp_2|\pp_3,\pp_4)\bbcro{\bar n_{\pp\sigma} \bar n_{\pp_2,-\sigma}  n_{\pp_3,-\sigma}  n_{\pp_4\sigma}- n_{\pp\sigma}  n_{\pp_2,-\sigma}  \bar n_{\pp_3,-\sigma}  \bar n_{\pp_4\sigma}}\\
+\frac{1}{2}\delta(\epsilon_{\pp\sigma}+\epsilon_{\pp_2\sigma}-\epsilon_{\pp_3\sigma}-\epsilon_{\pp_4\sigma})
 W_{\upa\upa}(\pp,\pp_2|\pp_3,\pp_4)\bbcro{\bar n_{\pp\sigma} \bar n_{\pp_2\sigma}  n_{\pp_3\sigma}  n_{\pp_4\sigma}- n_{\pp\sigma}  n_{\pp_2\sigma}  \bar n_{\pp_3\sigma}  \bar n_{\pp_4\sigma}}\Big)\Bigg)\\
+O(p_0/\pF)^3 \label{Icollsc}
\end{multline}
where \hk{$\epsilon_{\pp\sigma}=\meanv{\hat\epsilon_{\pp\sigma}}$ 
are the instantaneous quasiparticle energies}, and
\bea
W_{\sigma\sigma'}(\pp_1,\pp_2|\pp_3,\pp_4)=\bbcro{\mathcal{A}_{\sigma\sigma'}(\pp_1,\pp_2|\pp_3,\pp_4)}^2
\eea
are the collision probabilities.
Note that $W$ inherits exchange properties corresponding to detailed balance
and spin symmetry from \eqqrefs{pteA1}{pteA3}.

The collision integral $I_{\pp\sigma}$ is a non-linear functional of the quasiparticle distribution $n$.
If $\pp$ is inside the low-energy shell ($|p-\pF|<p_0$), then, by the conservation of energy and the absence of highly-excited quasiparticles,
so are
all the collision momenta\footnote{Energy conservation guarantees that one of the outgoing wavevector
is at low energy, say $|p_4-\pF|<p_0$. Then if the remaining wavevectors $\pp_2$ and $\pp_3$
are at high energy, they are necessarily on the same side of the Fermi level, $n_{p_2}^0=n_{p_3}^0$.
The collision in the state $\hat\varrho$ is then suppressed by the factor $n_{\pp_2\sigma'}\bar n_{\pp_3\sigma'}\underset{p_2,p_3\gg p_0}{\to}n_{p_2}^0 \bar n_{p_3}^0=0$ in the square bracket of \eqqref{Icollsc}.} 
$\pp_2$, $\pp_3$ and $\pp_4$. In other words, the low-energy space is stable under collisions.
The double summation over $\pp_2$ and $\pp_3$ (assuming that $\pp_4$ is fixed by momentum conservation)
is then restricted to a small interval $[\pF-p_0,\pF+p_0]$ about the Fermi momentum,
which allows us to estimate
\be
I_{\pp\sigma}=O\bb{\frac{p_0}{\pF}}^2 \label{Icollmagn}
\ee

\paragraph{Thermal lifetime}
In its general form, the kinetic equation \eqqref{Icollsc} is a nonlinear differential equation
where we cannot single-out the lifetime of quasiparticles in mode $\pp\sigma$. 
We now linearize the dynamics about
the thermal equilibrium state, which we
approximate\footnote{At low temperatures, $n_{\pp}^{\rm eq}$ differs from the zero-temperature Fermi-Dirac distribution $n_\pp^0$ by a $O(T)$.
In omitting $\hat H_4$ (and higher order terms) from $\hat{\varrho}_{\rm eq}$ we commit a small error, of order $O(T^2)$ on $n_{\pp}^{\rm eq}$.} 
by the matrix density
\be
\hat{\varrho}_{\rm eq}=\frac{1}{\mathcal{Z}}\eee^{- (\hat H_2-\mu\hat N)/T} \label{rhoeq}
\ee
Here $\mu$ is the chemical potential, 
$\mathcal{Z}=\text{Tr}(\eee^{-(\hat H_2-\mu\hat N)/T})$ is the low-temperature approximation of the partition function, and
$\hat N$ is the number of quasiparticles (see \eqqref{NNgamma}).
The state $\hat{\varrho}_{\rm eq}$ populates the quasiparticle modes according to the Fermi-Dirac distribution
\be
n_{\pp}^{\rm eq}(T)=n_{\rm eq}(\epsilon_\pp)\equiv\text{Tr}(\hat{\varrho}_{\rm eq} \hat n_{\pp\sigma})=\frac{1}{1+\eee^{(\epsilon_{\pp}-\mu)/T}}
\ee
It then fullfills the low-energy condition \eqqref{conditionBE} with $\pF\gg p_0\approx T/\vF$.

We then excite the quasiparticle in mode $\pp\sigma$, leaving the rest of the gas in the thermal state, which amounts to
preparing the initial distribution
\be
\meanv{ \hat n_{\pp'\sigma'}}=
\begin{cases}
n_{\pp}^{\rm eq}(T)+\delta n_{\pp\sigma}, \ \pp'\sigma'=\pp\sigma \\ n_{\pp'}^{\rm eq}(T), \ \pp'\sigma'\neq\pp\sigma
\end{cases}
\ee
As long as it remains much below $\pF$, the excited quasiparticle does not need to be inside the thermal window.

The kinetic \eqqref{Icollsc} then describes the thermal relaxation of $\delta n_{\pp\sigma}$:
\be
\partial_t \delta n_{\pp\sigma}=- \Gamma({\pp}) \delta n_{\pp\sigma}
\ee
The thermal damping rate is independent of $\sigma$; it is given by Fermi's golden rule\footnote{\hk{The deviations of the quasiparticle
energies are negligible in this case $\delta\epsilon_{\pp\sigma}=O(1/L^3)$.}}
\be
\Gamma({\pp})=\frac{2\pi}{L^6}\sum_{\pp_2\pp_3\pp_4\in\mathcal{D}} {W_+(\pp,\pp_2|\pp_3,\pp_4)}
 \delta_{\pp+\pp_2}^{\pp_3+\pp_4}\delta(\epsilon_{\pp}+\epsilon_{\pp_2}-\epsilon_{\pp_3}-\epsilon_{\pp_4})
\bbcro{n_{\pp_2}^{\rm eq}\bar{n}_{\pp_3}^{\rm eq}\bar{n}_{\pp_4}^{\rm eq} + \bar{n}_{\pp_2}^{\rm eq}n_{\pp_3}^{\rm eq}n_{\pp_4}^{\rm eq}}
\label{GammaCheqtransport}
\ee
with the spin-averaged collision probability
\be \label{Wplus}
W_+(\pp_1,\pp_2|\pp_3,\pp_4)=\frac{1}{2}\bbcro{W_{\upa\dwa}(\pp_1,\pp_2|\pp_3,\pp_4)+W_{\upa\dwa}(\pp_1,\pp_2|\pp_4,\pp_3)+W_{\sigma\sigma}(\pp_1,\pp_2|\pp_3,\pp_4)}
\ee
Integrating over energies and angles (see Section.~\ref{sec:icoll}) we recover the standard result for $\Gamma$:
\be
\Gamma({\pp})=\bb{\frac{m^\ast}{2\pi}}^3\meanvlr{\frac{W_+(\theta,\phi)}{2\cos\frac{\theta}{2}}}_{\theta,\phi}\bbcro{\pi^2 T^2+(\epsilon_{\pp}-\mu)^2} \label{Gammaexplicite}
\ee
We have reparametrized the probability $W$
in terms of the two angles $\theta=(\widehat{\pp_1,\pp_2})$ and $\phi=$ $(\widehat{\pp_1-\pp_2,\pp_3-\pp_4})$ that locate the four momenta
$\pp_1$, $\pp_2$, $\pp_3$, $\pp_4$ on the Fermi surface: $W_+(\pp_1,\pp_2|\allowbreak\pp_3,\pp_4)=$ $W_+(\theta,\phi)$ for $p_i=\pF$. We have then introduced the
average over solid angles 
\be
\meanv{f}_{\theta,\phi}=\frac{1}{4\pi}\int_0^\pi \int_0^{2\pi} f(\theta,\phi)\sin\theta \dd\theta\dd\phi \label{anglesolides}
\ee 
Since $|\epsilon_{\pp}-\mu|$ and $T$ are both below $\vF p_0$, \eqqref{Gammaexplicite} illustrates the $O(p_0/\pF)^2$ scaling of the collision integral.

We will see in section \ref{sec:thermalcorr} that there are corrections of order $O(T^3)$ or $O(p-\pF)^3$
to expression \eqref{Gammaexplicite} of $\Gamma(\pp)$. 

\subsection{Linearized transport equation at nonzero temperature}
\label{sec:transportT}

\paragraph{Linear response approximation}

To describe transport, we imagine that the system is driven out-of-equilibrium by an external field $U_\sigma$ coupled to quasiparticle density operators $\hat  \gamma^\dagger \hat \gamma$
\begin{equation}
    \hat{H}_{\rm ext} = \sum_{\pp\in\mathcal{D}, \sigma} U_{\sigma}(\qq,t) \hat{n}^{\qq}_{\pp \sigma} \label{Hext}
\end{equation}
where we use Anderson's notations \cite{Anderson1958} for the quasiparticle-quasihole excitation operator
\be
\hat n_{\pp\sigma}^\qq(t)=\hat{\gamma}^{\dagger}_{\pp-\qq/2, \sigma}\hat{\gamma}_{\pp+\qq/2,\sigma}
\ee
One can also see $\hat{H}_{\rm ext}$ as driving the density of quasiparticles in real space:
\be
\hat{H}_{\rm ext}=\int \dd^3 r \sum_\sigma U_\sigma(\rr,t) \hat\psi_{\gamma,\sigma}^\dagger(\rr) \hat\psi_{\gamma,\sigma}(\rr) \label{x}
\ee
where $\hat\psi_{\gamma,\sigma}(\rr)={L^{-3/2}}\sum_{\pp\in\mathcal{D}} \eee^{\ii\pp\cdot\rr}\hat\gamma_{\pp\sigma}$ is the field operator
associated to the quasiparticles, and $U_\sigma(\rr)=U_\sigma(\qq) \eee^{-\ii\qq\cdot\rr} $. 
Placing ourselves in the linear response regime, we assume a weak driving compared to the temperature
\be
|U_\sigma|\ll T
\ee
and we decompose the state of the system at time $t$ as
\be
\hat{\varrho}(t) = \hat{\varrho}_{\rm eq}+ \delta \hat{\varrho}(t) \label{rhot}
\ee
with $\delta \hat{\varrho}=O(U_\sigma/T)$.
In the linear response regime, the fluctuations about the thermal state are small $\hat n_{\pp\sigma}-n_\pp^{\rm eq}=O(U)$,
and we approximate the contribution of the drive to the equation of motion using
\be
    [\hat{n}_{\pp\sigma}^{\qq},\hat{H}_{\rm ext}]=U_\sigma(\qq) \bb{\hat n_{\pp-\frac{\qq}{2},\sigma}-\hat n_{\pp+\frac{\qq}{2},\sigma}}=U_\sigma(\qq) \bb{n_{\pp-\frac{\qq}{2}}^{\rm eq}- n_{\pp+\frac{\qq}{2}}^{\rm eq}}+O(U^2)
\ee

\paragraph{Quantum Boltzmann equation}
The first contribution to the Boltzmann equation is the streaming term\footnote{We have used the property $f_{\sigma\sigma}(\pp+{\qq}/{2},\pp-{\qq}/{2})=0$, valid for $v_{\rm F}q\ll \Lambda$, which guarantees that $\hat{n}_{\pp\sigma}^{\qq}$ commutes with $\delta \hat\epsilon_{\pp\pm\qq/2,\sigma}$}:
\be
    [\hat{n}_{\pp\sigma}^{\qq},\hat{H}_2+\hat{H}_4^{\rm d}] =\bb{\hat\epsilon_{\pp+\qq/2,\sigma}-\hat\epsilon_{\pp-\qq/2,\sigma}}\hat n_{\pp\sigma}^\qq \label{npHssH4x}
\ee
Then, the contribution of $\hat H_4^x$ is no longer antihermitian 
\begin{align}
     [\hat{n}_{\pp\sigma}^{\qq},\hat{H}_4^{\rm x}]=\frac{1}{2L^3} \sum_{\substack{\pp_4,\pp_2\neq\pp_3\\\sigma'=\upa\dwa}} 
     s_{\sigma\sigma'}\Bigg[&\mathcal{A}_{\sigma\sigma'}\bb{\pp+\frac{\qq}{2},\pp_2|\pp_3,\pp_4}
    \delta_{\pp+\frac{\qq}{2}+\pp_2}^{\pp_3+\pp_4} 
    \hat{\gamma}^{\dagger}_{\pp-\frac{\qq}{2}, \sigma}\hat{\gamma}^{\dagger}_{\pp_2 \sigma'} \hat{\gamma}_{\pp_3\sigma'}\hat{\gamma}_{\pp_4\sigma}\notag \\
    -& 
   \mathcal{A}_{\sigma\sigma'}\bb{\pp-\frac{\qq}{2},\pp_2|\pp_3,\pp_4}
    \delta_{\pp-\frac{\qq}{2}+\pp_2}^{\pp_3+\pp_4} 
     \hat{\gamma}^{\dagger}_{\pp_4 \sigma}\hat{\gamma}^{\dagger}_{\pp_3 \sigma'} \hat{\gamma}_{\pp_2\sigma'}\hat{\gamma}_{\pp+\frac{\qq}{2}\sigma} \Bigg]\label{npH4x}
\end{align}
where $s_{\upa\dwa}=1$ and $s_{\upa\upa}=1/2$. We treat this contribution using the cumulant expansion (\eqqref{expcumulant}) to obtain: 
\be
     [\hat{n}_{\pp\sigma}^{\qq},\hat{H}_4^{\rm x}]=\bb{n^{\rm eq}_{\pp-\frac{\qq}{2}}-n^{\rm eq}_{\pp+\frac{\qq}{2}}}\frac{1}{L^3} \sum_{\pp',\sigma'} \mathcal{A}_{\sigma\sigma'}\bb{\pp+\frac{\qq}{2},\pp'-\frac{\qq}{2}|\pp'+\frac{\qq}{2},\pp-\frac{\qq}{2}} \hat{n}^{\qq}_{\pp',{\sigma'}} 
+\ii \hat I_{\pp\sigma}
\label{commnH4x}
\ee
where $\hat I_{\pp\sigma}$ is the cumulant part of \eqqref{npH4x}.
Restricting to leading order in $U$, we have replaced the average values in the partially contracted terms by thermal averages: 
\be
\meanv{\hat \gamma_{\pp\sigma}^\dagger \hat \gamma_{\pp'\sigma'}}=\delta_{\pp\pp'}\delta_{\sigma\sigma'} n_\pp^{\rm eq}+O(U/T)
\ee
To recognize the Vlasov force in those terms, we use \eqqref{limdalphaf} and the condition $ \vF q\ll\Lambda$:
\be
\mathcal{A}_{\sigma\sigma'}\bb{\pp+\frac{\qq}{2},\pp'-\frac{\qq}{2}|\pp'+\frac{\qq}{2},\pp-\frac{\qq}{2}}= f_{\sigma\sigma'}(\pp,\pp') +O( \vF q/ \Lambda)
\ee
Note that the partial contractions also replace the instantaneous energies in \eqqref{npHssH4x} by their thermal value $\meanv{\hat \epsilon_{\pp\sigma}}=\meanv{\hat \epsilon_{\pp\sigma}}_{\rm eq}+O(U)$.

\paragraph{Collision integral}

Following the steps discussed in Sec.~\ref{sec:lifetime}, we compute the collision integral $\hat I_{\pp\sigma}$
in the Born-Markov approximation. 
Restricting to leading order in $\vF q/T$, 
we obtain the transport equation
\be
    \left(\ii\partial_t - \vF q u \right)\hat{n}^{\qq}_{\pp\sigma} +\vF q u\frac{\dd n_{\rm eq}}{\dd \epsilon}\Big\vert_{\epsilon=\epsilon_{\pp}} \left( U_{\sigma}(\qq)+\frac{1}{L^3}\sum_{\pp'\sigma'}f_{\sigma \sigma'}(\pp,\pp')\hat{n}^{\qq}_{\pp'\sigma'}\right) = \ii \hat{I}^{\rm lin}_{\pp\sigma}\bbcro{\hat n^\qq} \label{transportI}
\ee
where $u=\text{cos}(\widehat{\pp,\qq})$ and $n_{\rm eq}(\epsilon)=1/(1+\eee^{(\epsilon-\mu)/T})$.
The collision integral linearized about the equilibrium state takes the form
\begin{multline}
\hat I_{\pp\upa}^{\rm lin}[\hat n^\qq]=-\frac{2\pi}{L^6}\sum_{\beta,\gamma,\delta\in\mathcal{D}}\delta_{\pp+\beta}^{\gamma+\delta}\delta(\epsilon_\pp+\epsilon_\beta-\epsilon_\gamma-\epsilon_\delta)\\
\times\Bigg\{
\bbcro{n_\beta^{\rm eq} \bar n_\gamma^{\rm eq} \bar n_\delta^{\rm eq}+\bar n_\beta^{\rm eq}  n_\gamma^{\rm eq}  n_\delta^{\rm eq}}\bb{W_{\upa\dwa}(\pp,\beta|\gamma,\delta)+\frac{1}{2}W_{\sigma\sigma}(\pp,\beta|\gamma,\delta)}\hk{\bb{\hat n_{\pp\upa}^\qq-\frac{\dd n_{\rm eq}}{\dd \epsilon} \hat \epsilon_{\pp\upa}^{\qq}}}\\
+\bbcro{n_\pp^{\rm eq} \bar n_\gamma^{\rm eq} \bar n_\delta^{\rm eq}\!+\!\bar n_\pp^{\rm eq}  n_\gamma^{\rm eq}  n_\delta^{\rm eq}}\bb{\!W_{\upa\dwa}(\pp,\beta|\gamma,\delta)\hk{\bb{\!\hat n_{\beta\dwa}^\qq\!-\frac{\dd n_{\rm eq}}{\dd \epsilon} \hat \epsilon_{\beta\dwa}^{\qq}}}\!+\!\frac{1}{2}W_{\sigma\sigma}(\pp,\beta|\gamma,\delta)\hk{\bb{\!\hat n_{\beta\upa}^\qq-\frac{\dd n_{\rm eq}}{\dd \epsilon} \hat \epsilon_{\beta\upa}^{\qq}\!}\!}}\\
-\bbcro{n_\delta^{\rm eq} \bar n_\pp^{\rm eq} \bar n_\beta^{\rm eq}\!+\!\bar n_\delta^{\rm eq}  n_\pp^{\rm eq}  n_\beta^{\rm eq}}\bb{\!W_{\upa\dwa}(\pp,\beta|\gamma,\delta)\hk{\bb{\!\hat n_{\gamma\dwa}^\qq\!-\frac{\dd n_{\rm eq}}{\dd \epsilon} \hat \epsilon_{\gamma\dwa}^{\qq}}}\!+\!\frac{1}{2}W_{\sigma\sigma}(\pp,\beta|\gamma,\delta)\hk{\bb{\!\hat n_{\gamma\upa}^\qq-\frac{\dd n_{\rm eq}}{\dd \epsilon} \hat \epsilon_{\gamma\upa}^{\qq}\!}\!}}\\
-\bbcro{n_\gamma^{\rm eq} \bar n_\pp^{\rm eq} \bar n_\beta^{\rm eq}+\bar n_\gamma^{\rm eq}  n_\pp^{\rm eq}  n_\beta^{\rm eq}} \bb{W_{\upa\dwa}(\pp,\beta|\gamma,\delta)+\frac{1}{2}W_{\sigma\sigma}(\pp,\beta|\gamma,\delta)} \hk{\bb{\hat n_{\delta\upa}^\qq-\frac{\dd n_{\rm eq}}{\dd \epsilon} \hat \epsilon_{\delta\upa}^{\qq}\!}}\Bigg\}
\label{Icolllin}
\end{multline}
\hk{where 
\be
 \hat \epsilon^{\qq}_{\pp\sigma}= \frac{1}{L^3}\sum_{\pp'\sigma'} f_{\sigma\sigma'}(\pp,\pp')\hat n_{\pp'\sigma'}^\qq\label{nloc}
\ee
is the fluctuation of the energy of the quasiparticle.}

We may reinterpret the linearized transport equation
in real space by performing an inverse Wigner transform \hk{of $\hat n_{\pp\upa}^\qq$}
\be
n_{\sigma}(\pp,\qq)\equiv\meanv{\hat n_{\pp\sigma}^{\qq}}=\frac{1}{\sqrt{L^3}}\int \dd^3 r \eee^{-\ii\qq\cdot\rr} \bb{n_{\sigma}(\pp,\rr) \hk{-n_{\rm eq}(\epsilon_\pp)}}
\label{defnpq}
\ee
\hk{
Similarly, we convert $\hat \epsilon_{\pp\sigma}^{\qq}$ into ${\epsilon_{\sigma}(\pp,\rr) -\epsilon_\pp}$}.
The Wigner transform of $\meanv{\hat I_{\pp\sigma}^{\rm lin}}$ is then interpreted
as a collision integral linearized for small spatial fluctuations: 
\be
 \meanv{\hat I_{\pp\sigma}^{\rm lin}(\rr)}=I_{\pp\sigma}\hk{\bbcro{n_{\rm eq}+\delta n(\rr)}} +O(\delta n^2)
\ee
where $I_{\pp\sigma}[n]$ is defined by \eqqref{Icollsc} and $\delta n=n-n_{\rm eq}$ for the rest of this section. \hk{The linearization acts on the Fermi-Dirac gain-loss factors (between square brackets in \eqqref{Icollsc}) and takes into account the non-equilibrium values of the energies appearing in the Dirac functions via
$n_\sigma(\pp,\rr)=n_{\rm eq}(\epsilon_\pp)+\delta n_\sigma(\pp,\rr)=n_{\rm eq}(\epsilon_\sigma(\pp,\rr))+\delta n_\sigma(\pp,\rr)-\frac{\dd n_{\rm eq}}{\dd \epsilon}\delta\epsilon_\sigma(\pp,\rr)$}.
The transport equation may now be written in real space:
\be
    \partial_t {n}_{\sigma}+  \frac{\partial\epsilon_{\pp}}{\partial\pp}\cdot \frac{\partial n_{\sigma}}{\partial\rr} -  \frac{\dd n_{\pp}^{\rm eq}}{\dd\pp} \cdot \frac{\partial}{\partial\rr} \bb{\frac{1}{L^3}\sum_{\pp'\sigma'} f_{\sigma\sigma'}(\pp,\pp')  n_{\sigma'}(\pp')+U_\sigma(\rr)}=I_{\pp\sigma}[n(\rr)]
    \label{transportreel}
\ee

\paragraph{The hydrodynamic-collisionless crossover}

Let us discuss the different regimes of transport by assuming that the driving force $U_\sigma$
imposes a characteristic evolution frequency $\omega$ and a characteristic wavenumber $q$
to the perturbations of the fluid. Then, the left-hand-side of \eqqref{transportreel} is of order
$\omega$ or $\vF q$, two energy scales which we assumed to be comparable.
Meanwhile, the right-hand side of \eqqref{transportreel}, the collision integral,
is of the order of the damping rate $\Gamma({\pp})$ for $p\approx\pF$.
Let us define the characteristic collision time $\tau$ as
\be
\frac{1}{\tau}\equiv\frac{\Gamma({\pF})}{\pi^2}=\bb{\frac{m^\ast}{2\pi}}^3 T^2\meanvlr{\frac{W_+(\theta,\phi)}{2\cos\frac{\theta}{2}}}_{\theta,\phi}
\ee
such that the typical energy of the collision integral is $1/\tau$.

The ratio between $\omega,\vF q$ and $1/\tau$ can take arbitrary values,
even in the low-temperature regime of Fermi liquids. In the collisionless
regime
\be
\omega\tau,\vF q \tau\gg 1
\ee
the left-hand-side dominates \eqqref{transportreel}, the dynamics is nearly integrable and
the collision integral acts as a small perturbation that restores thermal equilibrium on time scales
long compared to the integrable dynamics. 
We study this regime for the contact Fermi gas in Ref.~\cite{VolumeII}.
In the opposite hydrodynamic regime
\be
\omega\tau,\vF q \tau\ll 1
\ee
the collision integral dominates \eqqref{transportreel}, and brings all the components
of $n_{\sigma}$ that it affects back to 0 within a time $\approx\tau$ short compared to the variation of $U_{\sigma}$. Only the conserved
quantities, unaffected by the collision integral, can fluctuate significantly. Their evolution
is describes by the Navier-Stokes equations, as we explain in Ref.~\cite{devvisco}.

\subsection{Transport equation at $T=0$}
\label{sec:transportT0}

We now let $T\to 0$ at fixed $q$ and $\omega$. We describe the state of the system
in the presence of $\hat H_{\rm ext}$ by
\be
\hat\varrho(t)=\FS\bra{{\rm FS}}+\delta\hat\varrho(t)
\ee
The linear response regime (\textit{i.e.} the absence of second harmonic
generation) ensures that only the momenta
that differ from $\pF$ by $q$ are excited. The fluctuations
of the quasiparticle distribution thus remain zero for $|p-\pF|\gg q$, and
the wavenumber $q$ acts as the small parameter $p_0$ 
of the low-energy expansion. The linearized transport equation is then
\begin{multline}
    \bb{\ii\partial_t - {\epsilon_{\pp+\qq/2}+\epsilon_{\pp-\qq/2}} }\hat{n}^{\qq}_{\pp\sigma} + \bb{n_{\pp+\frac{\qq}{2}}^0-n_{\pp-\frac{\qq}{2}}^0} \left( U_{\sigma}(\qq)+\frac{1}{L^3}\sum_{\pp'\sigma'}f_{\sigma \sigma'}(\pp,\pp')\hat{n}^{\qq}_{\pp'\sigma'}\right) \\= \ii \hat{I}^{\rm lin}_{\pp\sigma}[\hat n^\qq,T=0] \label{transportT0}
\end{multline}
The zero-temperature collision integral
$\hat{I}^{\rm lin}_{\pp\sigma}(\hat n^\qq,T=0)$ is obtained by replacing $n_{\pp}^{\rm eq}\to n_{\pp}^0$ in the nonzero temperature
expression \eqref{Icolllin}; it is of order $O(q/\pF)^2$ to leading order (see \eqqref{Icollmagn}), with corrections of order $O(q/\pF)^3$. 
To leading order in $q/\pF$ the transport equation \eqref{transportT0} then reduces to its collisionless left-hand side.

%% file: heff_icoll.tex
\section{The collision integral of a Fermi liquid} \label{sec:icoll}

This section is devoted to the mathematical treatment of the collision kernel.
The linearized transport equation \eqref{transportI} is a linear integral equation with the distribution
$n(\pp)$ as the unknown function, and a source term proportional 
to the driving potential $U_\sigma$. Solving this integral equation requires
a careful treatment of the linear kernels it contains. The first kernel, the function
$f_{\sigma\sigma'}(\pp,\pp')$ appearing in the energy fluctuation $\delta \epsilon$ 
is readily diagonalized by expanding over Legendre polynomials. The second kernel contained in the linearized collision integral \eqqref{Icolllin}
is harder to handle: it combines an energy and an angular dependence, and it possesses several
zero-energy eigenfunctions.

Here, we show how this collision kernel factorizes into an angular part and an energy part, and we relate
its zero-eigenfunctions to the conserved quantities of the fluid. This results in a dimensionless
version of the transport equation (\eqqref{equlowT}).

\subsection{Collision kernel}
We begin by expressing the linearized collision integral \eqqref{Icolllin} in terms of a collision kernel\footnote{
Note the transposed order of $\pp'\sigma'$ and $\pp\sigma$ in $\mathcal{N}$, \eqqref{ImathcalN}.}
\begin{align}
    I_{\sigma}^{\rm lin}(\pp,n(\qq))&=\meanv{\hat I_{\pp\sigma}^{\rm lin}[\hat n^\qq]}=\frac{1}{L^3}\sum_{\pp'\sigma'} \mathcal{N}_{\sigma'\sigma}(\pp',\pp) \hk{\bb{n_{\sigma'}(\pp',\qq)-\frac{\dd n_{\rm eq}}{\dd\epsilon}\delta\epsilon_{\sigma'}(\pp',\qq)}}  \label{ImathcalN} \\
    \text{ with }\mathcal{N}_{\sigma\sigma'}(\pp,\pp') &= -\Gamma(\pp) L^3\delta_{\sigma\sigma'}\delta_{\pp\pp'}-{E_{\sigma\sigma'}(\pp,\pp')+ S_{\sigma \sigma'}(\pp,\pp')} \label{IlinES} 
\end{align}
The diagonal part of $\mathcal{N}$ is given by the quasiparticles damping rate \eqqref{GammaCheqtransport} and the off-diagonal part involves four subkernels:
    \begin{align}
    E_{\sigma\sigma}(\pp,\pp')&=\frac{2\pi}{L^3}\sum_{\pp_3,\pp_4\in\mathcal{D}} \frac{W_{\sigma\sigma}(\pp',\pp|\pp_3,\pp_4)}{2}
     \delta_{\pp+\pp'}^{\pp_3+\pp_4}\delta(\epsilon_{\pp}+\epsilon_{\pp'}-\epsilon_{\pp_3}-\epsilon_{\pp_4})
      N_{\pp_3\pp_4}^{\pp'} \\
    E_{\upa\dwa}(\pp,\pp')&=\frac{2\pi}{L^3}\sum_{\pp_3,\pp_4\in\mathcal{D}} W_{\upa\dwa}(\pp',\pp|\pp_3,\pp_4)
     \delta_{\pp+\pp'}^{\pp_3+\pp_4}\delta(\epsilon_{\pp}+\epsilon_{\pp'}-\epsilon_{\pp_3}-\epsilon_{\pp_4})
       N_{\pp_3\pp_4}^{\pp'} \\ 
S_{\sigma\sigma}(\pp,\pp')&=\frac{2\pi}{L^3}\!\!\!\!\!\sum_{\pp_2,\pp_4\in\mathcal{D}} \!\!\!\!\bbcro{W_{\sigma\sigma}(\pp',\pp_2|\pp_4,\pp)+W_{\upa\dwa}(\pp',\pp_2|\pp_4,\pp)}
     \delta_{\pp+\pp_4}^{\pp_2+\pp'}\delta(\epsilon_{\pp}+\epsilon_{\pp_4}-\epsilon_{\pp'}-\epsilon_{\pp_2}) N_{\pp'\pp_2}^{\pp_4} \\
S_{\upa\dwa}(\pp,\pp')&=\frac{2\pi}{L^3}\sum_{\pp_2,\pp_4\in\mathcal{D}} W_{\upa\dwa}(\pp',\pp_2|\pp,\pp_4)
     \delta_{\pp+\pp_4}^{\pp'+\pp_2}\delta(\epsilon_{\pp}+\epsilon_{\pp_4}-\epsilon_{\pp'}-\epsilon_{\pp_2}) N_{\pp'\pp_2}^{\pp_4} 
    \end{align}
where 
\be
N_{\pp_1\pp_2}^{\pp_3} ={n_{\pp_1}^{\rm eq} {n}_{\pp_2}^{\rm eq}\bar{n}_{\pp_3}^{\rm eq} + \bar{n}_{\pp_1}^{\rm eq}\bar{n}_{\pp_2}^{\rm eq}n_{\pp_3}^{\rm eq}}
\ee
is the Fermi-Dirac gain loss factor.
The collisions kernels $E_{\sigma\sigma'}$ describe the coupling between quasiparticles in mode $\pp\sigma$ and $\pp'\sigma'$ through processes where $\pp$ and $\pp'$ are  on the same side of the collision (either incoming or outgoing). Conversely, $S_{\sigma \sigma'}$ describes the couplings where $\pp$ and $\pp'$ are on opposite sides.
\subsection{Conservation laws}
Collisions obey a few conservation laws which play a prominent role in transport phenomena: the numbers of spin $\upa$ and $\dwa$ particles,
the momentum and the energy are the same before and after each collision. In this respect, our isolated fermionic fluid
differs from the standard situation of electron in solids, which exchange momentum and energy with impurities and with the ionic lattice.
In mathematical terms, the conserved quantities correspond to 6 zero eigenfunctions of the collision kernel $\mathcal{N}$ (counting the 3 components of the momentum).
Since the kernel is not symmetric, it has distinct left and right eigenfunctions.

Let us illustrate this by contracting $\mathcal{N}$
by some arbitrary functions {$n_\sigma(\pp)$} to the left and $\nu_\sigma(\pp)$ to the right:
\begin{multline}
    \sum_{\substack{\pp\pp'\in\mathcal{D}\\\sigma\sigma'=\upa\dwa}}  n_{\sigma}(\pp) \mathcal{N}_{\sigma\sigma'}(\pp,\pp')\nu_{\sigma'}(\pp') = \frac{2\pi}{L^6}\sum_{\substack{\pp_1\pp_2\pp_3\pp_4\in\mathcal{D}\\\sigma=\upa\dwa}}\delta_{\pp_1+\pp_2}^{\pp_3+\pp_4}   \delta(\epsilon_{\pp_1}+\epsilon_{\pp_2}-\epsilon_{\pp_3}-\epsilon_{\pp_4}) N_{\pp_3\pp_4}^{\pp_2} \\
\times      \Bigg[\frac{1}{2}W_{\upa\upa}(\pp_1,\pp_2|\pp_3,\pp_4) \bb{{\nu_\sigma(\pp_1)+\nu_\sigma(\pp_2)-\nu_\sigma(\pp_3)-\nu_\sigma(\pp_4)}} \\+ W_{\upa\dwa}(\pp_1,\pp_2|\pp_3,\pp_4){(\nu_\sigma(\pp_1)+\nu_{-\sigma}(\pp_2)-\nu_{-\sigma}(\pp_3)-\nu_\sigma(\pp_4))}\Bigg] n_\sigma(\pp_1)
\end{multline}
The 6 functions $\nu_\sigma$ which cancel this expression for all $n_\sigma$, \textit{i.e.} the right zero-energy eigenfunctions, are $\nu_\sigma(\pp)=\delta_{\sigma,\upa}$, $\delta_{\sigma,\dwa}$, $p_x$, $p_y$, $p_z$ and $\epsilon_\pp$. 
The conserved quantities are the projection of $n(\pp)$ onto these eigenfunctions:
\bea
\delta\rho_\sigma&=&\frac{1}{L^3}\sum_{\pp\in\mathcal{D}} n_\sigma(\pp) \label{deltarho}\\
{m} \vv &=&\frac{1}{N}\sum_{\pp\in\mathcal{D},\sigma} \pp n_\sigma(\pp) \label{vv}\\
\delta e &=&\frac{1}{L^3}\sum_{\pp\in\mathcal{D},\sigma} \epsilon_\pp n_\sigma(\pp)  \label{deltae}
\eea
where we recognize the density fluctuations $\delta\rho_\sigma$, the macroscopic velocity $\vv$ and the energy density $\delta e$
Unsurprisingly, opposite spin collisions (with probability $W_{\upa\dwa}$) are responsible for the absence of conservation of the velocity imbalance $\vv_{\upa}-\vv_\dwa$ and energy imbalance $e_{\upa}-e_\dwa$.

\subsection{Total density and polarization}
In our unpolarized Fermi liquid, fluctuations of the density $n_+ = n_{\uparrow}+n_{\downarrow}$ and polarisation $n_- =n_{\uparrow}-n_{\downarrow}$ are decoupled, by the transport equation in general,
and by the collision integral in particular.
The corresponding collision kernels are: 
\begin{equation}
    \mathcal{N}_{\pm}(\pp,\pp') = -\Gamma(\pp)  L^3\delta_{\sigma\sigma'}\delta_{\pp\pp'} -E_{\pm}(\pp,\pp')+2S_{\pm}(\pp,\pp'),\qquad I_{\pm}^{\rm lin}(\pp)=I_{\upa}^{\rm lin}(\pp)\pm I_{\dwa}^{\rm lin}(\pp)
    \label{Npm}
\end{equation}
with
\bea
    E_{\pm}(\pp,\pp') &=& \frac{1}{L^3}\sum_{\pp_3\pp_4} W_{ E \pm}(\pp,\pp'|\pp_3,\pp_4) \ \delta_{\pp+\pp'}^{\pp_3+\pp_4}\delta(\epsilon_{\pp}+\epsilon_{\pp'}-\epsilon_{\pp_3}-\epsilon_{\pp_4})
    N_{\pp_3\pp_4}^{\pp'} \label{Epm}
\\
    S_{\pm}(\pp,\pp') &=& \frac{1}{L^3}\sum_{\pp_2\pp_4}  W_{ S \pm}(\pp,\pp_2|\pp',\pp_4) \ \delta_{\pp+\pp_2}^{\pp'+\pp_4}\delta(\epsilon_{\pp}+\epsilon_{\pp_2}-\epsilon_{\pp'}-\epsilon_{\pp_4})
    N_{\pp'\pp_4}^{\pp_2} \label{Spm}
\eea
We have defined the anti-symmetrized probabilities\footnote{The definition of $W_{S-}$ in Ref.~\cite{devvisco} 
has a minus sign compared to the definition here in the case $W_{\upa\upa}=0$.} $W_{\rm E -}$ and $W_{\rm S -}$ :
\begin{align}
W_{E -}(\pp_1,\pp_2|\pp_3,\pp_4) &= \frac{1}{2}W_{\uparrow\uparrow}(\pp_1,\pp_2|\pp_3,\pp_4) - \frac{1}{2}\left(W_{\uparrow\downarrow}(\pp_1,\pp_2|\pp_3,\pp_4) + W_{\uparrow\downarrow}(\pp_1,\pp_2|\pp_4,\pp_3)\right) \label{WEpm}\\
W_{S -}(\pp_1,\pp_2|\pp_3,\pp_4) &= \frac{1}{2}W_{\uparrow\uparrow}(\pp_1,\pp_2|\pp_3,\pp_4) - \frac{1}{2}\left(W_{\uparrow\downarrow}(\pp_1,\pp_2|\pp_3,\pp_4) - W_{\uparrow\downarrow}(\pp_1,\pp_2|\pp_4,\pp_3)\right) \label{WSpm}
\end{align}
The symmetrized probabilities $W_{E+}=W_{S+}$ simply coincide with the spin-average collision probability $W_+$ (\eqqref{Wplus})
which also enters in the expression of $\Gamma(\pp)$. All the probabilities verify the same symmetry properties (inherited from \eqqrefs{pteA2}{pteA1}):
\bea
W(\pp_4,\pp_3|\pp_2,\pp_1)=W(\pp_2,\pp_1|\pp_4,\pp_3)=W(\pp_1,\pp_2|\pp_3,\pp_4) \label{pteW1}
\eea

Among the conserved quantities \eqqrefs{deltarho}{deltae}, $\mathcal{N}_{+}$ inherits $\delta\rho_\upa+\delta\rho_\dwa$, $\vv$ and $\delta e$, while $\mathcal{N}_{-}$
inherits only $\delta\rho_\upa-\delta\rho_\dwa$.

\subsection{Quasiparticle distribution in the thermal window}

To focus on the thermal energy window, to which the fluctuations of $n(\pp)$ are limited,
we reparametrized the quasiparticle distributions as  
\begin{align}
        n_\pm(\pp) 
    &= -\frac{U_\pm(\qq)}{T} g(y) \ \nu_\pm(y,\theta),  \qquad \text{with} \qquad g(y)=\frac{1}{4\text{ch}^2(y/2)} \\
    \hk{ n_\pm(\pp)-\frac{\dd n_{\rm eq}}{\dd \epsilon} \delta\epsilon_\pm(\pp)}
    &\hk{= -\frac{U_\pm(\qq)}{T} g(y) \ \Phi_\pm(y,\theta),}
        \label{changmtvar}
\end{align}
We have parametrized the 3D momentum $\pp $ with $y=(\epsilon_{\pp\sigma}-\mu)/T$, $\theta=(\widehat{\pp,\qq})$ and an azimuthal angle $\phi$, of which
$\nu$ is independent due to the rotational invariance about $\qq$. 
\hk{We have also introduced the spin symmetric/antisymmetric part of $\delta\epsilon$ (see \eqqref{energielocale}):}
\be
\hk{\delta\epsilon_\pm(\pp)= \frac{1}{L^3}\sum_{\pp'}\bb{f_{\upa\upa}(\pp,\pp')\pm f_{\upa\dwa}(\pp,\pp')} n_\pm(\pp')}
\ee
Finally, in the spirit of linear response theory, 
we have scaled the distributions $\nu$ and $\Phi$ to the intensity 
\be
U_{\pm}=U_{\uparrow}\pm U_{\downarrow}
\ee 
of the drive.

By taking out the thermal broadening function $ {\dd n_{\rm eq}}/{\dd \epsilon}=-{g(y)}/{T} $, the change of variable \eqqref{changmtvar} smoothens the dependence of $\nu_\pm$ \hk{and $\Phi_\pm$} on $y$. 
It also transposes\footnote{This can be seen by writing $\frac{g(y')}{g(y)}=\frac{\bar n^{\rm eq}_{\pp'} n^{\rm eq}_{\pp'}}{\bar n^{\rm eq}_{\pp} n^{\rm eq}_{\pp}}$ and using
$\bar n^{\rm eq}_{\pp} n^{\rm eq}_{\pp} N_{\pp_3\pp_4}^{\pp'}=\bar n^{\rm eq}_{\pp'} n^{\rm eq}_{\pp'} N_{\pp_3\pp_4}^{\pp}$ for 4 wavectors $\pp$, $\pp'$, $\pp_3$ and $\pp_4$
constrained by energy-momentum conservation.} the collision kernels
\be
\mathcal{N}(\pp',\pp)\frac{g(y')}{g(y)}=\mathcal{N}(\pp,\pp')
\ee 
and similarly for $E_\pm$ and $S_\pm$. In terms of \hk{$\Phi$}, the collision integral becomes (notice the transposition of $\pp$ and $\pp'$ compared to \eqqref{ImathcalN})
\be
    I_{\pm}^{\rm lin}(\pp,n)=-\frac{U_\pm}{T}\frac{\bar n^{\rm eq}_{\pp} n^{\rm eq}_{\pp}}{L^3}\sum_{\pp'} \mathcal{N}_{\pm}(\pp,\pp') \hk{\Phi_{\pm}(\pp')} \label{Inu}
\ee

\subsection{Angular parametrization of 4 momentum-conserving wavevectors of the Fermi surface}

\begin{figure}
(a)\includegraphics[width=0.45\textwidth]{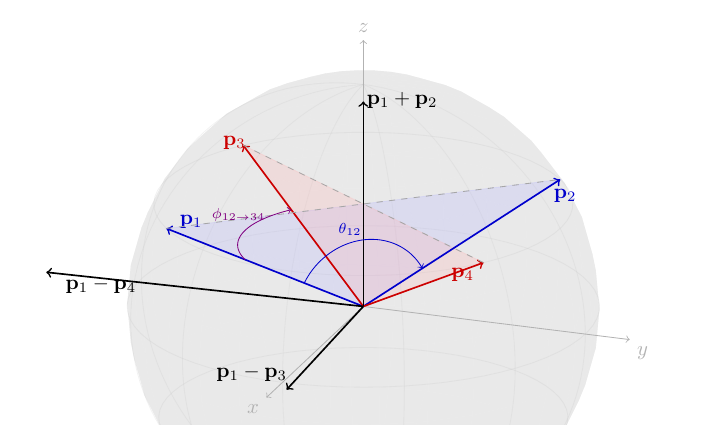}
(b)\includegraphics[width=0.45\textwidth]{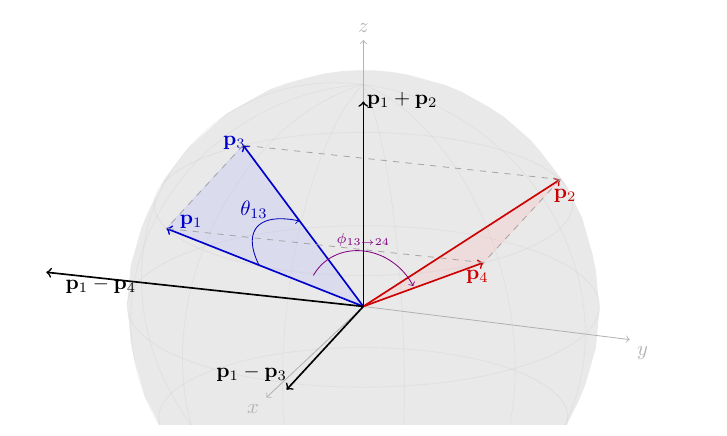}
\caption{(a) The angular parametrization where $\pp_1+\pp_2$ is chosen as the polar axis of the spherical frame. This parametrization is used for $W_{E}$ in \eqqref{wethetaphi}.
(b) The angular parametrization where $\pp_1-\pp_3$ is chosen as the polar axis of the spherical frame. This parametrization is used for $W_{S}$ in \eqqref{wsthetaphi} \label{fig123}.
The last parametrization where $\pp_1-\pp_4$ is chosen as the polar axis is not shown here.}
\end{figure}

To leading order in temperature, collisions of wavenumbers within the thermal window depend solely on the angles between these wavevectors.
Four wavevectors of the Fermi surface constrained by momentum conservation $\pp_1+\pp_2=\pp_3+\pp_4$ are
advantageously expressed in the orthogonal frame made of ($\pp_1+\pp_2$, $\pp_1-\pp_3$, $\pp_1-\pp_4$).
Depending on which vector is chosen as the $z$ axis frame, this leaves three different ways of parametrizing the angles,
two of which are depicted on Fig.~\ref{fig123}. Since $\pp$ and $\pp'$ play the role of $\pp_1$ and $\pp_2$ in
$E$, we use the parametrization of Fig.~\ref{fig123}a for this kernel:
\be
W_{E\pm}(\theta_{12},\phi_{12\to34})\equiv W_{E\pm}(\pp_1,\pp_2|\pp_3,\pp_4) \text{ with } \begin{cases} \theta_{12}\equiv(\widehat{\pp_1,\pp_2}) \\  \phi_{12\to34}\equiv(\pp_1-\widehat{\pp_2,\pp_3}-\pp_4) \\ \cos\theta_3=\cos\theta_4=\cos\frac{\theta_{12}}{2}, \theta_i=(\widehat{\pp_1+\pp_2,\pp_i})  \end{cases} 
\label{wethetaphi}
\ee
where the third line is the angular version of the momentum conservation constraint. For $S$ in which $\pp$ and $\pp'$ play the role of $\pp_1$ and $\pp_3$
we use the parametrization of Fig.~\ref{fig123}b:
\be
W_{S\pm}(\theta_{13},\phi_{13\to24})\equiv W_{S\pm}(\pp_1,\pp_2|\pp_3,\pp_4) \text{ with } \begin{cases} \theta_{13}\equiv(\widehat{\pp_1,\pp_3}) \\  \phi_{13\to24}\equiv(\pp_1+\widehat{\pp_3,\pp_2}+\pp_4) \\ \cos\theta_2=-\cos\theta_4=\sin\frac{\theta_{13}}{2}, \theta_i=(\widehat{\pp_1-\pp_3,\pp_i})  \end{cases} 
\label{wsthetaphi}
\ee

Since the collision amplitudes $\mathcal{A}_{\sigma\sigma'}$ are more readily
expressed in terms of the angles $\theta_{ij}$
between $\pp_i$ and $\pp_j$ (see \eqqrefs{Aupdw}{Aupup} for the contact Fermi gas), we use geometrical relations to convert those
angles to the angle $\theta$ and $\phi$ of a given parametrization. For example for the parametrization of Fig.~\ref{fig123}a:
\bea
\text{sin}^2\frac{\theta_{13}}{2}=\text{sin}^2\frac{\theta_{12}}{2} \text{sin}^2\frac{\phi_{12\to34}}{2} \\
\text{sin}^2\frac{\theta_{14}}{2}=\text{sin}^2\frac{\theta_{12}}{2} \text{cos}^2\frac{\phi_{12\to34}}{2} 
\eea

The angular integration in different parametrizations are related by the change of variable
\begin{equation}
\int \frac{\sin\theta_{13}\dd\theta_{13}\dd\phi_{13\to24}}{2\sin\frac{\theta_{13}}{2}}\tilde W(\theta_{13},\phi_{13\to24})=\int  \frac{\sin\theta_{12}\dd\theta_{12}\dd\phi_{12\to34}}{2\cos\frac{\theta_{12}}{2}}W(\theta_{12},\phi_{12\to34}) \label{anglesEanglesS}
\end{equation}
for any function $\tilde W(\theta_{13},\phi_{13\to24})=W(\theta_{12},\phi_{12\to34})$.
\subsection{Low temperature factorization of the kernel}

Among the fluids described by a Boltzmann equation, Fermi liquid have a remarkable property:
their collision kernel $\mathcal{N}(\pp,\pp')$ can be factorized into a radial (or energy) dependance
and an angular dependence on
\be
\alpha=(\widehat{\pp,\pp'}),\quad \cos\alpha=\cos\theta \cos\theta'+\sin\theta \sin\theta' \cos\phi
\ee
This a consequence of the restriction of both the collision probabilities and energy-conservation constraint
to the Fermi surface, such that the only remaining energy dependence in the kernel stems from the thermal
populations $n^{\rm eq}$.

We illustrate this decoupling in the calculation of $S_\pm$:
\begin{multline}
S_\pm(\pp,\pp')= \frac{(m^*)^2  T}{4\pi\pF |\sin\frac{\alpha}{2}|} \int_{-\infty}^{+\infty}  N_{y',y+y_2-y'}^{y_2} \dd y_2\int \frac{ \dd\Omega_2}{2\pi}\\\times W_{S\pm}(\pp,\pp_2|\pp',\pp-\pp'+\pp_2)\delta\bb{\cos\theta_2+\sin\frac{\alpha}{2}}
+O(T^2)
\label{exempleSpm}
\end{multline}
with $N_{y_1y_2}^{y_3}=n(y_1)n(y_2)\bar n(y_3)+\bar n(y_1)\bar n(y_2) n(y_3)$ and $n(y)=1/(1+\eee^y)$. From the original expression \eqref{Spm}, we have eliminated $\pp_4$ using momentum conservation, and switched the radial integration from $p_2$ to $y_2$ using the relation, valid for a function $h(\pp_2)$ peaked about $\pF$:
\begin{equation}\label{basseTchangevariable}
    \int \frac{\textrm{d}^3 \pp_2}{(2\pi)^3}\ h(\pp_2) = \frac{m^*  p_{\rm F}T}{(2\pi)^2}\int^{+\infty}_{-\infty} \textrm{d}y_2  \int \frac{\textrm{d}\Omega_2}{2\pi} h(y_2,\theta_2,\phi_2) +O(T)
\end{equation}
where the solid angle $\textrm{d}\Omega_2 = \sin{\theta_2}\textrm{d}\theta_2\textrm{d}\phi_2$ locates $\pp_2$ in the spherical frame of axis $\pp-\pp'$,
as depicted by Fig.~\ref{fig123}b (with $\pp=\pp_1$, $\pp'=\pp_3$). To leading order in $T$, the resonance
condition is
\be
\epsilon_\pp+\epsilon_{\pp_2}-\epsilon_{\pp'}-\epsilon_{\pp+\pp_2-\pp'}=\frac{2\pF^2\sin(\alpha/2)}{m^*}\bb{\sin\frac{\alpha}{2}+\cos\theta_2} + O(T) \label{resonance}
\ee
and allows us to integrate over $\theta_2$ in \eqqref{exempleSpm}. Recognizing the angles of \eqqref{wsthetaphi}, we
replace $W_{S\pm}(\pp,\pp_2|\pp',\pp-\pp'+\pp_2)$ by $W_{S\pm}(\alpha,\phi_2)$, and there remains to integrate separately
over the energy coordinate $y_2$ and the angle $\phi_2$.
The same calculation for $E$ leads to an expression similar to \eqqref{exempleSpm}
with $\pp_3$ playing to role of $\pp_2$.

We thus obtain the factorized kernels 
\bea
    E_{\pm}(y,y',\alpha) &=& \frac{(m^{\ast})^2 T}{2\pi p_{\rm F}}\mathcal{S}(y,-y')\Omega_{E \pm}(\alpha) +O(T^2)\\
    S_{\pm}(y,y',\alpha) &=& \frac{(m^{\ast})^2 T}{2\pi p_{\rm F}}\mathcal{S}(y,y')\Omega_{S\pm}(\alpha) +O(T^2)
\eea
Here, $\mathcal{S}$ is an energy kernel independent of the collision probabilities and thus universal to all Fermi liquids:
\begin{equation}
    \mathcal{S}(y,y') = \frac{y-y'}{2}\frac{1}{\sinh{\frac{y-y'}{2}}}\frac{\cosh{\frac{y}{2}}}{\cosh{\frac{y'}{2}}}
\end{equation}
The angular kernel $\Omega(\alpha)$ follows from an azimuthal averge of $W$ in the appropriate spherical frame
\bea
\Omega_{E \pm}(\alpha)&=&\int_{0}^{2\pi} \frac{\dd\phi}{2\pi}\frac{W_{E\pm}(\alpha,\phi)}{2|\text{cos}\frac{\alpha}{2}|} \label{OmegaEalpha}\\
\Omega_{S \pm}(\alpha)&=&\int_{0}^{2\pi} \frac{\dd\phi}{2\pi}\frac{W_{S\pm}(\alpha,\phi)}{2|\text{sin}\frac{\alpha}{2}|} \label{OmegaSalpha}
\eea

Changing the summation over $\pp'$ into integrals over $y'$ and $\theta'$, we express the collision integral in \eqqref{Inu} as
\be
\begin{split}
    & I_{\pm}^{\rm lin}(y,\theta)= \\ &\frac{\ii}{\tau}g(y)\Bigg\{ \overline{\Gamma}(y)\hk{\Phi}_{\pm}(y,\theta)  +\int \textrm{d}y'  \frac{\textrm{d}\Omega'}{2\pi}\left(\mathcal{S}(y,-y')\frac{\Omega_{\rm E \pm}(\alpha)}{\Omega_{\Gamma}}-2\mathcal{S}(y,y')\frac{\Omega_{\rm S\pm}(\alpha)}{\Omega_{\Gamma}}\right)\hk{\Phi}_{\pm}(y',\theta')  \Bigg\} \frac{U_\pm}{T}
\end{split}
\ee
We have extracted the typical collision time $\tau$ which gives the order of magnitude of the collision integral
\begin{equation}\label{collisiontime}
    \frac{1}{\tau}  = \frac{(m^*)^3 T^2}{(2\pi)^3} \Omega_\Gamma
\end{equation}
where
\be \label{OmegaGamma}
\Omega_\Gamma=\left\langle \frac{W_{+}(\theta,\phi)}{2\cos{\theta/2}}  \right\rangle_{\theta,\phi} 
\ee 
is the solid-angle average (see \eqqref{anglesolides}) of the spin-averaged collision probability.
Written with $\tau$, Expression \eqref{Gammaexplicite} of $\Gamma$ reads simply\footnote{
This result of Sec.~\ref{sec:lifetime} is recovered here using the number conservation law:
\be
\Gamma(\pp)=\frac{1}{L^3}\sum_{\pp'}S_+(\pp,\pp')=\frac{1}{\tau}\int_{-\infty}^{+\infty} \mathcal{S}(y,y')\dd y' \int_0^\pi\sin\alpha\dd\alpha\Omega_{S+}(\alpha)/\Omega_\Gamma
\ee
The energy integral is straightforward $\int_{-\infty}^{+\infty} \mathcal{S}(y,y')\dd y'=\bar \Gamma(y)/2$. Since $W_{S+}=W_{E+}$, the change of variable \eqqref{anglesEanglesS} shows that $\int_0^\pi\sin\alpha\dd\alpha\Omega_{S+}(\alpha)=2\Omega_\Gamma$.}:
\begin{equation}\label{barGamma}
    \Gamma(\pp) = \frac{1}{\tau}\bar{\Gamma}(y),\qquad \bar{\Gamma}(y) \equiv \pi^2+y^2
\end{equation}

\subsection{Transport equation in the thermal window}

We conclude this section by giving a dimensionless
form of the transport equation \eqref{transportI} in the thermal window.
Assuming a periodic driving $U_{\sigma}(\qq,t) = U_{\sigma}(\qq)\textrm{e}^{-\textrm{i}\omega t}$
and taking the average of \eqref{transportI} in $\hat\varrho=\hat\varrho_{\rm eq}(T)+\delta\hat\varrho(t)$,
we get:
\be
    \left(\omega - \vF q \cos\theta \right){n}_\sigma(\pp)+\vF q \cos\theta\frac{\partial n_{\rm eq}}{\partial \epsilon}\Big\vert_{\epsilon=\epsilon_{\pp}} \left( U_{\sigma}(\qq)+\frac{1}{L^3}\sum_{\pp'\sigma'}f_{\sigma \sigma'}(\pp,\pp'){n}_{\sigma'}(\pp')\right) = \ii {I}^{\rm lin}_{\sigma}(\pp,n) 
\ee
where the quasiparticle distribution $n(\pp,\qq,t)=n(\pp,\qq)\eee^{\ii\omega t}$ is defined by \eqqref{defnpq}.
Inserting the change of variable \eqqref{changmtvar} we obtain the explicit, dimensionless form:
\begin{equation}\label{equlowT}
\begin{split}
       &\frac{\omega}{\omega_0}\nu_{\pm}(y,\theta)-\cos{\theta}\bb{\Phi_{\pm}(y,\theta)-1 }= \\
    &-\frac{\ii}{\omega_0 \tau}\left\{ \overline{\Gamma}(y)\Phi_{\pm}(y,\theta) +\int_{-\infty}^{+\infty} \textrm{d}y'  \frac{\textrm{d}\Omega'}{2\pi}\left(\mathcal{S}(y,-y')\frac{\Omega_{\rm E \pm}(\alpha)}{\Omega_{\Gamma}}-2\mathcal{S}(y,y')\frac{\Omega_{\rm S\pm}(\alpha)}{\Omega_{\Gamma}}\right)\Phi_{\pm}(y',\theta')  \right\} 
    \end{split}
\end{equation}
where 
\be \label{omega0}
\omega_0={v_{\rm F}q}
\ee
is the typical excitation frequency, \hk{and
$\Phi$ is related to $\nu$ by the integral equation,
\be
\Phi_{\pm}(y,\theta)=\nu_{\pm}(y,\theta)+\int_{-\infty}^{+\infty}\dd y' \int  \frac{\dd\Omega'}{4\pi} F^\pm(\alpha)g(y')\nu_\pm(y',\theta')
\label{phinu}
\ee
The angular kernel here is the dimensionless symmetric/anti-symmetric Landau function\footnote{Restricting the definition
of $f_{\sigma\sigma'}$ to $p=p'=\pF$, we write $f_{\sigma\sigma'}(\pp,\pp')=f_{\sigma\sigma'}(\alpha)$ with $\alpha=(\widehat{\pp,\pp'})$.}
\begin{equation} \label{Falpha}
    F^{\pm}(\alpha) = \frac{m^* p_{\rm F}}{2\pi^2} \bb{f_{\uparrow\uparrow}(\cos\alpha)\pm f_{\uparrow \downarrow}(\cos\alpha)}
\end{equation}}

%% file: heff_BCS.tex
\section{Superfluid pairing of Landau quasiparticles}\label{SuperfluidFLT} \label{sec:BCS}

\begin{figure}[htb]
\begin{center}
\includegraphics[width=0.5\textwidth]{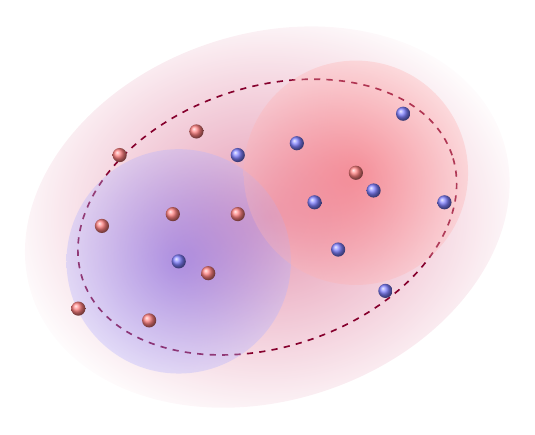}
\vspace{-0.4cm}
\end{center}
\caption{\label{fig:paireqp} Cooper pairs (dashed ellipse) in a Fermi superfluid are pairs of $\upa$ and $\dwa$ Landau quasiparticles (red and blue clouds). To first order,
the spin-$\upa$ quasiparticle  can be seen as a cloud of spin-$\dwa$ particles (blue dots) surrounding the original spin-$\upa$ particle (red dot).}
\end{figure}

In this section, we use the Landau quasiparticles, and their effective Hamiltonian \eqqref{HFS},
to describe the superfluid phase from the superfluid instability down to $T=0$.
This description is rigorous provided superfluidity remains a weak phenomenon in the sense that
\be\label{TcEF}
\Delta,\ T_c\ll \EF
\ee
where $\Delta$ is the superfluid order parameter and $T_c$ is the critical temperature.
In this regime, fermionic superfluids can be viewed as condensates of quasiparticle pairs \cite{Wolfle1990},
schematically depicted by Fig.~\ref{fig:paireqp}; this is a substantial improvement from the pairs of bare particles 
interacting via the bare interaction (as described by BCS theory), or even from the frequent picture
of bare particles interacting via a screened interaction.
The interactions between the pairs $(\pp,-\pp)$ and $(\pp',-\pp')$ of counterpropagating quasiparticles, 
described by the function $g_{\sigma\sigma'}$ 
favor the pairing instability and the appearance of a nonzero pairing field.

To describe superfluidity in the quasiparticle picture we exploit the separation of scales \eqqref{TcEF}
to insert the cutoff $\Lambda$ between the two extremes
\be
\Delta,\ T_c\ll \Lambda \ll \EF
\ee
In this way, $\Lambda$ acts first as infrared cutoff for the calculation
of the effective Hamiltonian (in particular for the regularization of the
pairing interaction $g_{\sigma\sigma'}$), then as a UV cutoff to regularize \textit{e.g.}
the divergence of the gap equation at energies $\gg\Delta$.
The IR and UV (logarithmic) divergences compensate, such that $\Delta$ and $T_c$
 depend only on the regularized function $G_{\sigma\sigma'}$ and no longer on $\Lambda$.
The angular average of $G_{\sigma\sigma'}$ must be negative 
(\textit{i.e.} the quasiparticle pair interactions must be attractive) to ensure the existence of a superfluid phase.

\subsection{Dynamics of the pairing field}
We formulate an evolution 
equation that captures the onset of quasiparticle pairing in the normal phase \cite{Sauls2022}, as the system approaches the critical temperature $T \rightarrow T_{\rm c}^+$.
This equation is to the quasiparticle pairing field $\hat\gamma_{\sigma}\hat\gamma_{\sigma'}$ what the transport equation is to
the density field $\hat \gamma^\dagger_\sigma\hat \gamma_\sigma$. Although pairing is in principle not restricted to the singlet spin wavefunction
(as in \textit{e.g.} the A-phase of ${}^3$He), we have in mind here the case of ultracold fermions, where
the interactions among opposite spin quasiparticles $\mathcal{A}_{\upa\dwa}$ dominate 
and favor the formation of $\upa\dwa$ pairs. We thus restrict to spin-singlet pairs; the corresponding quantum
pairing field in momentum space is
\begin{equation}
    \hat{d}^{\qq}_{\pp} = \hat{\gamma}_{-\pp+\qq/2,\downarrow}\hat{\gamma}_{\pp+\qq/2,\uparrow}
\end{equation}
This operator effectively annihilates a pair of $\pp\upa,-\pp\dwa$ quasiparticles with a center-of-mass momentum $\qq$.
By definition, its expectation value vanishes in an equilibrium state of the normal phase $\langle \hat{d}^{\qq}_{\pp}\rangle_{\rm eq}=0$ for $T>T_c$.
However, fluctuations of $\hat d$ are possible for example under the influence
of an external potential. The pair susceptibility, or pair response function,
then quantifies the magnitude of these fluctuations with respect to the drive intensity.
In accordance with Thouless criterion, we are looking for a divergence of the pair susceptibility, that would signal that the normal phase becomes unstable, and the system undergoes a phase transition.

To compute the pair susceptibility, we introduce an external perturbation $\hat{H}_{\rm ext}$ that couples directly to the pair field:
\begin{equation}
    \hat{H}_{\rm ext} =  \sum_{\pp} \phi(\qq,t) \left(\hat{d}^{\qq}_{\pp}\right)^{\dagger} + \textrm{h.c.}
\end{equation}
where the external pairing source oscillates at frequency $\omega$, $\phi(\qq,t)=\phi(\qq)\eee^{-\ii\omega t}$, causing $\hat d^{\qq}_\pp$ to oscillate at frequency $\omega-2\mu$.
We expand the state of the system about a thermal quasiparticle state $\hat\varrho=\hat\varrho_{\rm eq}(T)+\delta\hat\varrho$ (see
\eqqref{rhoeq} for the definition of $\hat\varrho_{\rm eq}(T)$). Within linear response, the deviation from equilibrium is controlled by the drive intensity $\delta\hat\varrho=O(\phi/\EF)$.
We then evolve $\hat{d}^{\qq}_{\pp}$ according to the Heisenberg equation of motion
\begin{equation}
    \ii\partial_t \hat{d}^{\qq}_{\pp} = [\hat{d}^{\qq}_{\pp},\hat{H}+\hat{H}_{\rm ext}]
\end{equation}

The derivation proceeds analogously to the derivation of the transport equation in Sec.~\ref{sec:demotransport}. The streaming term arises from the diagonal part of the Hamiltonian:
\begin{equation}
    [\hat{d}^{\qq}_{\pp},\hat{H}_2+\hat{H}_4^{\rm d}+\hat{H}_{\rm ext}] = \left(\hat{\epsilon}_{\pp+\qq/2,\uparrow}+\hat{\epsilon}_{-\pp+\qq/2,\downarrow}\right)\hat{d}^{\qq}_{\pp} + \bbcro{1-n^{\rm eq}_{\pp+\qq/2}-n^{\rm eq}_{\pp-\qq/2}}{\phi}(\qq)+O(\phi)^2
\end{equation}
with $\hat{\epsilon}_{\pp,\sigma}={\epsilon}_{\pp}$ to leading order in $T/\TF$ and $\phi/\epsilon_{\rm F}$.
In the quartic terms stemming from $\hat{H}_4^{\rm x}$, we inject the cumulant expansion \eqqref{expcumulant} :
\begin{equation}\label{dH4x}
    [\hat{d}^{\qq}_{\pp},\hat{H}_4^{\rm x}]   = \bbcro{1-n^{\rm eq}_{\pp+\qq/2}-n^{\rm eq}_{\pp-\qq/2}} \frac{1}{L^3}\sum_{\pp'\in\mathcal{D}} \mathcal{A}_{\uparrow\downarrow}^\Lambda\left(\pp+\frac{\qq}{2},-\pp+\frac{\qq}{2}\Big|-\pp'+\frac{\qq}{2},\pp'+\frac{\qq}{2}\right) \hat{d}^{\qq}_{\pp'}+\hat J_{\pp}
\end{equation}
We have regrouped the quartic cumulants $(\hat a\hat b\hat c\hat d)_c$ in a collision integral $\hat J_{\pp}$ which is negligible for the calculation of $T_c$. Note that the interaction between same-spin quasiparticles $\mathcal{A}_{\sigma\sigma}$ contributes to $\hat J_{\pp}$
but not to the partially contracted terms in \eqqref{dH4x}. This is specific to the normal phase where the anomalous averages $\meanv{\hat \gamma_{\dwa} \hat \gamma_{\upa}}_{\rm eq}$ vanish.
The pair transport equation of the Landau quasiparticles is then:
\begin{equation}\label{eq:paircoll}
    \begin{split}
    \left(\omega - \epsilon_{\pp+\qq/2} - \epsilon_{\pp-\qq/2} +2\mu  \right) & \hat{d}^{\qq}_{\pp}   =   \bbcro{1-{n}^{\rm eq}_{\pp+\qq/2}-n^{\rm eq}_{\pp-\qq/2}}\\ & \times \left \{ \frac{1}{L^3}\sum_{\pp'\in\mathcal{D}} \mathcal{A}_{\uparrow\downarrow}\left(\pp+\frac{\qq}{2},-\pp+\frac{\qq}{2}\Big|-\pp'+\frac{\qq}{2},\pp'+\frac{\qq}{2}\right) \hat{d}^{\qq}_{\pp'} + \phi(\qq)\right \}
    \end{split}
\end{equation}

\subsection{Pair susceptibility}

According to Thouless criterion, we expect the pair susceptibility to diverge for $\omega=0$ and $q=0$.
Restricting our pairing equation \eqref{eq:paircoll} first to $q=0$, and taking its average value in $\hat\varrho(t)$, we have: 
\begin{equation}\label{eq:omega0q0}
   \left(\omega -2(\epsilon_{\pp}-\mu)\right) d(\pp) = (1-2n^{\rm eq}_{\pp}) \left \{ \frac{1}{L^3}\sum_{\pp'\in\mathcal{D}}\mathcal{A}_{\uparrow\downarrow}^{\Lambda}(\pp,-\pp|-\pp',\pp')\ d(\pp') + {\phi} \right\}
\end{equation}
where $d(\pp)=\meanv{\hat d_\pp^\zero}$ is the homogeneous pair distribution function. When $p=p'=\pF$, we have
(by definition, see Section \ref{BSsec}) $\mathcal{A}_{\uparrow\downarrow}^\Lambda(\pp,-\pp|-\pp',\pp')=g_{\upa\dwa}(\cos\alpha,\Lambda)$
with $\alpha=(\widehat{\pp,\pp'})$.
We assume that $\mathcal{A}(\pp,-\pp|-\pp',\pp')$ has no sharp variations in the narrow momentum window about $\pF$ where 
pairing effects are sensible, \textit{i.e.} where $d(\pp)$ is appreciably nonzero. However, to regularize the
UV divergence of the momentum integral in \eqqref{eq:omega0q0}, we recall that the off-diagonal elements of our effective
Hamiltonian (including the pairing interaction) are restricted to energy differences lower than $\Lambda$. We thus
replace the amplitude $\mathcal{A}$ using
\begin{equation}
\mathcal{A}_{\uparrow\downarrow}^\Lambda(\pp,-\pp|-\pp',\pp') = g_{\upa\dwa}(\cos\alpha,\Lambda)\Pi_{\Lambda}(2(\epsilon_{\pp'}-\epsilon_{\pp}))
\end{equation}
where $\Pi_\Lambda$ (\eqqref{PiLambdaHRG}) filters out the energy differences larger than $\Lambda$.

When superfluidity occurs in a high partial wave, $d$ may have a non trivial dependence on the angle between $\pp$ and a reference direction; we focus
here on s-wave pairing, for which the pair distribution is isotropic $d(\pp)=d(p)$.
The angular integration in \eqqref{eq:omega0q0} then simply selects the $l=0$ component of $g_{\sigma\sigma'}$:
\begin{equation}
    g_{\upa\dwa}^0(\Lambda) = \frac{1}{2}\int^{\pi}_0 \textrm{d}\alpha \sin{\alpha} \, g_{\upa\dwa}(\cos\alpha,\Lambda)
\end{equation}
For the remaining radial dependence, we introduce a change of variable:
\begin{equation}\label{eq:CdVdtoD}
    d(p) = T\frac{ 1-2n^{\rm eq}_{\pp}}{{\omega -2(\epsilon_{\pp}-\mu)}}D(y)=  \frac{\text{tanh}(y/2)}{\omega/T-2y} \ D(y),\qquad y=\frac{\epsilon_\pp-\mu}{T}
\end{equation}
This reparametrization 
may seem analogous to the change of variable
$\delta n\to \nu$ (see \eqqref{changmtvar}) performed on the density field to focus on the low-energy region.
It extracts a prefactor that depends rapidly on energy from the unknown function $d$, and
we may expect $D$ to be a smooth function of $y$.
Note however that the prefactor $\text{tanh}(y/2)/(\omega/T-2y)$ does not vanish exponentially at large $y$. 

The integral equation on the function $D(y)$ reads\footnote{
We convert the summation over $\pp'$ to an integral over $y$, $\alpha$ and an azimuthal angle $\varphi$ using:
\be
\frac{1}{L^3}\sum_{\pp'} \Pi_{\Lambda}(2(\epsilon_{\pp'}-\epsilon_{\pp})) \to \frac{m^* T \pF}{(2\pi)^3}\int_{-\infty}^{\infty} \Pi_\Lambda\bb{2T(y-y')}  \dd y' \int_{0}^{\pi} \int_{0}^{2\pi}\sin\alpha \dd\alpha \dd\varphi 
\ee}
\begin{equation}\label{Dy}
    D(y) =  4\pi\bar g_{\upa\dwa}^0(\Lambda) \int_{-\frac{\Lambda}{2T}+y}^{\frac{\Lambda}{2T}+y}   \frac{\dd y'}{\omega/T-2y'} \ D(y') \tanh{\frac{y'}{2}}  + {\phi}+O\bb{\frac{T}{\TF}}
\end{equation}
where we recall the nondimensionalization $\bar g_{\upa\dwa}=(m^\ast \pF/(2\pi)^3)g_{\upa\dwa}$. The only remaining dependence on $y$ of the
right-hand side is the integration interval $[-\frac{\Lambda}{2T}+y,\frac{\Lambda}{2T}+y]$ whose centre is shifted from 0 by $y=O(1)$. To leading order
in $T/\Lambda$, we
can then approximate the pair field $D$ by a constant
\begin{equation}
    D(y) = D_0 + O\bb{\frac{yT}{\Lambda}}
\end{equation}

The integral equation is now trivial, and yields the pair susceptibility
\begin{equation}
    \chi_{\rm pair}(\omega,T) \equiv  \frac{D_0}{{\phi}}  = \frac{1}{1+4\pi\bar  g_{\upa\dwa}^0(\Lambda) \mathcal{N}_{\Lambda}(\omega,T)}
\end{equation}
with $\mathcal{N}_{\Lambda}$ defined as:
\begin{equation}
    \mathcal{N}_{\Lambda}(\omega,T) = \int^{\Lambda/2T}_{-\Lambda/2T} \frac{\textrm{d}y'}{2y'-\omega/T} \ \tanh{\frac{y'}{2}}
\end{equation}

\subsection{Critical temperature from Thouless criterion}
The critical temperature is finally determined by applying Thouless’ criterion to the pair susceptibility:
\begin{equation}\label{Thoulesscrit}
   {\chi_{\rm pair}^{-1}(\omega=0,T=T_c)} = 0 \iff 1+4\pi \bar g_{\upa\dwa}^0(\Lambda) \mathcal{N}_{\Lambda}(\omega,T_c)=0
\end{equation}
In the limit where $\Lambda/T \gg 1$, the integral $\mathcal{N}_{\Lambda}(0,T)$ diverges logarithmically 
\begin{equation}
    \mathcal{N}_{\Lambda}(0,T) = \ln{\left(\frac{\Lambda}{\pi T}\right)} + \gamma + O\bb{\frac{T}{\Lambda}}
\end{equation}
where $\gamma \approx 0.577$ is the Euler-Mascheroni constant. 
This divergence exactly compensates the logarithmic suppression of the s-wave pairing amplitude
$g_{\upa\dwa}^0(\Lambda)$ in \eqqref{tildeg}, such that the critical temperature
is independent of the cutoff $\Lambda$, and set by the regularized parameter $G_{\upa\dwa}^0$
\begin{equation}\label{Tcexacte}
    {\frac{T_{\rm c}}{\TF} = \frac{\textrm{e}^{\gamma}}{\pi} \ \textrm{e}^{1/4\pi G_{\upa\dwa}^0}}
\end{equation}
This relation is valid generically in Fermi liquids subject to a weak superfluid instability.
It is non-perturbative and exact if the effective parameter $G_{\upa\dwa}^0$ is known exactly; 
this parameter must be negative, \textit{i.e.} the effective pairing interaction must be attractive,
to trigger the superfluid instability.
It must also remain small in absolute value (\textit{i.e.} the effective pairing interaction
must remain weak) to maintain the validity of the quasiparticle picture,
through the inequality $T_c\ll\TF$.

\subsection{Gap equation at $T=0$}

Extending our low-energy effective theory further into the superfluid phase, we now calculate the order parameter at $T=0$,
through the gap equation:
\be
\Delta(\pp)=-\sum_{\pp'}\mathcal{A}^\Lambda_{\upa\dwa}(\pp,-\pp|-\pp',\pp') \frac{\Delta(\pp')}{2\sqrt{(\epsilon_{\pp'}-\mu)^2+\Delta^2(\pp')}}
\ee
Computing the integral restricted to the low-energy region 
and regularizing $g_{\upa\dwa}$ into $G_{\upa\dwa}$
as before, we obtain
\be \label{Delta}
    \boxed{\frac{\Delta}{\EF}=\textrm{e}^{1/4\pi G_{\upa\dwa}^0}}
\ee
The ratio $\Delta/T_c=\pi/\eee^\gamma\simeq1.764$ found by BCS theory is thus
universal to all superfluids made of Landau quasiparticles \cite{Popov1987-III10}; it is well verified
in superfluid ${}^3$He \cite{Meisel2000}, even though the fluid is strongly interacting ($F_0^+>10$). Deviations from
the BCS ratio (as \textit{e.g.} in a unitary Fermi gas \cite{Ketterle2008,Zwierlein2012}) may then be interpreted
as evidences of a non-Fermi liquid behavior.

%% file: heff_thermalcorr.tex
\section{Thermal corrections to the quasiparticle lifetime} \label{sec:thermalcorr}

In this final section, we extend our effective theory
in the direction opposite to the superfluid phase, that is, towards higher temperatures,
focusing on the calculation of the term of order $T^3$
in quasiparticle lifetime $\Gamma(\pp)$.
This result can been seen as a preliminary to
a wider calculation of thermal corrections to the transport
coefficients (that is, terms of order $T^3$ in $1/\eta $ and $1/D$,
and the term of order $T^2$ in $1/\kappa$).
Knowing the thermal corrections to the transport coefficients should facilitate the comparisons to experimental
results, which are often limited to $T/\TF\gtrsim 0.1$, and therefore
rely on a low-temperature extrapolation for comparisons with Fermi liquid theory \cite{yaleexp}.
Conceptually, this is also an incursion into the edge of the strongly-correlated
regime where the dynamics of the Fermi fluid is described by a BBGKY hierarchy.

The literature on thermal corrections to Fermi liquid theory is vast and often unsettled.
The corrections to the static properties, such as the specific heat or the spin susceptibility
are best understood \cite{BaymPethick,Vojta1997,Maslov2003,Glazman2005,Glazman2005thermo,Maslov2007}.
In 3D, they depend logarithmically on $T$, with \textit{e.g.} a (relative) correction proportional to $T^2 \text{ln}\, T$ to
the specific heat \cite{Engelsberg1966}.
Concerning the thermal corrections to the dynamical properties -- the primary focus of this section --
the literature remains incomplete and contradictory. Pethick and 
coauthors (see \cite{Pethick1969FiniteT,DyPethick1969} and Sec.~1.4.3 in \cite{BaymPethick})
predicted corrections of order $T^3$ to $\Gamma(\pp)$, $1/\eta$ and $1/D$
by considering only ``small-momentum transfers'', \textit{i.e.} processes in which $\qq_{4}=\pp-\pp_4$
is small in \eqqref{GammaCheqtransport}. Their result is however incomplete (by the authors' own admission \cite{Pethick1969FiniteT})
as it is ``inconsistent [with] the weak-coupling problem'' in which the corrections to $\Gamma(\pp)$
``depend on the Fourier transform of the potential for a wave number $2\pF$'',
 \textit{i.e.} on processes where $\qq_{3}=\pp-\pp_3$ is small. Meanwhile,
 Rainwater and Möhling \cite{Mohling1976} obtained thermal corrections to $1/\eta$ and $1/\kappa$ by excluding
the frontal collisions for which $\qq_2=(\pp+\pp_2)/2$ is small.
 Their results applied to the contact Fermi gas also fail to reproduce the thermal
 corrections to $\Gamma(\pp)$ of the weakly-interacting Fermi gas:
 their correction is proportional to $a T^3$, and therefore vanishes in the weakly-interacting
 limit. This surprising conclusion
 contradicts the numerical results
from Ref.~\cite{vtemp}. 

We solve this controversy here by arguing
that none of the three channels (small $q_2$, small $q_3$ and small $q_4$, \textit{i.e.} Hartree, Fock and Bogoliubov)
can be omitted; on the contrary, the thermal correction to $\Gamma(\pp)$ comes from an interaction
between them.

\bigskip

We recall the general expression \eqref{GammaCheqtransport} of the quasiparticle damping rate.
To access the thermal corrections, the strategy of the calculation is to express $\Gamma$ as an integral
over the ``bosonic'' momenta $\qq_4=\pp-\pp_4$, $\qq_2=(\pp+\pp_2)/2$ and
energies $\omega_4=\epsilon_{\pp_4}-\epsilon_{\pp}$ and $\omega_2=\epsilon_\pp+\epsilon_{\pp_2}-2\mu$.
For simplicity, we will assume that $p$ is exactly at the Fermi level, that is:
\be
p=\pF,\qquad \epsilon_\pp=\mu
\ee
We use momentum conservation to integrate over $\pp_3$, and replace
$\dd^3 p_4$ by $\dd^3 q_4$. We then convert the integral
over $p_2$ to an integral over the bosonic energy\footnote{There are corrections of order $O(T)$ to this relation. Their
contribution to $\Gamma$ is however zero due to the quasiparticle-quasihole symmetry} 
$p_2^2\dd p_2=m^\ast \pF \dd\omega_2$, and the polar
integral $\sin\theta_{\pp\pp_2}\dd\theta_{\pp\pp_2}$ by an integral over the bosonic momentum using $q_2\simeq\pF\cos\theta_{\pp\pp_2}$.
We have so far
\begin{multline}
\!\!\!\!\!\!\Gamma(\pp)=\frac{m^\ast  }{4\pi^4\pF}\int_{-\infty}^{\infty}\!\dd \omega_2  \int_0^{\pF}  \! q_2\dd  q_2 \int \!\! \dd^3  q_4 \delta(2\mu+\omega_{2}-\epsilon_{\pp-\qq_4}-\epsilon_{\pp_2+\qq_4})  W_+(\pp,\pp_2|\pp_2+\qq_4,\pp-\qq_4) \\
\bbcro{n_{\rm eq}(\mu+\omega_2)\bar n_{\rm eq}(\epsilon_{\pp-\qq_4}-\mu) \bar n_{\rm eq}(\epsilon_{\pp_2+\qq_4}-\mu)+\bar n_{\rm eq}(\mu+\omega_2) n_{\rm eq}(\epsilon_{\pp-\qq_4}-\mu)  n_{\rm eq}(\epsilon_{\pp_2+\qq_4}-\mu)} \label{Gamma1}
\end{multline}
We introduce the bosonic energy $\omega_4$ through a splitting of the Dirac function $\delta(\epsilon_\pp+\epsilon_{\pp_2}\allowbreak-\epsilon_{\pp_3}-\epsilon_{\pp_4})=\allowbreak\int_{-\infty}^\infty \dd\omega_4\delta(\epsilon_\pp-\epsilon_{\pp_4}-\omega_4)\delta(\epsilon_{\pp_2}-\epsilon_{\pp_3}+\omega_4)$. Since the Fermi-Dirac gain-loss factor restricts all energies to thermal values, we introduce:
\be
\tilde\omega_2=\frac{\omega_2}{T}, \qquad \tilde\omega_4=\frac{\omega_4}{T}, \qquad  \bar T=\frac{T}{\vF\pF}
\ee
where $\bar T$ is the small parameter of the expansion. The bosonized expression of the damping rate reads
\begin{multline}
\Gamma(\pp)=\frac{4 m^\ast T^2 }{(2\pi)^4}\int_{-\infty}^{\infty} \dd\tilde\omega_2 \dd \tilde\omega_4 \int_0^{\pF} q_2\dd q_2  \int_0^\infty q_4^2 \dd q_4 \dd\Omega_{\qq_4} \delta\bb{\epsilon_{\pp_2+\qq_4}-\mu- {\omega_{2}+\omega_{4}}}  
\\\times \delta\bb{\epsilon_{\pp-\qq_4}-\mu- \omega_{4}}  \bbcro{n(\tilde\omega_2)\bar n(\tilde\omega_4) \bar n(\tilde\omega_2-\tilde\omega_4)+\bar n(\tilde\omega_2) n(\tilde\omega_4)  n(\tilde\omega_2-\tilde\omega_4)} W_+(\omega_2,\omega_4,q_2,q_4) \label{Gamma2}
\end{multline}
One can use the Dirac functions to integrate over  the solid angle $\Omega_{\qq_4}$ locating $\qq_4$. The on-shell
collision probability is then a function of the 4 remaining integration variables, \textit{i.e.} the bosonic momenta and energies:
\be
W_+(\pp,\pp_2|\pp_2+\qq_4,\pp-\qq_4)=W_+(\omega_2,\omega_4,q_2,q_4)
\ee

\begin{figure}[htb]
\begin{center}    
    \includegraphics[width=0.75\linewidth]{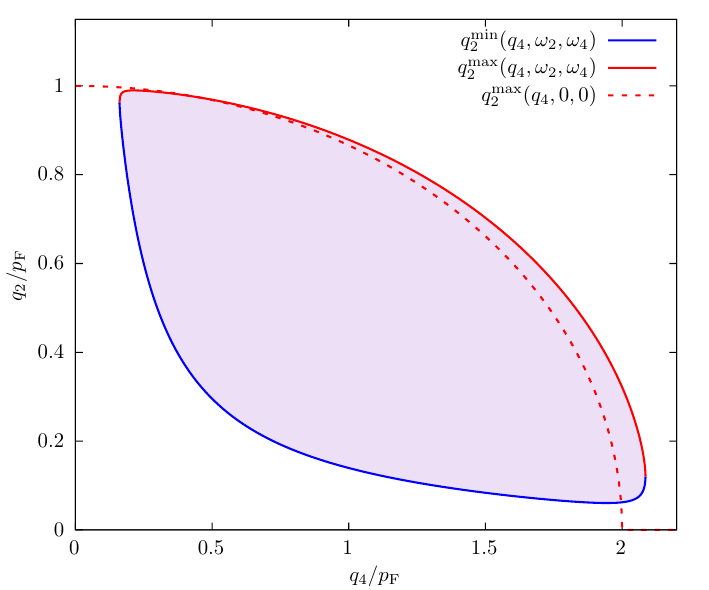} 
    \caption{The integration domain over $q_{2}$ and $q_{4}$ in \eqqref{DeltaWm} 
    is shown as a purple zone between $q_{2}^{\rm min}(q_{4},\omega_{2},\omega_{4})$ (blue solid curve) 
    and $q_{2}^{\rm max}(q_{4},\omega_{2},\omega_{4})$ (red solid curve) for bosonic frequencies
    $\tilde\omega_{2}=1.1$ and $\tilde\omega_{4}=1.5$, and $\bar T=0.1$. For comparison, we show the integration
    domain for $\omega_{2}=\omega_{4}=0$, that is the quarter-ellipse defined by $0<\bar q_{2}<\sqrt{1-\bar q_{4}^2/4}$ (red dashed curve).
    \label{domaineint}}
 \end{center}
\end{figure}

The standard and leading-order result  amounts to setting $\omega_2=\omega_4=0$ in $W$ and in the Dirac functions; 
this decouples the energy integrals (acting only on the Fermi-Dirac gain-loss factor) from the averaging
of the collision probability $W_+$:
\be
W_m\equiv\frac{1}{2\pi} \int_0^{1} \dd\bar q_{2} \int_0^2\dd\bar q_{4} {W_+(0,0,q_{2},q_{4})} \frac{\Theta\bb{f_0(q_2,q_4)}}{\sqrt{f_0(q_2,q_4)}}
\label{Wm}
\ee
(we recall that $\bar q=q/\pF$).
The function
\be
f_0(q_2,q_4)=1-\bar q_2^2-\frac{\bar q_4^2}{4}=1-\frac{||\qq_2+\qq_4/2||^2}{\pF^2}
\ee
makes sure that the vector $\qq_2+\qq_4/2=(\pp+\pp_3)/2$ remains inside
the Fermi sea. The integration domain in the $(q_2,q_4/2)$ plane
is then a quarter-circle (see the dashed red line in  Fig.~\ref{domaineint}).
The average \eqref{Wm} of $W_+$ over the bosonic momenta naturally
coincides with the solid-angle average $\meanvlr{{W_+(\theta,\phi)}/{2\cos(\theta/2)}}_{\theta,\phi}$, such that
we may rewrite the standard result of Fermi liquid theory \eqqref{Gammaexplicite} as
\be
\Gamma_{\rm FL}(\pF)=\frac{(m^\ast)^3 T^2}{8\pi} W_m
\ee

To compute the next-to-leading term, we should in principle compute the dependence of $W_+$ on
$\omega_2$ and $\omega_4$, which become non-negligible (for $\omega_2\approx\omega_4\approx T$)
when $q_2$, $q_3$ or $q_4$ are small.
These dependencies are often viewed as a non-analytic behavior of the collision amplitude;
they can be derived by extending the Bethe-Salpeter equations \eqqref{BS}--\eqref{BetheSalpeterpaire}
to energies slightly off the Fermi surface. Here, we propose a partial calculation
where these non-analyticities are omitted:
\be
W_+(\omega_2,\omega_4,q_2,q_4)\approx W_+(0,0,q_2,q_4)\equiv W_+(q_2,q_4)
\ee
This assumption is valid for example in the weakly-interacting regime,
when the fermions interact through a bare potential, whose variations
for $\omega_2\approx\omega_4\approx T$ are typically negligible.
As we shall see, this simple case is enough to invalidate Refs.~\cite{BaymPethick,Mohling1976}. 

When $\omega_2= T\tilde\omega_2$ and $\omega_4= T\tilde\omega_2$ are reintroduced
in the Dirac functions, this modifies the integration domain in $q_2$ and $q_4$ over which 
$W_+$ is averaged.
In the limit $\bar T\to0$, the largest deformations occur at the corners of the integration domain (see Fig.~\ref{domaineint}),
that is for $q_2q_4\approx m^\ast T$ or $q_2(2\pF-q_4)\approx m^\ast T$. These corners
can be seen as crossed channels: Bogoliubov$\times$Hartree or Bogoliubov$\times$Fock.
This simple observation invalidates the calculation by Rainwater and Möhling (where the frontal
collisions are neglected) \cite{Mohling1976} and Pethick \cite{BaymPethick}
where the thermal corrections are attributed exclusively to the forward (\textit{i.e.} Hartree) channel,
that is, to the region $\vF q_4\approx T$, much narrower than the left corner of Fig.~\ref{domaineint}.
The deviation of the average over the new integration domain from the leading-order average \eqqref{Wm} is
given by
\begin{multline}
\Delta W_m=\frac{1}{2\pi}\int_0^{+\infty}\dd\bar q_{4} \dd\bar q_2\bbcro{\frac{\Theta\bb{f( q_2, q_4, \omega_2, \omega_4)}}{f( q_2, q_4, \omega_2, \omega_4)}-\frac{\Theta\bb{f_0(q_2,q_4)}}{\sqrt{f_0(q_2,q_4)}}}W_+(q_2,q_4)\\=-\frac{\pi}{4}\bbcro{\vphantom{\frac{1}{2}}W_+(0,0)|\tilde\omega_4|+W_+(0,2\pF)\bb{|\tilde\omega_3|-\tilde\omega_2}}\bar T+O(\bar T^2) \label{DeltaWm}
\end{multline}
with $\omega_3=\omega_2-\omega_4$. The function $f$ contains the correction to the integration domain up to $O(T^2)$:
\be
f(q_2,q_4,\omega_2,\omega_4)=f_0(q_2,q_4)+\frac{\bar T \bar q_4^2\tilde \omega_2}{4}+\bar T^2\bbcro{\frac{\tilde \omega_2 }{4\bar q_2^2}\bb{{2\tilde \omega_4}(1-\bar q_2^2)-\tilde\omega_2 \bar q_4^2}-(1- \bar q_2^2)\frac{\tilde \omega_4^2}{\bar q_2^2 \bar q_4^2}}
\ee
To perform the average in \eqqref{DeltaWm} we have assumed that $W_+$ does not vary significantly in the corners of the
integration domain, that is for $\bar q_2, \bar q_4\approx\bar T^{1/2}$ or $\bar q_2, |2-\bar q_4|\approx\bar T^{1/2}$.
This is likely not true in a Fermi liquid due to the logarithmic behavior of the frontal collision amplitude, consequence
of the Bethe-Salpeter equation \eqref{BetheSalpeterpaire}. This is however true in a weakly-interacting
fluid where the collision amplitude coincides with the bare potential, and varies over momentum scales
$\approx \pF$.

The presence of the absolute value of $\omega_3$ and $\omega_4$ in \eqqref{DeltaWm} is essential: it
guarantees that the thermal correction will not vanish under particle-hole symmetry
(\textit{i.e.} under the exchange $\omega\leftrightarrow-\omega$), contrarily to the term $\propto\omega_2$.
Multiplying by the Fermi-Dirac gain-loss factor and integrating over energies, we obtain finally:
\be
{\frac{\Delta\Gamma({{\pF}})}{\Gamma_{\rm FL}(\pF)}=-7\zeta(3)\frac{T}{\vF\pF}\frac{W_+(0,0)+W_+(0,2\pF)}{W_m}+o(T)}
\label{DeltaGamma}
\ee 
where $\Delta\Gamma=\Gamma-\Gamma_{\rm FL}$.
The correction thus depends on the collision probability for  momentum transfers $q_4=0$ and $2\pF$, but
more importantly for center-of-mass momentum $q_2=0$. In a more advanced calculation
where the logarithmic suppression of $W$ for $q_2\to0$ is accounted for, the correction
linear in $T$ should turn into a $T/\text{ln}\, T$.

In the case of an isotropic collision probability ($W_+=$ cte), our result gives 
\be
\frac{\Delta\Gamma({{\pF}})}{\Gamma_{\rm FL}(\pF)}=-14\zeta(3)\frac{T}{\vF\pF}
\ee 
This result is exact (the assumption of smoothly varying collision probability needed for \eqqref{DeltaGamma} is of course valid for an isotropic probability),
and contradicts the result of Pethick et al. \cite{BaymPethick} by a crucial factor 2.

%% file: heff_Conclusion.tex
\label{sec:concl}

Using an hamiltonian renormalisation scheme, we have formulated an intuitive and controlled
construction of the Landau quasiparticles and the effective Hamiltonian 
governing their dynamics.
Instead of the usual momentum cutoff, we introduce an energy cutoff 
$\Lambda$ that separates resonant from off-resonant couplings.
In this framework, we interpret the quasiparticle annihilation operator $\hat \gamma$  
as the bare operator $\hat a$ dressed only by the 
off-resonant couplings. This dressing is implemented through a 
unitary transformation, which becomes a Continuous Unitary Transformation (CUT) in the limit of infinitesimal variations of $\Lambda$.
Truncated to terms quadratic in the fluctuations $\delta(\hat\gamma^\dagger\hat\gamma)$ of the quasiparticle density field about the Fermi sea average, 
our effective Hamiltonian gathers the interaction function $f_{\sigma\sigma'}$ of Fermi liquid theory, 
the BCS pairing amplitude $g_{\sigma\sigma'}$, and the finite-angle collision amplitude 
in a single function $\mathcal{A}$ regularized by $\Lambda$;
it thereby unifies ingredients that are usually treated separately.
Both $f_{\sigma\sigma'}$ and $\mathcal{A}$ reach stationary values for $\Lambda\ll\EF$,
which makes them good effective parameters that can be used to express the physical observables.
Conversely, the pairing amplitude retains a logarithmic dependence on $\Lambda$. We explain
how to regularize this dependence to form the renormalized pairing strength $G_{\sigma\sigma'}$,
which is independent of $\Lambda$.

From the flow equations of the effective Hamiltonian,
we derive key relations between its matrix elements.
In the limits of small scattering angles, the smallness
of either the transferred or the center-of-mass wavenumber $q$
introduces a small characteristic energy $\vF q\ll\EF$, to which the 
renormalization flow is sensitive. In the Hartree channel
(for small transferred momentum), the integration of the
flow from $\Lambda=\vF q$ to $0$ recovers the Bethe-Salpeter
relation \cite{Landau1959} between the forward scattering amplitude 
$\mathcal{A}_{\sigma\sigma'}(\pp,\pp'|\pp',\pp)$ and the interaction
function $f_{\sigma\sigma'}$.
More importantly, we demonstrated a second Bethe-Salpeter
equation by integrating the renormalization flow
in the Bogoliubov channel, that is for small total momentum.
This equation relates the amplitude of frontal 
collisions $\mathcal{A}_{\sigma\sigma'}(\pp,-\pp|\pp',-\pp')$
to the pairing amplitude $g_{\sigma\sigma'}$, through
an integral equation with a logarithmic angular kernel.
To the best of our knowledge, this relation between
was never derived in the context of Fermi liquid theory.

After the renormalization flow, the second side of our description of Fermi fluids is
the study of quantum kinetics restricted to low-energy transitions.
For the quasiparticle density field $\hat\gamma^\dagger\hat\gamma$, this results in a transport equation,
which we carefully derive from the effective Hamiltonian
exploiting the validity of the Born-Markov approximation in the quasiparticle picture.
For the pairing field $\hat \gamma\hat \gamma$, this results in the pair susceptibility;
applying Thouless' criterion, we relate the superfluid critical temperature $T_c$ to the renormalized
pairing strength  $G_{\sigma\sigma'}$.
Finally, we present a calculation of the thermal corrections to the quasiparticle damping rate $\Gamma$.
Solving a longstanding controversy of the literature, we explain that $T^3$ corrections arise from
collisions that are at frontal and forward at the same, \textit{i.e.}, the transferred \textit{and} center-of-mass
momentum of the colliding quasiparticles vanish simultaneously. 

In the second part of this study, we will apply the effective picture 
to an atomic Fermi gas with contact interactions, and show
how switching to Landau quasiparticles systematically 
improves the weak-coupling approximations, in particular the RPA approximation on the speed of zero sound,
and the BCS approximation on the superfluid gap and critical temperature.

%% file: heff_formulaire.tex
\pagebreak
\section{List of symbols} \label{formulaire}
\renewcommand{\arraystretch}{1.2}
{\centering \textbf{\textit{General symbols}} \par}
\begin{tabular}{p{8.6cm}p{9.cm}}
\hline
\hline
$\hbar=k_{\rm B}=1$ & everywhere \tabularnewline
$\mathcal{V}=L^3$ & Volume of the fluid \tabularnewline
$\rho$ & Total equilibrium density \tabularnewline
$\mu$ & Chemical potential \tabularnewline
$m,m^\ast$ & Bare and effective mass \tabularnewline
$\kF=\pF=(3\pi^2\rho)^{1/3}$ & \tabularnewline
$\vF=\pF/m^*$ &  Fermi units \tabularnewline 
$\EF=\TF=\kF^2/2m$ &  \tabularnewline
$a$ or $\bar a=\kF a$ & s-wave scattering length \tabularnewline
$T$ or $\bar T=T/\vF\pF$ & Temperature \tabularnewline
$g=4\pi a/m$ or $\bar g=\frac{m\pF}{(2\pi)^3}g$ & Renormalized coupling constant\tabularnewline
$U_\sigma(\rr,t)$ & Driving field\tabularnewline
$\Lambda$ or $\bar \Lambda=\Lambda/\vF\pF$ & Energy cutoff on the transitions $E_i\to E_f$\tabularnewline
$\omega_0=\vF q$&  Typical frequency at wavenumber $q$ \tabularnewline
$c=\omega/\omega_0$& Reduced phase velocity \tabularnewline
$\mathcal{D}$ & The set of modes of $\hat H_0$ \tabularnewline
\end{tabular}

{\centering \textbf{\textit{Angles}} \par}
\begin{tabular}{p{8.6cm}p{9.cm}}
\hline
\hline
$\theta_{ij}=(\widehat{\pp_i,\pp_j})$ & \tabularnewline
$c_{ij}=\cos\frac{\theta_{ij}}{2}$ & \tabularnewline
$s_{ij}=\sin\frac{\theta_{ij}}{2}$ & \tabularnewline
$\meanv{W}_{\theta\phi}=\int\dd\Omega W(\theta,\phi)/4\pi$ & Solid-angle average 
\end{tabular}

{\centering \textbf{\textit{Quantum formalism}} \par}
\begin{tabular}{p{8.cm}p{9.6cm}}
\hline
\hline
$\meanv{\hat O}=\text{Tr}(\hat\varrho\hat O)$ & Average-value in the state $\hat\varrho$ of the fluid \tabularnewline
$\hat a_{\pp\sigma}$/$\hat\gamma_{\pp\sigma}$ & Particle/quasiparticle annihilation operator \tabularnewline
$\FS$/$\FS_0$ & Particle/quasiparticle Fermi sea \tabularnewline
$\delta(\hat \gamma_{\pp\sigma}^\dagger \hat \gamma_{\pp'\sigma})=\hat \gamma_{\pp\sigma}^\dagger \hat \gamma_{\pp'\sigma}-\delta_{\pp\pp'} n_\pp^0$ & Quantum fluctuations about the Fermi sea \tabularnewline
$n_{\pp\sigma}^{\qq}=\hat{\gamma}_{\pp-\frac{\qq}{2}\sigma}^\dagger \hat{\gamma}_{\pp+\frac{\qq}{2}\sigma}$ \text{ and } $n_{\sigma}(\pp,\qq)=\meanv{\hat n_\pp^{\qq}}$ & Quasiparticle density field
\end{tabular}

\begin{tabular}{p{8.cm}p{9.6cm}}
$\hat{d}^{\qq}_{\pp} = \hat{\gamma}_{-\pp+\frac{\qq}{2},\downarrow}\hat{\gamma}_{\pp+\frac{\qq}{2},\uparrow}$ \text{ and } $d(\pp,\qq)=\meanv{\hat d_\pp^{\qq}}$ & Quasiparticle pair field\tabularnewline
$\hat\epsilon_{\pp\sigma}^{\qq}=\frac{1}{L^3}\sum_{\pp'\sigma'}f_{\sigma\sigma'}(\pp,\pp')\hat n_{\pp'\sigma'}^\qq$ \\ and
$\delta\epsilon_{\sigma}(\pp,\qq)=\meanv{\hat\epsilon_{\pp\sigma}^{\qq}}$& The fluctuation of the quasiparticle energy \tabularnewline
$\delta n_\sigma(\pp,\rr)/\delta \epsilon_\sigma(\pp,\rr)$ & Wigner transform of $n_\sigma(\pp,\qq)$/$\delta \epsilon_\sigma(\pp,\qq)$, see \eqqref{defnpq} \tabularnewline
$\mathcal{P}_\Lambda\bb{\frac{1}{E}}$ & The $\Lambda$-principal part $=\begin{cases}1/E\text{ if }|E|>\Lambda \\ 0 \text{ else} \end{cases}$ \tabularnewline
$\Pi_\Lambda\bb{{E}}$ & $=\begin{cases}E\text{ if }|E|>\Lambda \\ 0 \text{ else} \end{cases}$ \tabularnewline
\end{tabular}

{\centering \textbf{\textit{Effective parameters of Fermi liquids}} \par}
\begin{tabular}{p{8.cm}p{9.6cm}}
\hline
\hline
$\omega_\pp=p^2/2m$  & Kinetic energy \tabularnewline
$\epsilon_\pp$ or $\bar\epsilon_\pp=\epsilon_\pp/\vF\pF$ or $y=(\epsilon_\pp-\mu)/T$  & Quasiparticle eigenenergy \tabularnewline
$\Gamma(\pp)$ or $\bar \Gamma(y)=\tau\Gamma(\pp)=\pi^2+y^2$ & Quasiparticle damping rate \tabularnewline
$f_{\sigma\sigma'}(\pp,\pp')\underset{p=p'=\pF}{=}f_{\sigma\sigma'}(\cos(\widehat{\pp,\pp'}))$ & Landau interaction function  ($\sigma,\sigma'=\upa$ or $\dwa$) \tabularnewline
$F^{\pm}(\alpha) = \frac{m^* p_{\rm F}}{2\pi^2} \bb{f_{\uparrow\uparrow}(\cos\alpha)\pm f_{\uparrow \downarrow}(\cos\alpha)}$ & Dimensionless symmetric/antisymmetric interaction function \tabularnewline
$F_l^\pm$ & The symmetric/antisymmetric Landau parameters \tabularnewline
$g_{\sigma\sigma'}(\pp,\pp',\Lambda)\underset{p=p'=\pF}{=}g_{\sigma\sigma'}(\cos(\widehat{\pp,\pp'}),\Lambda)$ & Pair interaction function \tabularnewline
$G_{\sigma\sigma'}(\cos(\widehat{\pp,\pp'}))$ or $G_{\sigma\sigma'}^l$ & Renormalized pair interaction, see \eqqrefs{gl}{tildeg}\tabularnewline
$\mathcal{B}_{\sigma\sigma'}(\pp_1,\pp_2|\pp_3,\pp_4)$ & Coefficient of $\delta(\hat\gamma_{\pp_1\sigma}^\dagger \hat\gamma_{\pp_4\sigma})\delta(\hat\gamma_{\pp_2\sigma'}^\dagger \hat\gamma_{\pp_3\sigma'}) $ in $\hat H_{\rm eff}$, \eqqref{H4x} \tabularnewline
$\mathcal{A}_{\sigma\sigma'}(\pp_1,\pp_2|\pp_3,\pp_4)$ \newline $=\mathcal{V}\bra{f}\hat\gamma_{\pp_1\sigma}^\dagger \hat\gamma_{\pp_2\sigma'}^\dagger \hat\gamma_{\pp_3\sigma'} \hat\gamma_{\pp_4\sigma}\ket{i}$ & Collision amplitude   \tabularnewline
$\bar f_{\sigma\sigma'}=\frac{m^\ast \pF}{(2\pi)^3} f_{\sigma\sigma'}$, \quad  $\bar g_{\sigma\sigma'}=\frac{m^\ast \pF}{(2\pi)^3} g_{\sigma\sigma'}$  & Dimensionless interaction functions \tabularnewline
$\bar{\mathcal{A}}_{\sigma\sigma'}=\frac{m^\ast \pF}{(2\pi)^3}{\mathcal{A}}_{\sigma\sigma'}$ & Dimensionless amplitudes \tabularnewline
$W_{\sigma\sigma'}=|\mathcal{A}_{\sigma\sigma'}|^2$ & Collision probability \tabularnewline
$W_+=W_{E+}=W_{S+}$ & Spin-symmetric collision probability \newline see \eqqref{Wplus} \tabularnewline
$W_{E\pm}(\theta,\phi)=W_{E\pm}(\pp_1,\pp_2|\pp_3,\pp_4)$ & Particle-particle parametrization of $W$: $\theta=(\widehat{\pp_1,\pp_2})$ and $\phi=(\widehat{\pp_1-\pp_2,\pp_3-\pp_4})$ 
\end{tabular}

\begin{tabular}{p{8.cm}p{9.6cm}}
$W_{S\pm}(\theta,\phi)=W_{S\pm}(\pp_1,\pp_2|\pp_3,\pp_4)$ & Particle-hole parametrization of $W$: $\theta=(\widehat{\pp_1,\pp_3})$ and $\phi=(\widehat{\pp_1+\pp_3,\pp_2+\pp_4})$ \tabularnewline
${1}/{\tau} =\frac{\Gamma({\pF})}{\pi^2}=\bb{\frac{m^\ast}{2\pi}}^3 T^2\meanvlr{\frac{W_+(\theta,\phi)}{\cos\frac{\theta}{2}}}_{\theta,\phi}$ &Mean collision rate\tabularnewline
$1/\tau_\sigma={8ma^2T^2}$ & Mean collision rate of the weakly-interacting Fermi gas ($\tau\underset{\bar a \to0}{=}4\pi\tau_\sigma$) 
\end{tabular}

\pagebreak
{\centering \textbf{\textit{Fermi-Dirac distribution}} \par}
\begin{tabular}{p{8.6cm}p{9.cm}}
\hline
\hline
$n_{\pp}^{0}=\Theta(\pF-p)$ & $T=0$ Fermi-Dirac distribution \tabularnewline
$n_{\pp}^{\rm eq}=n_{\rm eq}(\epsilon_\pp)=\frac{1}{1+\eee^{(\epsilon_{\pp}-\mu)/T}}$ & $T\neq0$ Fermi-Dirac distribution \tabularnewline
$n(y)=\frac{1}{1+\eee^{y}}$& Dimensionless Fermi-Dirac distribution \tabularnewline
$\bar n_\pp=1-n_\pp$, or $\bar n_{\rm eq}=1-n_{\rm eq}$ & Hole occupation number\tabularnewline
$N_{\pp_1\pp_2}^{\pp_3} ={n_{\pp_1}^{\rm eq} {n}_{\pp_2}^{\rm eq}\bar{n}_{\pp_3}^{\rm eq} + \bar{n}_{\pp_1}^{\rm eq}\bar{n}_{\pp_2}^{\rm eq}n_{\pp_3}^{\rm eq}}$ & Fermi-Dirac gain-loss factor\tabularnewline
$g(y)=-T(\dd n_{\rm eq}/\dd\epsilon)=\frac{1}{4\text{ch}^2(y/2)}$ & Density of available states 
\end{tabular}
\vspace{0.4cm}

{\centering \textbf{\textit{Conserved quantities}} \par}
\vspace{0.4cm}

\begin{tabular}{p{8.6cm}p{9.cm}}
\hline
\hline
$\delta\rho_\sigma=\frac{1}{L^3}\sum_{\pp\in\mathcal{D}} n_\sigma(\pp)$ & Fluctuation of the spin $\sigma$ density \tabularnewline
${m} \vv =\frac{1}{\rho L^3}\sum_{\pp\in\mathcal{D}} \pp n_\sigma(\pp) $ & Total velocity of the fluid \tabularnewline
$\delta e =\frac{1}{L^3}\sum_{\pp\in\mathcal{D}} \epsilon_\pp n_\sigma(\pp) $ & Fluctuation of the energy density 
\end{tabular}
\vspace{0.4cm}

{\centering \textbf{\textit{Decomposition over orthogonal polynomials}} \par}
\vspace{0.4cm}

\begin{tabular}{p{9.6cm}p{8.cm}}
\hline
\hline
$n_{+}(\pp,\qq)=n_{\upa}(\pp,\qq)+ n_{\dwa}(\pp,\qq)$ & Quasiparticle density distribution  \tabularnewline
$n_{-}(\pp,\qq)=n_{\upa}(\pp,\qq)- n_{\dwa}(\pp,\qq)$ & Quasiparticle polarisation distribution  \tabularnewline
$\delta\epsilon_{\pm}(\pp,\qq)=\delta\epsilon_{\upa}(\pp,\qq)\pm \delta\epsilon_{\dwa}(\pp,\qq)$ & Spin symmetric/antisymmetric fluctuation of $\epsilon$  \tabularnewline
$P_l$ & Legendre polynomials\tabularnewline
$Q_n$ & Orthogonal polynomials for the energy dependence \tabularnewline
$\nu_\pm(y,\theta)$ & Rescaled quasiparticle distribution, see \eqref{changmtvar} \tabularnewline
$\Phi_\pm(y,\theta)=\nu_\pm(y,\theta)+\int\dd y'g(y')$ $\int\frac{\dd\Omega'}{4\pi} F_\pm(\alpha)\nu_\pm(y',\theta')$ & Rescaling of $n-(\dd n_{\rm eq}/\dd\epsilon)\delta\epsilon$, see \eqref{changmtvar}  \tabularnewline
$\nu_\pm^l(y)$ & Component of $\nu_\pm(y,\theta)$ over $P_l(\cos\theta)$  see \eqref{decomponul} \tabularnewline
$\nu_{n\pm}^l$ & Component of $\nu_\pm(y,\theta)$ over $P_l(\cos\theta)Q_n(y)$  see \eqref{decomp_nunl} \tabularnewline
\end{tabular}
\renewcommand{\arraystretch}{1}

%% file: heff_appflotEN.tex
\section{Flow equations for collisions on the Fermi sphere} \label{app:flot}

Here we present a general form of the flow equation for $\mathcal{A}$,
valid both at finite scattering angles (Section \ref{anglesnonnuls}) and in the narrow-angle
limit (Section \ref{BSsec}).

We consider a generic intermediate process $(\pp_3,\pp_4)\to(\pp_1,\pp_2)$, where two of the $\pp_i$
belong to the external momenta $\alpha$, $\beta$, $\gamma$, $\delta$, while the other two
are intermediate momenta.
We characterize the amplitude of this process by three angles
adapted to each diagram: we choose $\qq_{\alpha\gamma}$, $\qq_{\alpha\delta}$, or $\PP_{\alpha\beta}$ as the $z$ axis of the coordinate system
for the Hartree, Fock, and Bogoliubov diagrams, respectively. 
For each channel x=h, f, or b, we then define a function $\mathcal{A}^{\rm x}$ of two polar angles
$\theta_i^{\rm x}$ and $\theta_j^{\rm x}$, and one azimuthal angle:
\be
\mathcal{A}_{\sigma\sigma'}(\pp_1,\pp_2|\pp_3,\pp_4)=\mathcal{A}_{\sigma\sigma'}^{\rm x}(\theta_i^{\rm x},\theta_j^{\rm x},\phi_j^{\rm x}-\phi_i^{\rm x})
\ee
where $i,j=1,3$ for x=h, and $i,j=1,4$ for x=f or b.
The strict resonance condition \eqqref{resstricte} on the external momenta
translates into the equality of the polar angles:
\be
\epsilon_\alpha+\epsilon_\beta=\epsilon_\gamma+\epsilon_\delta\implies \theta_\alpha^{\rm h}=\theta_\gamma^{\rm h},\quad\theta_\alpha^{\rm f}=\theta_\delta^{\rm f}, \quad\text{and}\quad\theta_\alpha^{\rm b}=\theta_\delta^{\rm b}
\ee
see Eqs. \eqref{thetapi2Hartree} and \eqref{thetapi2Bog}.
By contrast, the azimuthal offset $\phi_i^{\rm x}-\phi_j^{\rm x}$ remains arbitrary at resonance, which is naturally a major difference between the 3D and the 2D cases.
The $\Lambda$-resonance condition in Eq.~\eqref{dAud} (generically $\epsilon_1+\epsilon_2-\epsilon_3-\epsilon_4=\pm\Lambda$)
instead allows the intermediate momenta to deviate in latitude from the external momenta.

We write the dimensionless flow equation
using an operator $\mathcal{L}$:
\be
\frac{\dd\bar{\mathcal{A}}^{\Lambda}_{\sigma\sigma'}}{\dd\bar\Lambda}=\mathcal{L}_{\rm h}[\mathcal{A}_{\sigma\sigma'}^{\Lambda}]+\mathcal{L}_{\rm f}[\mathcal{A}_{\sigma\sigma'}^{\Lambda}]+\mathcal{L}_{\rm b}[\mathcal{A}_{\sigma\sigma'}^{\Lambda}] \label{flotL}
\ee
The operator $\mathcal{L}$ contains the sum over the intermediate spins
$\sigma_a$, $\sigma_b$ and the integration over the intermediate momentum $\pp$,
parameterized by its reduced energy $e$ (\eqqref{energiereduite}), and the angles $\theta$, $\phi$
adapted to each diagram:
\begin{multline}
\mathcal{L}_{\rm x}\bbcro{\mathcal{A}_{\sigma\sigma'}}(\alpha\beta|\gamma\delta)=\sum_{\sigma_a\sigma_b}\int_{0}^{1}\dd e \int\dd\Omega \bar{\mathcal{A}}^{\Lambda,{\rm x}}_{\sigma\sigma_a}(\theta_\alpha^{\rm x},\theta,\phi)(\mathcal{S}^{\rm x})^{\sigma\sigma'}_{\sigma_a\sigma_b}\bar{\mathcal{A}}^{\Lambda,{\rm x}}_{\sigma_b\sigma'}(\theta,\theta_j^{\rm x},\phi_j^{\rm x}-\phi)
\\ \times\Big[\delta\bb{E_{\rm x}(e,\theta,\bar\Lambda)+\bar\Lambda}
+\delta\bb{E_{\rm x}(e,\theta,-\bar\Lambda)-\bar\Lambda}\Big]
\label{Lx}
\end{multline}
(here $j=\gamma$ for x=h and $j=\delta$ for x=f or b). Only the spin structure $\mathcal{S}^{\rm x}$ and the resonance conditions\footnote{We have combined the positive-energy integration domain ($\Lambda>\epsilon_\pp>0$) and the negative-energy domain ($-\Lambda<\epsilon_\pp<0$) through the change of variable
$e\leftrightarrow -e$ together with the property $E_{\rm x}(-e,\theta,\bar\Lambda)=E_{\rm x}(e,\theta,-\bar\Lambda)$.}
$E_{\rm x}$ appearing in the $\delta$ function depend
on the diagram under consideration:
\begin{alignat}{5}
E_{\rm h}&=&{\bar\epsilon_{\pp-\qq_{\alpha\delta}}-\bar\epsilon_\pp}&=2s_{\alpha\delta}&\bbcro{s_{\alpha\delta}-(1+e\bar\Lambda) \cos\theta}&+O(\Lambda^2)\label{EHartree}\\
E_{\rm f}&=&{\bar\epsilon_{\pp-\qq_{\alpha\gamma}}-\bar\epsilon_\pp}&=2s_{\alpha\gamma}&\bbcro{s_{\alpha\gamma}-(1+e\bar\Lambda) \cos\theta}&+O(\Lambda^2) \label{EFock}\\
E_{\rm b}&=&{2\bar\mu-\bar\epsilon_{2\PP_{\alpha\beta}-\pp}-\bar\epsilon_\pp}{}&=2c_{\alpha\beta}&\bbcro{(1+e\bar\Lambda) \cos\theta-c_{\alpha\beta}}&-2e\bar\Lambda+O(\Lambda^2) \label{EBogolioubov}
\end{alignat}

\begin{figure}[htb]
\begin{center}
\includegraphics[width=0.8\textwidth]{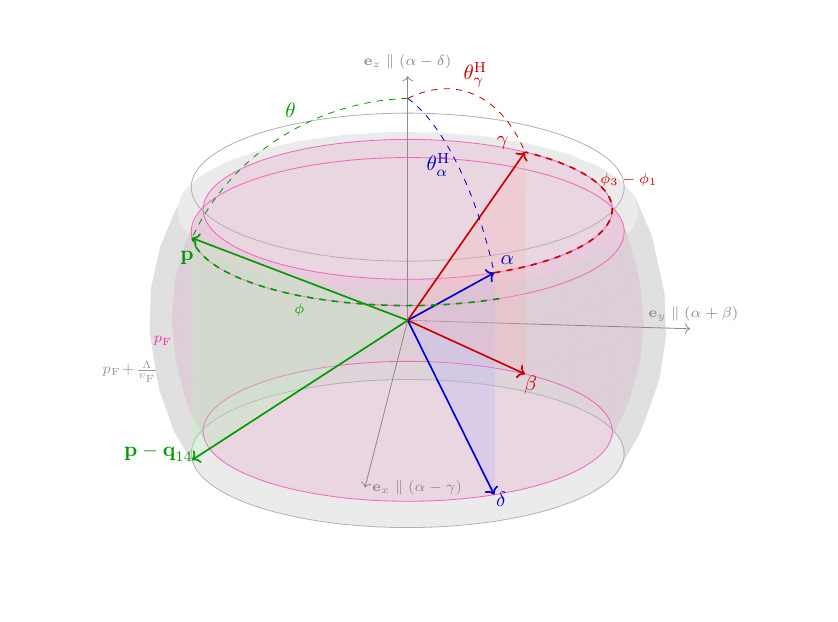}
\end{center}
\caption{Polar coordinate system adapted to the Hartree diagram, with $\textbf{e}_z\parallel \alpha-\delta$. The external momenta lie
on the Fermi sphere (magenta surface). Since they are strictly resonant, they
have the same absolute latitude ($\theta_{\alpha}^{\rm h}=\theta_{\gamma}^{\rm h}=\pi-\theta_{\beta}^{\rm h}=\pi-\theta_{\delta}^{\rm h}$).
The intermediate momenta $\pp$ and $\pp-\qq_{\alpha\delta}$, however, may deviate from the Fermi sphere
up to $\pF+\Lambda/\vF$ (gray surface), which slightly relaxes their latitude. This freedom is generally restricted
to $\theta-\theta_{\alpha}^{\rm h}=O(\Lambda)$, but extends over the entire sphere in the forward-scattering limit $\theta_{\alpha}^{\rm h},\theta_{\delta}^{\rm h}\to\pi/2$. \label{figHartree}}
\end{figure}

\begin{figure}[htb]
\begin{center}
\includegraphics[width=0.8\textwidth]{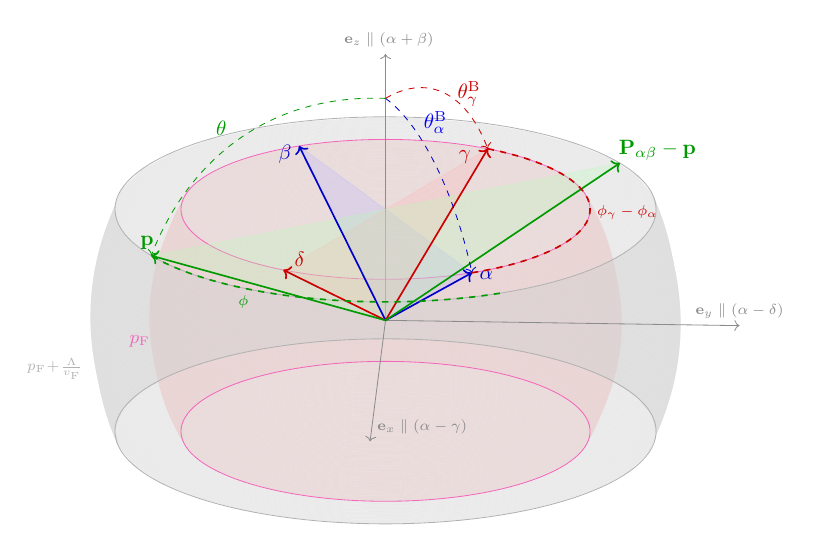}
\end{center}
\caption{Polar coordinate system adapted to the Bogoliubov diagram, with $\textbf{e}_z\parallel \alpha+\beta$. The external momenta lie
on the Fermi sphere (magenta surface) and have the same latitude ($\theta_{\alpha}^{\rm b}=\theta_{\beta}^{\rm b}=\theta_{\gamma}^{\rm b}=\theta_{\delta}^{\rm b}$).
The intermediate momenta $\pp$ and $2\PP_{\alpha\beta}-\pp$, however, may deviate from the Fermi sphere
up to $\pF+\Lambda/\vF$ (gray surface), which slightly relaxes their latitude. This freedom is generally restricted
to $\theta-\theta_{\alpha}^{\rm b}=O(\Lambda)$, but extends over the entire sphere in the forward-scattering limit $\theta_{\alpha}^{\rm b},\theta_{\beta}^{\rm h}\to\pi/2$.
\label{figBogo}}
\end{figure}

In the generic case, that is, when none of the vectors $\alpha$, $\beta$, $\gamma$, or $\delta$ are collinear,
the energy constraint restricts the polar angle $\theta$ to a window of width $\approx\bar\Lambda$.
To leading order in $\Lambda$, the condition $E_{\rm x}=\pm\bar\Lambda$ fixes $\theta=\theta_\alpha^{\rm x}$ (that is,
$\cos\theta=s_{\alpha\delta},\, s_{\alpha\gamma},\, c_{\alpha\beta}$ for the Hartree, Fock, and Bogoliubov diagrams, respectively).
At subleading order, it allows for a small deviation $\theta-\theta_\alpha^{\rm x}=O(\bar\Lambda)$, illustrated
in the three-dimensional Figs.~\ref{figHartree}--\ref{figBogo}. Since the integrations over the reduced energy $e$ and the azimuthal angle $\phi$
extend over $O(1)$ domains, we conclude that the variations of $\mathcal{A}^{\Lambda}$ from $\Lambda_0\ll\EF$ down to $\Lambda_f=0$
remain small, see Eq.~\eqref{varAanglesnonnuls}.

